\documentstyle[12pt]{article}
\newcommand{\mathbf}[1]{\mbox{\boldmath $ #1 $}}
\newcommand{\beq}{\begin{equation}}
\newcommand{\eeq}{\end  {equation}}

\newcommand{\beqar}{\begin{eqnarray}}
\newcommand{\eeqar}{\end  {eqnarray}}

\newcommand{\benum}{\begin{enumerate}}
\newcommand{\eenum}{\end  {enumerate}}

\newcommand{\bfig}{\begin{figure}}
\newcommand{\efig}{\end  {figure}}

\newcommand{\btab}{\begin{table}}
\newcommand{\etab}{\end  {table}}

\newcommand{\bold}[1]{\mbox{\boldmath $#1$}}    

\newcommand{\p}{\bold{p}}       
\newcommand{\q}{\bold{q}}       
\newcommand{\r}{\bold{r}}       

\newcommand{\veC}[1]{\stackrel{\rightarrow}{#1}}        
\newcommand{\Vec}[1]{\stackrel{\leftarrow}{#1}}         

\newcommand{\eg}{{\em e.g.}}    
\newcommand{\etal}{{\em et al.}}
\newcommand{\ie}{{\em i.e.}}    

\newcommand{\MeV}{{\rm MeV}}            
\newcommand{\GeV}{{\rm GeV}}            
\newcommand{\fm}{{\rm fm}}              
\begin{document}
\begin{titlepage}

\noindent{\sl Annals of Physics}\hfill LBL-36359\\[8ex]

\begin{center}
{\large {\bf
Spin-Isospin Modes in Heavy-Ion Collisions\\
       {\normalsize {\bf I: Nuclear Matter at Finite Temperatures}}
$^*$}}\\[8ex]
{\sl Johan Helgesson and J\o rgen Randrup}\\[1ex]
Nuclear Science Division, Lawrence Berkeley Laboratory\\
University of California, Berkeley, California 94720\\[4ex]

December 5, 1994\\[4ex]	
{\sl Abstract:}
\end{center}
With a view towards implementation in microscopic transport simulations
of heavy-ion collisions,
the properties of spin-isospin modes are studied in nuclear matter
consisting of nucleons and $\Delta$ isobars
that interact by the exchange of $\pi$ and $\rho$ mesons.
For a standard $p$-wave interaction
and an effective $g'$ short-range interaction,
the dispersion relations for the spin-isospin modes,
and the associated amplitudes,
are calculated at various nuclear densities and temperatures,
within the random-phase approximation.
Quantities of physical interest are then extracted,
including the total and partial $\Delta$ decay widths
and the $\Delta$ cross sections in the nuclear medium.
The self-consistent inclusion of the $\Delta$ width
has a strong effect on the $\Delta$ cross sections at
twice normal nuclear density,
as compared with the result of ignoring the width.
Generally,
the obtained quantities exhibit a strong density dependence,
but are fairly insensitive to the temperature, at least up to $T=25\ \MeV$.
Finally,
it is described how these in-medium effects may be consistently included
into microscopic transport simulations of nuclear collisions,
and the improvements over previous approaches are discussed.

\vfill
{\small\noindent
$^*$This work was supported by the Swedish Natural Science Research Council,
by the National Institute for Nuclear theory at the University of Washington
in Seattle,
    and by the Director, Office of Energy Research, Office of High Energy
    and Nuclear Physics, Nuclear Physics Division of the U.S. Department of
    Energy under Contract No.\ DE-AC03-76SF00098.
}
\end{titlepage}

\section{Introduction}

In recent years,
large efforts have been devoted to the understanding of hadronic matter
at high densities and temperature,
created in collisions between two heavy nuclei at bombarding energies
from a few hundred MeV per nucleon up to several GeV per nucleon.
A large number of energetic particles are produced in such collisions
and they may be used as probes of the hot and dense phase of the reaction,
since they were not present initially \cite{Metag,Cassing,Mosel}.

Particle production in heavy-ion reactions has been fairly successfully
described by microscopic transport models,
such as BUU and QMD \cite{Cassing,Mosel,Wolf,Aichelin}.
In these transport models,
the nucleons propagate in an effective one-body field
while subject to direct two-body collisions.
Sufficiently energetic nucleon-nucleon collisions may agitate
one or both of the colliding nucleons to a nucleon resonance,
especially $\Delta(1232)$, $N^*(1440)$, and $N^*(1535)$.
Such resonances propagate in their own mean field
and may collide with nucleons or other nucleon resonances as well.
Furthermore, the nucleon resonances may decay by meson emission
and these decay processes constitute the main mechanisms
for the production of energetic mesons \cite{Mosel}.

The by far most important nucleon resonance is the $\Delta(1232)$ isobar.
It contributes substantially to the production of
pions, kaons, anti-protons, and di-leptons,
either directly via its decay,
or indirectly as an intermediate particle \cite{Mosel}.
The $\Delta$ isobar is also very important
for the thermal equilibration of the reaction,
by absorbing the kinetic energies of the nucleons.
The transport descriptions normally employ
the vacuum properties of the resonances and mesons,
\ie\ the needed cross sections, decay widths, and dispersion relations
are taken according to their values in vacuum \cite{Wolf}.
However, it is well known that the pion changes its dispersion relation in the
nuclear medium,
and that the $\Delta$ changes its decay width \cite{EricsonWeise}.
These in-medium properties have proven to be important to incorporate
in the description of various reactions involving-isospin degrees of freedom,
for example $\pi$-nucleus reactions \cite{Oset-Weise,Strottman}
and (p,n) and ($^3$He,t) charge-exchange reactions \cite{Gaarde,Osterfeld}.

It is therefore reasonable to expect in-medium properties of $\pi$ mesons and
$\Delta$ isobars to play an important role in transport descriptions of
heavy-ion collisions.
Some in-medium modifications have already been employed in calculations
of heavy-ion collisions,
both qualitatively \cite{Weise}
and by transport simulations \cite{Bertsch,Giessen,Texas}.
Although the particular properties considered in those studies
may not change much in more refined treatments \cite{Bertsch,Giessen,Texas},
other properties can change significantly.
More refined models have been used in various contexts,
for example the self-consistent coupling of pions and $\Delta$-hole states
in nuclear matter \cite{Siemens}
and the calculation of the $\Delta$ width in nuclear matter \cite{Oset,PaJh}.
The purpose of the present investigation
is to develop and employ such a refined model to derive
several in-medium quantities that are useful for transport models,
and to discuss their meaning and how they can be implemented consistently in
dynamical simulations.

It is important to recognize that the physical scenarios in heavy-ion
collisions differ from those in charge-exchange and $\pi$-nucleus reactions,
and therefore the effects of in-medium properties may differ as well.
Charge-exchange and $\pi$-nucleus reactions occur in a cold medium
at relatively low density (in the surface).
As larger densities are probed in heavy-ion collision,
the in-medium effects are expected to be enhanced.
On the other hand,
there are other mechanisms that may reduce the in-medium effects:
the dense medium is usually hot as well,
many individual baryon-baryon collisions and resonance decays are involved,
and many of the observed mesonic particles
have been created at the surface of the reaction zone
and are thus less sensitive to the dense interior region.

In infinite nuclear matter, a system of interacting $\pi$ mesons, nucleons,
and $\Delta$ isobars will couple to form spin-isospin modes.
Some of these modes are collective and correspond to pion-like states
(quasipions),
while other modes are non-collective in their character.
While a free pion has only a pion component in its wave function,
a collective pionic mode contains components from many
$\Delta N^{-1}$ and $N N^{-1}$ states.
The strength of the various components will vary
with the momentum of the collective mode,
and the energy will differ substantially
from the energy of the unperturbed pion or $\Delta N^{-1}$ states.
On the other hand,
non-collective modes are dominated by a single component
of a  baryon-hole state ($N N^{-1}$ or $\Delta N^{-1}$)
and its energy is close to that
of the corresponding unperturbed baryon-hole state.

Our aim is to calculate the properties of the spin-isospin modes in infinite
nuclear matter at various nuclear densities, $\rho_N$, and temperatures, $T$.
{}From these properties,
we will also deduce a number of other physical quantities,
such as decay widths and cross sections, which will depend on $\rho_N$ and $T$.
In order to obtain a consistent description,
it is important to calculate simultaneously the dispersion relation,
the $\Delta$ width, and the cross sections.
As new decay channels for the $\Delta$ are present in the nuclear medium,
the $\Delta$ width will differ from its vacuum value\cite{Oset,PaJh}.
However, some of these decay channels correspond in a transport description to
$\Delta$-nucleon collisions
and should be excluded from the $\Delta$ width in a transport description.
Instead, it is important to take into account the in-medium properties
for calculating $\Delta$-nucleon cross sections.
It is also important to calculate the dispersion relations
and the $\Delta$ width simultaneously,
because these quantities are interdependent.
We have done this with an iterative procedure.
In a subsequent paper we will present results
from microscopic transport simulations of heavy-ion collision
with the presently calculated in-medium properties having been implemented
by means of a local density approximation,
$\rho_N =  \rho_N(\mathbf{r})$ and $T = T(\mathbf{r})$.

The effects of including the in-medium properties of pions and $\Delta$ isobars
in descriptions of heavy-ion collisions have been studied previously
but only some particular aspect was considered,
such as the $\Delta$ production cross section \cite{Bertsch},
the production of pionic modes \cite{Weise},
or the changed dispersion relations of the pionic modes \cite{Giessen,Texas},
rather than treating all of the aspects consistently within the same model
(dispersion relations, $\Delta$ width, and cross sections),
as is done in the present paper.
Furthermore,
we have improved upon a number of approximations done in previous works.
While the $\Delta$ width was excluded from a number of quantities
in refs.\ \cite{Bertsch,Giessen,Texas},
we have included the $\Delta$ width in a self-consistent manner.\footnote{We
	include the $\Delta$ width by an iterative procedure
	and refer to this as the self-consistent case.
	This nomenclature is not intended to suggest that the entire
	set of coupled equations for the interacting hadons are solved
	self-consistently.
	Such truly self-consistent calculations were recently carried out
	with the $\Delta$ and $\pi$ degrees of freedom included,
	but ignoring the $NN^{-1}$ excitations \cite{Siemens}.}
Some of the previous treatments excluded the nucleon-hole channel and used
a simplified expression for describing the continuum of $\Delta$-hole states
\cite{Bertsch,Giessen,Texas},
and the in-medium properties were calculated in cold nuclear matter
in refs.\ \cite{Bertsch,Weise,Giessen,Texas}.
The avoidance of these approximations has important consequences
for the dispersion relations, the $\Delta$ width, and the cross sections.
This will be further discussed in sec.\ \ref{sec_Res},
where we also will emphasize the importance of treating collective
and non-collective spin-isospin modes in a consistent way within the model
and will compare to previous treatments.

Although our present study constitutes a more consistent way of incorporating
in-medium effects into a dynamical transport description than what has been
done previously,
our treatment does rely on certain approximations.
Thus, we limit our considerations to systems consisting of
interacting nucleons, $\Delta$ isobars, and $\pi$ and $\rho$ mesons,
and we assume that only relatively few $\Delta$ isobars and mesons are present.
In addition to the increasing role played by higher resonances,
there may also be other effects occurring at high densities
that are not taken into account in our model,
such as partial restoration of chiral symmetry.

In sec.\ \ref{sec_Model} we present the model.
While part of the formalism can be found in the literature
(\eg\ ref.\ \cite{EricsonWeise}),
there are also important differences from the traditional formalism.
Therefore, to make the presentation as clear as possible,
we include some steps that can be found elsewhere.
In sec.\ \ref{sec_Param},
we motivate and discuss our choice of parameter values.
The results are then presented in sec.\ \ref{sec_Res}
which also contains a discussion of the implementation of
the in-medium quantities in transport simulations,
and a comparison with previous works.
Finally, our results are summarized in sec.\ \ref{sec_Summary}.

\section{The model}					\label{sec_Model}
We will consider a system of interacting nucleons ($N$), delta isobars
($\Delta$), pi mesons ($\pi$), and rho mesons ($\rho$).
In order to investigate the matter properties of the interacting particles,
we employ a cubic box with side length $L$;
the calculated properties are not sensitive to the actual size,
so we need not take the limit $L\to\infty$ explicitly.

The in-medium properties are obtained by using the Green's function technique,
starting from non-interacting hadrons.
The non-interacting Hamiltonian can be written
\begin{equation}
 H_0 = \sum_k e_k \hat{b}^{\dagger}_k \hat{b}^{ }_k +
   \sum_l \hbar \omega_\pi(\mathbf{q_l})
          \hat{\pi}^{\dagger}_l \hat{\pi}^{ }_l +
   \sum_n \hbar \omega_\rho(\mathbf{q_n})
   \hat{\rho}^{\dagger}_n \hat{\rho}^{ }_n\ .
\label{eq_H0}
\end{equation}
Here the index $k=(\mathbf{p}_k; \, s_k, m_{s_k}; \, t_k, m_{t_k})$
represents the baryon momentum, spin, and isospin.
The spin and isospin quantum numbers, $s_k$ and $t_k$,
take the values $1\over2$ and $3\over2$ for $N$ and $\Delta$, respectively.
The energy of baryon $k$ moving in a (spatially constant) potential
is denoted $e_k$.
The baryon creation and annihilation operators,
$\hat{b}^{\dagger}_k$ and $\hat{b}^{ }_k$,
are normalized such that they satisfy the usual anti-commutation relation,
\begin{equation}
  \{ \hat{b}^{\dagger}_k, \: \hat{b}^{\mbox{ }}_{k'} \} = \delta_{k,k'}\ .
\end{equation}
In the pion part of $H_0$,
the index $l$ represents the pion momentum and isospin,
$l = (\mathbf{p}_l,\lambda_l=0,\pm~1)$,
while the index $n$ in the $\rho$ term also includes the spin $m_{s_n}$.
The meson energies are given by
$\hbar \omega_{\pi,\rho} =  [m_{\pi,\rho}^2 + \mathbf{q}^2]^{1/2}$
respectively, and the creation and annihilation operators of the pion are
normalized such that they satisfy the usual commutation relation,
\begin{equation}
  [ \hat{\pi}^{\dagger}_l, \: \hat{\pi}^{ }_{l'} ] = \delta_{l,l'}\ ,
\end{equation}
and analogously for the $\rho$ meson operators, $\hat{\rho}^{\dagger}_n$ and
$\hat{\rho}^{ }_n$.

Note that the $\Delta$ isobar described by $H_0$ has no decay width,
$\Gamma_\Delta=0$.
When the interactions are turned on,
the $\Delta$ width will emerge
and it will then automatically include also the free width.

\subsection{Basic interactions}
At the $N \pi N$ and $N \pi \Delta$ vertices we will use effective $p$-wave
interactions, $V_{N \pi N}$ and $V_{N \pi \Delta}$, which in the
momentum representation can be written as \cite{OTW}
\begin{eqnarray}
&~&   V_{N \pi N} = i c \: \frac{(\hbar c)^{1\over2}}{L^3} \:
                 \left[ \frac{2m_N c^2}{m_N c^2 + \sqrt{s}} \right]^{1\over2}
                 \frac{f^\pi_{NN}}{m_\pi c^2}\ F_{\pi}(q) \;
                 (\mathbf{\sigma} \mathbf{\cdot} \mathbf{q}_{cm}) \:
                 \veC{\tau} \mathbf{\cdot} \Vec{\phi}_\pi(\q) \label{eq_Vnpn}
                  \\
&~&   V_{N \pi \Delta} =
         i c \: \frac{(\hbar c)^{1\over2}}{L^3} \:
         \left[ \frac{2m_\Delta c^2}{m_\Delta c^2 + \sqrt{s}}\right]^{1\over2}
         \frac{f^\pi_{N\Delta}}{m_\pi c^2}\ F_{\pi}(q) \;
         (\mathbf{S^+} \mathbf{\cdot} \mathbf{q}_{cm}) \:
         \veC{T}^+ \mathbf{\cdot} \Vec{\phi}_\pi(\q)
         + {\rm h.c.} \label{eq_Vnpd}
\end{eqnarray}
In these expressions, $\sqrt{s}$ is the center-of-mass energy in the $N \pi$
system and $\mathbf{q}_{cm}$ is the pion momentum in the $N \pi$ center-of-mass
system, which in the non-relativistic limit is given by
\begin{equation}
  \mathbf{q}_{cm} \approx \frac{m_N c^2}{m_N c^2 + \hbar \omega}\ \mathbf{q}
                  - \frac{\hbar \omega}{m_N c^2 + \hbar \omega}\ \mathbf{p}_N\
,
\label{eq_qcm}
\end{equation}
where $\hbar \omega$ and $\mathbf{q}$ is the pion energy and momentum, and
$\mathbf{p}_N$ is the nucleon momentum in an arbitrary frame.
The Pauli spin
and isospin matrices are denoted $\mathbf{\sigma}$ and $\vec{\tau}$, and
 $\mathbf{S^+}$ and $\vec{T}^+$ are spin and isospin $1\over2$ to $3\over2$
transition operators normalized such that
$<{3\over2},{3\over2}|S^+_{+1}|{1\over2},{1\over2}>=1$.\footnote{For clarity,
        we generally employ bold-face characters to denote quantities with
        vector and tensor properties under ordinary spatial rotations,
        while arrows are employed to indicate the transformation properties
        under rotations in isospace.}
The momentum representation of the pion field is taken as
\begin{equation}
 \phi^\pi_\lambda(\mathbf{q}) =
    \frac{L^{3/2} \hbar c}{\sqrt{2 \hbar \omega_\pi(\mathbf{q})}}
    \left[ \hat{\pi}^{ }_\lambda(\mathbf{q}) +
           (-1)^\lambda  \hat{\pi}_{-\lambda}^\dagger(-\mathbf{q}) \right]\ .
\end{equation}
The interactions contain a monopole form factor,
\begin{equation}
 F_{\pi}(q) = \frac{\Lambda_\pi^2 - (m_\pi c^2)^2}{\Lambda_\pi^2 - (cq)^2}\ ,
\label{eq_Fpi}
\end{equation}
and the coupling constants are determined at $(cq)^2 = (\hbar \omega)^2 -  (c
\, \mathbf{q})^2 = (m_\pi c^2 )^2$ and $\sqrt{s} = m_N c^2$ or  $\sqrt{s} =
m_\Delta c^2$.
Apart from the factor  $2m_B c^2/[\, m_B c^2 + \sqrt{s} \, ]$
the interactions $V_{N \pi N}$ and $V_{N \pi \Delta}$ are obtained in the
non-relativistic limit from the phenomenological Lagrangians \cite{OTW}
\begin{eqnarray}
  {\cal L}_{N \pi N} & = & (\hbar c)^{3/2} \frac{f^\pi_{NN}}{m_\pi c^2}\
        F_{\pi}(q)\ \bar{\psi}_N \gamma^\mu \gamma_5 \veC{\tau} \psi_N
                           \partial_\mu \Vec{\phi}_\pi  \\
  {\cal L}_{N \pi \Delta} & = & (\hbar c)^{3/2}
                  \frac{f^\pi_{N\Delta}}{m_\pi c^2}\ F_{\pi}(q)\
                  \bar{\psi}^\mu_\Delta
                  \veC{T}^+ \psi_N \partial_\mu  \Vec{\phi}_\pi
                  + \mbox{ h.c}\ .
\end{eqnarray}
The factor $2m_B c^2/[\, m_B c^2 + \sqrt{s} \, ]$ is a relativistic correction
that takes into account that in relativistic calculations the energy
denominators usually appear in the form $2m_B c^2/[\, (m_B c^2)^2 - s \, ]$,
while in non-relativistic calculations the form
$[\, m_B c^2 - \sqrt{s}\, ]^{-1}$ usually appears \cite{OTW}.
With this correction,
the correct relativistic form of a Breit-Wigner resonance
is obtained for the free $\Delta$ resonance \cite{Oset-Weise}.

The interactions at $\rho$ meson vertices are less well determined than for
$\pi$ meson vertices. Here we choose a form analogous to eqs.\
(\ref{eq_Vnpn}) and (\ref{eq_Vnpd}),
\begin{eqnarray}
   V_{N \rho N} & = & i c \: \frac{(\hbar c)^{1\over2}}{L^3} \:
                 \left[ \frac{2m_N c^2}{m_N c^2 +\sqrt{s}} \right]^{1\over2}\;
                 \frac{f^\rho_{NN}}{m_\rho c^2} F_{\rho}(q) \;
                 (\mathbf{\sigma} \times \mathbf{q}_{cm}) \mathbf{\cdot} \:
                 [\veC{\tau} \mathbf{\cdot} \Vec{\mathbf{\phi}}_\rho(q)]
                 \label{eq_Vnrn}\\
   V_{N \rho \Delta} & = &
        i c \: \frac{(\hbar c)^{1\over2}}{L^3} \:
        \left[ \frac{2m_\Delta c^2}{m_\Delta c^2 +\sqrt{s}} \right]^{1\over2}\;
        \frac{f^\rho_{N\Delta}}{m_\rho c^2} F_{\rho}(q) \;
        (\mathbf{S^+} \times \mathbf{q}_{cm}) \mathbf{\cdot} \:
        [\veC{T}^+ \mathbf{\cdot} \Vec{\mathbf{\phi}}_\rho(q)]  \nonumber \\
        & + & \mbox{ h.c.} \label{eq_Vnrd}
\end{eqnarray}
Apart from the factor  $2m_B c^2/[\, m_B c^2 +\sqrt{s} \, ]$,
these interactions can also be obtained as the non-relativistic limit
of relativistic Lagrangians \cite{OTW}.

In addition we will also include effective short-range interactions at
nucleon-hole vertices, again written in momentum space,
\begin{equation}
  V_{NN^{-1},NN^{-1}} = \: \left( \frac{\hbar c}{L} \right)^3 \:
                        g_{NN}' \left( \frac{f^\pi_{NN}}{m_\pi} \right)^2
                        |F_g(q)|^2\ (\mathbf{\sigma_1 \cdot \sigma_2})
                        (\veC{\tau_1} \cdot \Vec{\tau_2})\ ,
\label{eq_Vg}
\end{equation}
and the corresponding interactions obtained
when one (or two) of the nucleons is replaced by a $\Delta$.
The strength of the short-range interactions is determined by the
correlation parameters $g_{NN}'$, $g_{N\Delta}'$, and $g_{\Delta \Delta}'$.
If the form factor
\begin{equation}
  F_g(q) = {\Lambda_g^2 \over \Lambda_g^2 - (cq)^2 }
\label{eq_Fg}
\end{equation}
is omitted, this interaction has vanishing spatial range,
$V\sim \delta(\mathbf{r}_1 - \mathbf{r}_2)$.


\subsection{Green's functions}

The interactions between $N$, $\Delta$, $\pi$ and $\rho$ will lead to the
formation of spin-isospin modes which will carry the quantum numbers of
the $\pi$ or $\rho$ mesons. The propagation of these spin-isospin modes can be
represented by an appropriate Green's function.
For this purpose, we define a Green's function
\begin{equation}
  i G(\alpha,\beta;t,t') =
    \frac{1}{\hbar}<\Theta \{ \hat{B}_\alpha^\dagger(t) \:
                                         \hat{B}^{ }_\beta(t') \}>\ ,
\label{eq_GSIM}
\end{equation}
where $\Theta$ is a time ordering operator, and the brackets $<\cdot>$
denote either the expectation value of the interacting ground state
(at zero temperature) or the thermal average (at finite temperatures).
The interacting ground state is an eigenstate of the Hamiltonian, $H$,
which consists of the noninteracting part $H_0$ (given in eq.\ (\ref{eq_H0}))
and an interacting part defined by the interactions in
eqs.\ (\ref{eq_Vnpn}--\ref{eq_Vg}).
The ``channel'' indices $\alpha$ and $\beta$ are used for convenience,
so that the creation operator
\begin{equation}
  \hat{B}_\alpha^\dagger(t) =
     {\rm e}^{iHt/\hbar} \hat{B}_\alpha^\dagger {\rm e}^{-iHt/\hbar}\ ,
\end{equation}
represents either a $\pi$ meson operator,
\begin{equation}
  \hat{B}_\alpha^\dagger =
         \hat{\pi}^\dagger_{\lambda_\alpha}(\mathbf{q}_\alpha)\ ,
         \quad \mbox{ or } \quad
         \hat{\pi}_{-\lambda_\alpha}^{ }(-\mathbf{q}_\alpha)
\label{eq_Bpi}
\end{equation}
a $\rho$ meson operator (defined analogously),
or a two-baryon operator,
\begin{equation}
\hat{B}_\alpha^\dagger = \hat{b}_{k_\alpha}^{\dagger} \hat{b}^{ }_{l_\alpha}\ ,
\label{eq_Bph}
\end{equation}
which in the zero-temperature limit
corresponds to creation of a $NN^{-1}$ or $\Delta N^{-1}$ state.

At zero temperature the Green's function in eq.\ (\ref{eq_GSIM}) can be
written in the energy representation as
\begin{eqnarray}
  i G(\alpha,\beta;\omega) & = &
       \int_{-\infty}^{\infty} d(t-t')\ {\rm e}^{i \omega (t-t')}\
       i G(\alpha,\beta;t,t')  \nonumber \\
 & = & i \sum_\nu \frac{ <\Psi_0| \hat{B}_\alpha^\dagger |\Psi_\nu>
                <\Psi_\nu| \hat{B}_\beta |\Psi_0> }
              {\hbar \omega - \hbar \omega_\nu + i \eta}  \nonumber \\
 & - & i \sum_\nu \frac{ <\Psi_0| \hat{B}_\beta |\Psi_\nu>
                <\Psi_\nu| \hat{B}_\alpha^\dagger |\Psi_0> }
              {\hbar \omega + \hbar \omega_\nu - i \eta}\  \nonumber \\
 & \approx & i \sum_\nu \frac{ X_\alpha^\nu (X_\beta^\nu)^* }
              {\hbar \omega - \hbar \omega_\nu + i \eta} -
       i \sum_\nu \frac{ (X_\beta^\nu)^T (X_\alpha^\nu)^\dagger }
              {\hbar \omega + \hbar \omega_\nu - i \eta}\ ,
\label{eq_Gabw}
\end{eqnarray}
where, as is usual,
we have approximated the excited states  $|\Psi_\nu>$ by excited
RPA states,
$|\Psi_\nu>~\approx~Q_\nu^\dagger~|\Psi_0>$,
generated by a generalized RPA operator,
\begin{equation}
  Q_\nu^\dagger = \sum_\alpha X_\alpha^\nu \hat{B}_\alpha^\dagger\ ,
\end{equation}
and approximated the matrix elements by
\begin{equation}
<\Psi_0| \hat{B}_\alpha^\dagger |\Psi_\nu>\
=\ <\Psi_0| [\hat{B}_\alpha^\dagger,Q_\nu^\dagger] |\Psi_0>\
\approx\ <\Phi_0| [\hat{B}_\alpha^\dagger,Q_\nu^\dagger] |\Phi_0>\ ,
\label{eq_0nuApp}
\end{equation}
with $|\Phi_0>$ denoting the non-interacting ground state.

\subsection{RPA approximation}
We want to calculate the spin-isospin mode Green's function defined in equation
(\ref{eq_GSIM}) within the RPA approximation, symbolically
\begin{equation}
  G^{\rm RPA}(\alpha,\beta;\omega) = G_{0}'(\alpha,\beta;\omega) +
     \sum_{\gamma,\kappa} G_{0}'(\alpha,\gamma;\omega)\
                          {\cal V}(\gamma,\kappa;\omega)\
                          G^{\rm RPA}(\kappa,\beta;\omega)\ .
\label{eq_GRPA}
\end{equation}
The spin-isospin modes, here represented by the Green's function $G^{\rm RPA}$,
will in this approximation be obtained as an infinite iteration of
(non-interacting) pion, $\rho$ meson, nucleon-hole, and $\Delta$-hole states,
represented by the diagonal Green's function $G_0'$, coupled with the
interactions specified in eqs.\ (\ref{eq_Vnpn}--\ref{eq_Vg}) which here
are summarized by the symbolical interaction ${\cal V}$.
$G^{\rm RPA}$ is graphically illustrated in fig.\ \ref{fig_GsimGraph}.
As the lowest-order Green's function we will,
in the case $\alpha$ corresponds to a meson index,
take $G_0' = G^{\pi,\rho}_+$ or  $G_0' = G^{\pi,\rho}_-$, with
\begin{eqnarray}
  G^{\pi,\rho}_+(\alpha,\beta;\omega) & = & \phantom{-} \left[
     \frac{1+n_{\pi,\rho}} {\hbar \omega - \hbar \omega_{\pi,\rho} + i \eta} +
     \frac{n_{\pi,\rho}} {\hbar \omega + \hbar \omega_{\pi,\rho} + i \eta}
  \right] \delta_{\alpha\beta} \label{eq_Gpipl} \\
  G^{\pi,\rho}_-(\alpha,\beta;\omega) & = & - \left[
     \frac{1+n_{\pi,\rho}} {\hbar \omega + \hbar \omega_{\pi,\rho} - i \eta} +
     \frac{n_{\pi,\rho}} {\hbar \omega - \hbar \omega_{\pi,\rho} - i \eta}
  \right] \delta_{\alpha\beta} \label{eq_Gpimi} \\
\end{eqnarray}
Note that the free $\pi$ or $\rho$ meson propagator, $D^{\pi,\rho}_0$,
is related to $G^{\pi,\rho}_+$ and $G^{\pi,\rho}_-$ by
\begin{equation}
  D_0^{\pi,\rho}(\alpha,\beta)  =
    \frac{1}{2 \hbar \omega_{\pi,\rho}(\mathbf{q_\alpha})}
    \left[ G^{\pi,\rho}_+(\alpha,\beta) +
           G^{\pi,\rho}_-(\alpha,\beta) \right]\ .
\label{eq_Dpi0}
\end{equation}

In nuclear collisions at beam energies up to about one GeV per nucleon,
which is the domain of application that we have in mind,
only relatively few mesons and isobars are produced
and so the associated quantum-statistical effects may be ignored.
Accordingly,
we assume $n_\Delta\approx0$, $n_\pi\approx0$, and $n_\rho\approx0$.
Since we consider thermal equilibrium at a specified temperature $T$,
the nucleon occupation probabilities are
\begin{equation}
n_N(k) = \frac{1}{ 1 + {\rm e}^{(e_k-\mu)/T} }\ .
\label{eq_n}
\end{equation}

In the case $\alpha$ corresponds to a two-baryon index we take
$G_0' = G^{N N^{-1}}$ or $G_0' = G^{\Delta N^{-1}}$, with
\begin{eqnarray}
  G^{N N^{-1}}(\alpha,\beta;\omega) & = &  \left[
     \frac{n_N(k_\alpha) - n_N(l_\alpha) }
              {\hbar \omega - e_{l_\alpha} + e_{k_\alpha} +
               i \eta \cdot \mbox{sign}(\omega)}
  \right] \delta_{\alpha\beta} \label{eq_GNhT} \\
  G^{\Delta N^{-1}}(\alpha,\beta;\omega) & = &
     \frac{n_N(k_\alpha)  }
              {\hbar \omega - e_{l_\alpha} + e_{k_\alpha} -
               \Sigma_{\Delta N^{-1} }(e_{k_\alpha}+\omega,l_\alpha) + i \eta}\
     \delta_{\alpha\beta}  \nonumber \\ & - &
     \frac{ n_N(l_\alpha)  }
              {\hbar \omega + e_{k_\alpha} - e_{l_\alpha}
              + \Sigma_{\Delta N^{-1} }(e_{l_\alpha}-\omega,k_\alpha) - i
\eta}\
     \delta_{\alpha\beta}\ ,
\label{eq_GDhT}
\end{eqnarray}
The Green's function $G^{\Delta N^{-1}}$ has been calculated from
\begin{eqnarray}
&~&     G^{\Delta N^{-1}}(\alpha,\beta;\omega)\ =\\ \nonumber
&~&     \delta_{\alpha\beta} \: \frac{1}{i} \:
     \int_{-\infty}^{\infty} \frac{dE}{2\pi}
        \left[ G^\Delta(l_\alpha,E+\omega)\ G_0^N(k_\alpha,E)
        +G_0^N(l_\alpha,E+\omega)\ G^\Delta(k_\alpha,E) \right]\ ,
\label{eq_GDh0}
\end{eqnarray}
where $G^\Delta$ is the full in-medium Green's function for the $\Delta$,
containing the $\Delta$ self energy $\Sigma_\Delta$.
We note that the quantity $\Sigma_{\Delta
N^{-1} }$ in eq.\ (\ref{eq_GDhT}) is identical to the $\Delta$ self-energy
$\Sigma_\Delta$ when $G^{\Delta N^{-1}}$ is calculated from
eq.\ (\ref{eq_GDh0}).
However, as a first approximation when calculating $G^{\rm RPA}$,
the quantity $\Sigma_{\Delta N^{-1} }$ in eq.\ (\ref{eq_GDhT}) can be
ignored. The $G^{\rm RPA}$ obtained with this approximation can then in turn
be used to calculate the $\Delta$ self-energy. It is therefore convenient to
use different notations in order to distinguish between the input
$\Sigma_{\Delta N^{-1}}$ in eq.\ (\ref{eq_GDhT}) and the calculated
$\Delta$ self-energy $\Sigma_\Delta$ using $G^{\rm RPA}$.

We will obtain two different, but equivalent, expressions for the Green's
function $G^{\rm RPA}$, as defined by eq.\ (\ref{eq_GRPA}). The first
expression is based on a summation of the Green's functions in eqs.\
(\ref{eq_Dpi0}--\ref{eq_GDhT}), according to the diagrams in fig.\
\ref{fig_GsimGraph}. The details of this derivation can be found in the
literature, for example in refs.\ \cite{Oset,PaJh}, so in this paper we will
only state the final expression in sec.\ \ref{sec_DeltaW}. This expression is
useful for calculating quantities like the total $\Delta$ width and different
cross sections, involving a $\Delta$ isobar, and will be used in section
\ref{sec_DeltaW}.

The second expression is based on an expansion in RPA eigenstates. For this
purpose we will derive a set of RPA equations,
equivalent to eq.\ (\ref{eq_GRPA}).
We will solve these equations to obtain eigenvectors and
eigenenergies for the different spin-isospin modes.
The eigenvectors will yield the amplitudes of the different components
($\pi$, $\rho$, $NN^{-1}$, $\Delta N^{-1}$)
forming the particular spin-isospin mode with the given eigenenergy.
We will show that the eigenvectors form a complete orthogonal set,
and we will expand the Green's function
$G^{\rm RPA}(\alpha,\beta;\omega)$ on this set.
This expression for $G^{\rm
RPA}$ will be useful for calculating partial contributions to the total
$\Delta$ width. Furthermore the RPA amplitudes of the different components will
contain important information about the nature of the different spin-isospin
modes. This will be further discussed in sec.\ \ref{sec_ResSIM}.

\subsubsection{Interactions and operators}
It is convenient to rewrite the total Hamiltonian
specified by eqs.\ (\ref{eq_H0}--\ref{eq_Vg}) in the form
\begin{eqnarray}
  H & = & H_0' + V^{(3)} + V^{(4)}\ ,\\
  H_0' & = & \sum_k E_k\ \hat{b}^{\dagger}_k \hat{b}^{ }_k +
            \sum_k \hbar \omega_\pi(\mathbf{p_k})\
         \hat{\pi}^{\dagger}_k \hat{\pi}^{ }_k +
            \sum_k \hbar \omega_\pi(\mathbf{p_k})\
         \hat{\rho}^{\dagger}_k \hat{\rho}^{ }_k   \nonumber \\
&~&     +   \frac{1}{4} \sum_{jkj'k'} \bar{v}_{jkj'k'}^{\Delta N^{-1}}\
\hat{b}^{\dagger}_j\hat{b}^{\dagger}_k\hat{b}^{ }_{k'}\hat{b}^{ }_{j'} \ ,\\
  V^{(3)} & = &
\sum_{jkj'}v^{\it l}_{jkj'}\ \hat{b}^{\dagger}_j\hat{\pi}^{ }_k\hat{b}^{
}_{j'}+
\sum_{jkj'}v^{\it t}_{jkj'}\ \hat{b}^{\dagger}_j\hat{\rho}^{ }_k\hat{b}^{
}_{j'}
+             \mbox{h.c. }\ , \\
  V^{(4)} & = &
            \frac{1}{4} \sum_{jkj'k'} \bar{v}_{jkj'k'}\
  \hat{b}^{\dagger}_j \hat{b}^{\dagger}_k \hat{b}^{ }_{k'} \hat{b}^{ }_{j'}
              \nonumber \\ & = &
            \frac{1}{4} \sum_{jkj'k'} \bar{v}_{jkj'k'}^{\it l}\
  \hat{b}^{\dagger}_j \hat{b}^{\dagger}_k \hat{b}^{ }_{k'} \hat{b}^{ }_{j'}
      +     \frac{1}{4} \sum_{jkj'k'} \bar{v}_{jkj'k'}^{\it t}\
  \hat{b}^{\dagger}_j \hat{b}^{\dagger}_k \hat{b}^{ }_{k'} \hat{b}^{ }_{j'}\ .
\end{eqnarray}
The interaction $V^{(3)}$ corresponds to the baryon-meson-baryon vertices in
eqs.\ (\ref{eq_Vnpn}), (\ref{eq_Vnpd}), (\ref{eq_Vnrn}), and (\ref{eq_Vnrd}).
In the spin-isospin summation the spin-longitudinal and the spin-transverse
channels are orthogonal and do not mix.
The interaction $V^{(4)}$
contains the effective short-range $g'$ interaction (\ref{eq_Vg})
and is separated into a spin-longitudinal part and a spin-transverse part.
In addition to these interactions,
we also include an effective interaction, $v^{\Delta N^{-1}}_{jkj' k'}$,
which for the $\Delta N^{-1}$ states includes the $\Delta$ self energy in our
RPA formalism in a way analogous to eqs.\ (\ref{eq_GRPA}) and (\ref{eq_GDhT}).
In the above expressions, $\bar{v}_{jkj'k'} = v_{jkj'k'} -v_{jkk'j'}$
is formally an anti-symmetrized two-body matrix element,
but the exchange term $v_{jkk'j'}$ will be neglected in the calculations.

The spin-isospin modes (or excited RPA states) $|\Psi_\nu>$ are created by an
operator $Q^{\dagger}_\nu$.
For specified values of
their momentum $\mathbf{q}$, isospin $\lambda$, and spin $\mu$
the associated energy can be obtained as a solution to the RPA equations.
In the spin-longitudinal channel the spin is zero,
while the spin-transverse channel has spin 1,
with two non-vanishing contributions, $\mu=\pm1$,
for the spin-projection along the $\hat{\mathbf{q}}$-axis.
We take
\begin{equation}
  Q^{\dagger}_{\it l}(\mathbf{q},\lambda) =
        \sum_{jk} X^{\it l}_{jk}(\mathbf{q},\lambda)
        \hat{b}^{\dagger}_j  \hat{b}^{ }_k +
        \sum_{k}  Z^{\it l}_k(\mathbf{q},\lambda) \, \hat{\pi}_k^\dagger -
        \sum_{k}  W^{\it l}_k(\mathbf{q},\lambda) \, \hat{\pi}^{ }_k
\label{eq_Qrpal}
\end{equation}
and
\begin{equation}
  Q^{\dagger}_{\it t}(\mathbf{q},\lambda,\mu) =
        \sum_{jk} X^{\it t}_{jk}(\mathbf{q},\lambda,\mu)
        \hat{b}^{\dagger}_j  \hat{b}^{ }_k +
        \sum_{k}  Z^{\it t}_k(\mathbf{q},\lambda,\mu) \, \hat{\rho_k}^\dagger -
        \sum_{k}  W^{\it t}_k(\mathbf{q},\lambda,\mu) \, \hat{\rho}^{ }_k\ .
\label{eq_Qrpat}
\end{equation}
In appendix \ref{sec_RPAsolu} we will restrict the summation over baryon and
meson states in eqs.\ (\ref{eq_Qrpal}) and (\ref{eq_Qrpat}), by taking
$X_{jk} \propto \delta_{\mathbf{p}_j, \mathbf{p}_k + \mathbf{q}}$,
$Z_k \propto \delta_{\mathbf{p}_k, \mathbf{q}} \delta_{\lambda_k, \lambda}$,
and
$W_k \propto \delta_{\mathbf{p}_k, -\mathbf{q}} \delta_{\lambda_k, -\lambda}$.

The RPA equations are obtained from the relation
\begin{equation}
  <[\delta Q,[H,Q^{\dagger}]]> = \hbar \omega <[\delta Q,Q^{\dagger}]>\ ,
\label{eq_RPA-gen}
\end{equation}
with
$\delta Q=\hat{b}^{\dagger}_k\hat{b}^{ }_j$, $\hat{\pi}_r$,
$\hat{\pi}_r^\dagger$, $\hat{\rho}_r$, or $\hat{\rho}_r^\dagger$,
and where the brackets $<\cdot>$ as previously denote the thermal average
(which at temperature zero becomes the expectation value in the interacting
ground state).

\subsubsection{RPA equations}
\label{sec_RPAeq}

The general structure of the RPA equations is similar in the spin-longitudinal
and spin-transverse channels. We will therefore not write out the symbols
${\it l}$ and ${\it t}$ in this section. Calculating the necessary commutation
relations using $<\hat{b}^{\dagger}_k \hat{b}^{ }_k > = n(k)$,
we obtain
\begin{equation}
 \left( \begin{array}{ccc}
            A^{(1)} +   A^{(2)}   & {\cal N}C & - {\cal N}C^* \\
            C^\dagger  & D & 0   \\
            -C^T       & 0 & D   \\
        \end{array} \right)
 \left( \begin{array}{c}
            {\cal N} X^\nu \\ Z^\nu \\ W^\nu \\
        \end{array} \right) = \hbar \omega_\nu
 \left( \begin{array}{c}
            {\cal N} X^\nu \\ Z^\nu \\ - W^\nu \\
        \end{array} \right)\ .
\label{eq_RPAMx}
\end{equation}
with
\begin{eqnarray}
  A^{(1)}_{j  k j' k'} & = &  [e_{j'}- e_{k'} + \Sigma_{j'k'}]
                               \delta_{j'j } \delta_{k'k} \\
  A^{(2)}_{j  k j' k'} & = &  [n(k) - n(j )]
                               v^{{\it l},{\it t}}_{j  k' k j'}  \\
  C_{j k r}            & = &  v_{j r k} \\
  D_{r r'}             & = &  \hbar \omega_\pi(\mathbf{q_r}) \delta_{rr'} \\
  {\cal N}_{j k j' k'} & = &  [n(k') - n(j')] \delta_{j'j } \delta_{k'k}\ ,
\end{eqnarray}
where the matrix $C$ has the properties $C_{j k r} = C_{k j r}$ and
$C_{j k r} = - C_{j k r}^*$ after the spin-isospin summations have been
performed.
Because we have
\begin{equation}
  \Sigma_{jk} =
     \left\{ \begin{array}{ll}
   0
&  (t_k = {1\over2},\ t_j = {1\over2})
\\
   \Sigma_{\Delta N^{-1}}(e_k+\hbar \omega, \mathbf{p}_k + \mathbf{q})
&  (t_k = {1\over2},\ t_j = {3\over2})\ ,
\\
   - \Sigma_{\Delta N^{-1}}(e_j-\hbar \omega, \mathbf{p}_j - \mathbf{q})
&  (t_k = {3\over2},\ t_j = {1\over2})
     \end{array} \right.
\end{equation}
the matrix $A^{(1)}$ is non-Hermitian
and so the usual RPA orthonormality relations do not hold.
In order to construct an orthonormal set,
we will use the solution of the equation
obtained by replacing $A^{(1)}$ by $(A^{(1)})^*$ .
This corresponds to employing a bi-orthonormal set.
The solutions of the auxiliary equation,
as well as other quantities related to the auxiliary equation,
will be denoted by a tilde symbol,
\begin{equation}
 \left( \begin{array}{ccc}
           \tilde{X}^\nu\ , & \tilde{Z}^\nu\ , & \tilde{W}^\nu
        \end{array} \right)^T\ .
\end{equation}

The RPA equations have the property that if $\omega_\nu$ is a solution
of eq.\ (\ref{eq_RPAMx}) then $\tilde{\omega}_\nu =\omega_\nu^*$
is a solution of the auxiliary equation.
The equations (\ref{eq_RPAMx}) also have the property that if
$({\cal N}X^\nu, \mbox{ } Z^\nu ,\mbox{ } W^\nu)^T$
is a solution of (\ref{eq_RPAMx}) with the eigenvalue $\omega_\nu$,
then
\begin{equation}
  \left( \begin{array}{c}
          [{\cal N} X^\mu]_{kl}(\omega_\mu) \\
          Z^\mu(\omega_\mu) \\ W^\mu(\omega_\mu)
  \end{array} \right) =
  \left( \begin{array}{c}
          [{\cal N} X^\nu]_{lk}(\omega_\nu) \\  W^\nu(\omega_\nu) \\
          Z^\nu(\omega_\nu)
  \end{array} \right)
\label{eq_omNegSol}
\end{equation}
is a solution of (\ref{eq_RPAMx}) with the eigenvalue $\omega_\mu=-\omega_\nu$.

Taking into account the strong $\omega$ dependence of $\Sigma_{\Delta N^{-1}}$
in the matrix $A^{(1)}$, but neglecting other weak $\omega$ dependences
due to form factors and relativistic corrections,
we obtain the orthonormality relation
\begin{equation}
  \delta_{\nu\mu}\mbox{ sign(Re}\ \omega_\nu) =
 \left( \begin{array}{ccc}
      \tilde{X}^\mu, & \tilde{Z}^\mu, & \tilde{W}^\mu
        \end{array} \right)^\dagger
 \left( \begin{array}{ccc}
            {\cal N}  & 0 &  0   \\
                0     & 1 &  0   \\
                0     & 0 &  1   \\
        \end{array} \right)
 \left( \begin{array}{ccc}
                \eta^{\mu,\nu}  & 0 &  0   \\
                0               & 1 &  0   \\
                0               & 0 &  1   \\
        \end{array} \right)
 \left( \begin{array}{c}
          X^\nu \\ Z^\nu \\ W^\nu
        \end{array} \right) \ .
\label{eq_Normz}
\end{equation}
Here the $\omega$ dependence of
$\Sigma_{\Delta N^{-1}}$ is contained in the factor
$\eta^{\mu,\nu}_{jk,j'k'}(t_j t_k, t_{j'} t_{k'})$,
which has the following non-vanishing elements
\begin{eqnarray}
      \eta^{\mu,\nu}_{jk,jk}(\frac{1}{2} \frac{1}{2},
                             \frac{1}{2}  \frac{1}{2})
& = &
        1\ ,
\\
      \eta^{\mu,\nu}_{jk,jk}(\frac{3}{2} \frac{1}{2},
                             \frac{3}{2}  \frac{1}{2} )
& = &
      1 - \frac{  \Sigma_{\Delta N^{-1}}(e_k+\hbar \omega_\mu, \mathbf{p}_j)
       - \Sigma_{\Delta N^{-1}}(e_{k}+\hbar \omega_\nu,\mathbf{p}_{j}) }
            { \hbar \tilde{\omega}_\mu^* - \hbar \omega_\nu }\ ,
\\
      \eta^{\mu,\nu}_{jk,jk}(\frac{1}{2} \frac{3}{2},
                             \frac{1}{2} \frac{3}{2})
& = &
      1 - \frac{ \Sigma_{\Delta N^{-1}}(e_j-\hbar \omega_\mu, \mathbf{p}_k)
       - \Sigma_{\Delta N^{-1}}(e_{j}-\hbar \omega_\nu,\mathbf{p}_{k}) }
            {\hbar \tilde{\omega}_\mu^* - \hbar \omega_\nu}\ .
\end{eqnarray}
While the $\omega$ dependence of
$\Gamma_{\Delta N^{-1}} = -2 {\rm Im}\ \Sigma_{\Delta N^{-1}}$ is strong,
the  $\omega$ dependence of {\rm Re}\ $\Sigma_{\Delta N^{-1}}$ is weak
and can be neglected.
It is convenient to take the quantity $\eta^{\mu,\nu}$ in the approximate form
$\eta^{\mu,\nu}_{jk,jk} \approx [\tilde{\eta}^\mu_{jk}]^*\eta^\nu_{jk}$ with
\begin{equation}
  \eta^\nu_{jk} =
     \left\{ \begin{array}{ll}
   1
&
   t_k = t_j = {1\over2}
\\
   \left[ 1 + i \partial
          \Gamma_{\Delta N^{-1}}(e_k+\hbar \omega_\nu,
                                 \mathbf{p}_k + \mathbf{q})
          / 2 \partial \hbar \omega \right]^{1/2}
&  t_k = {3\over2}, t_j = {1\over2}
\\
   \left[ 1 + i \partial
          \Gamma_{\Delta N^{-1}}(e_j-\hbar \omega_\nu,
                                 \mathbf{p}_j - \mathbf{q})
          / 2 \partial \hbar \omega \right]^{1/2}
&  t_k = {1\over2}, t_j = {3\over2}
     \end{array} \right. \ ,
\label{eq_etaApp}
\end{equation}
and $\tilde{\eta}^\mu_{jk} = (\eta^\mu_{jk})^*$.
With this approximation,
the resolution of unity takes the form
\begin{equation}
   1 = \sum_\nu \mbox{sign(Re } \omega_\nu)
       \left( \begin{array}{c} \eta^\nu X^\nu \\ Z^\nu \\ W^\nu \end{array}
       \right)
       \left( \begin{array}{ccc}
            [\tilde{\eta}^\nu \tilde{X}^\nu]^\dagger {\cal N}\ , &
            [\tilde{Z}^\nu]^\dagger\ , &
          - [\tilde{W}^\nu]^\dagger \end{array} \right)\ ,
\label{eq_ComplRel}
\end{equation}
and an arbitrary vector $F$ can be expanded as
\begin{equation}
   F = \sum_\nu f_\nu
       \left( \begin{array}{c} \eta^\nu X^\nu \\ Z^\nu \\ W^\nu \end{array}
       \right)\ .
\end{equation}
We have checked numerically that the orthonormality and completeness relations,
(\ref{eq_Normz}) and (\ref{eq_ComplRel}),
with the approximation (\ref{eq_etaApp}),
are satisfied within an error less than 3-4\%.

It is convenient to express the two-body Green's function as
\begin{equation}
  G^{\rm RPA}(\omega)= \left( \begin{array}{ccc}
            {\cal N}  & 0 &  0   \\
                0     & 1 &  0   \\
                0     & 0 &  1   \\
        \end{array} \right)
  G(\omega)
 \left( \begin{array}{ccc}
            {\cal N}  & 0 &  0   \\
                0     & 1 &  0   \\
                0     & 0 &  1   \\
        \end{array} \right)\ ,
\end{equation}
where $G(\omega)$ can be expanded on the RPA states as
\begin{equation}
  G(\omega) = \sum_{\nu \mu}
              \left( \begin{array}{c}
                     \eta^\nu X^\nu \\ Z^\nu \\ W^\nu \end{array}
          \right)
g^{\nu\mu}(\omega)
      \left( \begin{array}{ccc}
          [\tilde{\eta}^\mu \tilde{X}^\mu]^\dagger\ ,      &
          [\tilde{Z}^\mu]^\dagger\ ,                     &
          [\tilde{W}^\mu]^\dagger
      \end{array} \right)\ .
\end{equation}
When $\eta^\nu$ is unity\footnote{This is generally the case
        when $\Gamma_{\Delta N^{-1}}$ is independent of $\omega$,
        and in particular when $\Gamma_{\Delta N^{-1}}$ vanishes identically,
        as in our reference case.}
we can use eq.\ (\ref{eq_GRPA}) to determine the solution,
\begin{equation}
  g^{\nu\mu}(\omega) = \delta_{\nu,\mu}\
        \frac{ \mbox{sign(Re } \omega_\nu) }
             {\hbar \omega - \hbar \omega_\nu +
              i \delta \mbox{ sign(Re } \omega_\nu) }\ .
\end{equation}
After rewriting the expansion of $G^{\rm RPA}$
as a sum over positive frequencies by using eq.\ (\ref{eq_omNegSol}),
we arrive at the following expression for $G^{\rm RPA}$,
\begin{equation}
  G^{\rm RPA}(\alpha,\beta;\omega) =  \sum_{ {\rm Re}\ \omega_\nu > 0}
   \left[
   \frac{y_>^\nu(\omega_\nu)_\alpha \tilde{y}_>^\nu(\omega_\nu)_\beta^* }
        {\hbar \omega - \hbar \omega_\nu + i \eta }
    -    \frac{y_<^\nu(\omega_\nu)_\alpha \tilde{y}_<^\nu(\omega_\nu)_\beta^* }
         {\hbar\omega + \hbar \omega_\nu - i \eta } \right] \ ,
\label{eq_GrpaEx}
\end{equation}
where $y_>$ and $y_<$ are short notations for
\begin{equation}
   y_>^\nu(\omega_\nu) \equiv  \left( \begin{array}{c}
         [{\cal N} X^\nu(\omega_\nu)]_{k_\alpha, l_\alpha} \\
         Z^\nu(\omega_\nu) \\ W^\nu(\omega_\nu) \end{array}
     \right)
   \qquad \mbox{and } \qquad
   y_<^\nu(\omega_\nu) \equiv  \left( \begin{array}{c}
         [{\cal N} X^\nu(\omega_\nu)]_{l_\alpha, k_\alpha} \\
         W^\nu(\omega_\nu) \\
         Z^\nu(\omega_\nu) \end{array}
     \right)\ .
\end{equation}
At $T=0$ this is the same expansion as in eq.\ (\ref{eq_Gabw}),
with the approximation in eq.\ (\ref{eq_0nuApp}).
However, the expansion in eq.\ (\ref{eq_GrpaEx})
also holds at finite temperatures.

However, when $\eta^\nu$ is taken from eq.\ (\ref{eq_etaApp}),
with $\partial \Gamma_{\Delta N^{-1}}/\partial \hbar \omega \neq 0$,
the expansion
of $G^{\rm RPA}$ will no longer be diagonal in the channels $\nu, \mu$,
\ie\ $g^{\nu, \mu}(\omega) \neq g^\nu(\omega) \delta_{\nu\mu}$.
Therefore, in the present paper,
the RPA expansion of $G^{\rm RPA}$ will only be used when
$\Gamma_{\Delta N^{-1}}$ vanishes.

It is straightforward to obtain the explicit solution of the RPA equations for
the interaction specified by eqs.\ (\ref{eq_Vnpn}--\ref{eq_Vg}),
but the notation is somewhat tedious. We therefore present the essential steps
together with the final expressions in appendix \ref{sec_RPAsolu}.

\subsection{The $\Delta$ self energy and $\Delta$ cross sections}
\label{sec_DeltaW}
In this section we will derive expressions for the $\Delta$ self energy
$\Sigma_\Delta$ and the cross sections $\sigma(1+2 \to 3+4)$ where 1--4
represent baryons. These quantities can be expressed by an effective
interaction that is obtained by inserting $G^{\rm RPA}$ between two-baryon
states, as illustrated in fig.\ \ref{fig_EffInt}. We will denote this
effective  interaction with $M(34,12)$. It is convenient to write the
spin-isospin matrix elements as a separate factor and define the
quantity $\bar{M}^{l,t}(34,12)$ by
\begin{equation}
     M^{l,t}(34,12) =
        \vartheta^{l,t}(31) (\vartheta^{l,t}(24))^*
        \bar{M}^{l,t}(34,12)\ ,
\end{equation}
where the spin-isospin matrix elements, with
1=2=4=$N$ and 3=$\Delta$, in the spin-longitudinal ($l$) and
the spin-transverse ($t$) channels are written
\begin{eqnarray}
  \vartheta^l(31) (\vartheta^l(24))^* & = &
      <m_{s_3}|(\mathbf{S}^+ \mathbf{\cdot} \mathbf{q}_{31})|m_{s_1}>
      <m_{s_4}|(\mathbf{\sigma} \mathbf{\cdot} \mathbf{q}_{24})|m_{s_2}>
    \nonumber \\ & &
      <m_{t_3}|\veC{T}^+|m_{t_1}> \mathbf{\cdot} <m_{t_4}|\Vec{\tau}|m_{t_2}>
\label{eq_spinMxl}
    \\
  \vartheta^t(31) (\vartheta^t(24))^*  & = &
      <m_{s_3}|(\mathbf{S}^+ \times \mathbf{q}_{31})|m_{s_1}>  \mathbf{\cdot}
       <m_{s_4}|(\mathbf{\sigma} \times \mathbf{q}_{24})|m_{s_2}>
    \nonumber \\ & &
      <m_{t_3}|\veC{T}^+|m_{t_1}> \mathbf{\cdot}<m_{t_4}|\Vec{\tau}|m_{t_2}> \
{}.
\label{eq_spinMxt}
\end{eqnarray}

When $G^{\rm RPA}$ is expressed as a sum of non-interacting Green's functions,
the effective spin-isospin interaction $\bar{M}(34,12)$ becomes
\begin{eqnarray}
  \bar{M}^l(34,12) & = &
    \frac{ f^\pi_{31} f^\pi_{24} }{(m_\pi c^2)^2}
    \left[ D_\pi F_\pi^2 q^2_{\rm eff}(34,12) + F_g^2 g'_{\rm eff}(34,12)
    \right]
\label{eq_MbarlDia} \\
  \bar{M}^t(34,12)  & = &
    \frac{ f^\rho_{31} f^\rho_{24} }{(m_\rho c^2)^2}
    D_\rho F_\rho^2 q^2_{\rm eff}(34,12) +
    \frac{ f^\pi_{31} f^\pi_{24} }{(m_\pi c^2)^2} F_g^2 g'_{\rm eff}(34,12) \ .
\label{eq_MbartDia}
\end{eqnarray}
The expression of the quantities $D_{\pi,\rho}$, $q^2_{\rm eff}(34,12)$ and
$g'_{\rm eff}(34,12)$ are somewhat lengthy and are
therefore given in appendix \ref{sec_AppEffI}.

Alternatively $\bar{M}(34,12)$ can also be expressed using the RPA expansion in
eq.\ (\ref{eq_GrpaEx}),
\begin{eqnarray}
     \bar{M}^{l,t}(34,12) & = &
     \sum_{{\rm Re}\ \omega_\nu^{l,t} > 0}
     \left\{ \frac{ h^{l,t}(31;\nu) h^{l,t}(24;\nu) }
                  { \hbar \omega - \hbar \omega_\nu^{l,t} + i \eta }
       -     \frac{ h^{l,t}(31;\nu)  h^{l,t}(24;\nu) }
                  { \hbar \omega + \hbar \omega_\nu^{l,t} - i \eta }
       \right\} \nonumber \\
&~& +\  \frac{ f^\pi_{31} f^\pi_{24} }{(m_\pi c^2)^2} F_g^2 g'_{34,12} \ .
\label{eq_Mrpa}
\end{eqnarray}
The factor $h(jk,\nu)$ is obtained from the interactions at the vertex
consisting of baryons $j$ and $k$, and the spin-isospin mode $\nu$.
The interactions to be used depend on the non-interacting states
that the mode is made up of and must therefore be multiplied
by the amplitude of the corresponding state.
For example,
\begin{equation}
  h^l(jk,\nu) \vartheta^l(jk) = \frac{V_{j \pi k}}{\sqrt{2 \hbar \omega_\pi}}
                \, [Z^l(\nu)+W^l(\nu)]                     +
                \sum_{mn}  V_{jk,mn} \, X^l_{mn}(\nu) \ ,
\label{eq_hMotiv}
\end{equation}
where $V_{j \pi k}$ is defined in (\ref{eq_Vnpn}) and
(\ref{eq_Vnpd}), $V_{jk,mn}$ is  defined in (\ref{eq_Vg}), and
the amplitudes $X^l_{mn}$, $Z^l$, $W^l$ are defined in eq.\ (\ref{eq_Qrpal}).
{}From appendix \ref{sec_RPAsolu} we obtain the amplitudes
$(X_{mn},Z,W$) as the solution of the RPA equations.
These expressions are also somewhat lengthy and they are therefore
relegated to appendix \ref{sec_AppEffI}.

\subsubsection{The total $\Delta$ width}

The $\Delta$ self energy $\Sigma_\Delta$ is calculated according to the
diagrams in fig.\ \ref{fig_DseGraph}, by  taking into account all the diagrams
corresponding to the $\Delta$ decaying into a spin-isospin mode and a nucleon,
which then again form a $\Delta$. Since the spin-longitudinal and
spin-transverse channels are orthogonal we can treat them separately.
The spin-isospin summation gives,
for a spin average over the external $\Delta$ spin states,
a factor $1/3$ in the spin-longitudinal channel,
and a factor $2/3$ in the transverse channel.

The contribution to the $\Delta$ self energy in the spin-longitudinal channel
can be expressed as (see also ref.\ \cite{PaJh,Oset}),
\begin{equation}
 \Sigma^l_\Delta(E_\Delta,\mathbf{p}_\Delta)  =  \frac{i}{3}
     \left( \frac{\hbar c}{L} \right)^3
     \sum_{\mathbf{q}} \int_{-\infty}^{\infty} \frac{d \hbar \omega}{2 \pi} \:
     \bar{M}^l(\Delta N,N \Delta) \;
     G^N(E_\Delta- \hbar \omega,\mathbf{p}_\Delta-\mathbf{q})   \ .
\label{eq_DSE1}
\end{equation}
By writing the nucleon propagator $G_N$ as
\begin{eqnarray}
  G^N(E,\mathbf{p}_N) & = &
    \frac{ 1 - n_N(\mathbf{p}_N)}{E - e_N(\mathbf{p}_N) + i \eta}
    +  \frac{n_N(\mathbf{p}_N)}{E - e_N(\mathbf{p}_N) - i \eta}
    \nonumber \\  & = &
    \frac{1}{E - e_N(\mathbf{p}_N) + i \eta} +
    2 \pi i n_N(\mathbf{p}_N) \delta(E - e_N(\mathbf{p}_N)) \ ,
\label{eq_GN}
\end{eqnarray}
we can carry out the $\omega$ integration in (\ref{eq_DSE1}) by
performing a Wick-rotation \cite{Oset,Oset2}, and we obtain the expressions for
the $\Delta$ width, $\Gamma_\Delta = - 2 \mbox{Im } \Sigma_\Delta$, as
\begin{equation}
 \Gamma^l_\Delta(E_\Delta,\mathbf{p}_\Delta) = \mbox{Im } \frac{2}{3}
     \left( \frac{\hbar c}{L} \right)^3
     \sum_{\mathbf{q}} \,
     \left[ \Theta({\cal E}) - n(\mathbf{p}_\Delta-\mathbf{q}) \right]
     \bar{M}^l(\Delta N,N \Delta)\ .
\label{eq_DSE2}
\end{equation}
where the energy available for the spin-isospin mode is given by
\begin{equation}
   {\cal E} = E_\Delta - e_N(\mathbf{p}_\Delta-\mathbf{q})\ .
\end{equation}
In the same way we obtain the contribution to $\Gamma_\Delta$ from the
transverse channel,
\begin{equation}
 \Gamma^t_\Delta(E_\Delta,\mathbf{p}_\Delta)  = \mbox{Im } \frac{4}{3}
     \left( \frac{\hbar c}{L} \right)^3
     \sum_{\mathbf{q}} \,
     \left[ \Theta({\cal E}) - n(\mathbf{p}_\Delta-\mathbf{q}) \right]
     \bar{M}^t(\Delta N,N \Delta)\ .
\label{eq_DSEtr}
\end{equation}

\subsubsection{Specific $\Delta$ channels}

The total $\Delta$ width, $\Gamma_\Delta = \Gamma_\Delta^l + \Gamma_\Delta^t$,
gives the transition probability per unit time for the $\Delta$ resonance to
decay to any of its decay channels.
In a transport description one explicitly
allows the $\Delta$ resonance to decay into specific final particles.
Consequently, one needs not only the total $\Delta$ width
(which is the sum of all decay channels)
but also the partial widths governing the decay into specific RPA channels.
These decay channels consists
of a nucleon and one of the spin-isospin modes.
Since we have access to all the amplitudes of a given spin-isopsin mode
on the different unperturbed states,
it is possible to derive an expression for the partial contribution to
$\Gamma_\Delta$ from the $\Delta$ decay to a specific mode $\nu$.
The right-hand side of fig.\ \ref{fig_DseGraph}
shows a diagrammatic representation of such a process.
The partial $\Delta$ width for a $\Delta$ decay to a nucleon
and a spin-longitudinal mode $\nu$ becomes
\begin{eqnarray}
  \Gamma_\Delta^\nu(E_\Delta,\mathbf{p}_\Delta)
    & = & \int \frac{d^3p_N}{(2\pi)^3} \,
          \frac{d^3q}{(2\pi)^3} \;
          | \frac{V_{\Delta \pi N}}{\sqrt{2 \hbar \omega_\pi}}
            \cdot [Z^l(\nu)+W^l(\nu)] +
            \sum_{mn}  V_{\Delta N,mn} \cdot X^l_{mn}(\nu)|^2
    \nonumber \\ & ~& \times
            \bar{n}_N(\mathbf{p}_N) \,
            (2\pi)^3 \delta( \mathbf{p}_\Delta - \mathbf{p}_N - \mathbf{q} )
            2\pi \delta( E_\Delta - e_N - \hbar \omega ) \nonumber \\
     & = &   {\cal M}_\Gamma \int \frac{d^3q}{(2\pi)^3} \:
            | h^l(\Delta N,\omega_\nu) |^2 \,
            \bar{n}_N(\mathbf{p}_\Delta - \mathbf{q}) \,
            2\pi \delta( E_\Delta - e_N - \hbar \omega_\nu )
\label{eq_GnuReal}
\end{eqnarray}
where $\bar{n}_N = 1-n_N$, and the spin-isospin summation gives a factor ${\cal
M}_\Gamma = 1/3$ in the spin longitudinal channel and 2/3 in the
spin-transverse.
This expression is identical to the contribution from one of
the $\nu$ terms in eq.\ (\ref{eq_DSE2}), if the RPA form (\ref{eq_Mrpa}) is
used for $\bar{M}(34,12)$ and Im $\omega_\nu = 0$.
If the $\Gamma_{\Delta N^{-1}}$ is taken to be zero in the calculation
of the  spin-isospin modes the energies $\omega_\nu$ will be real.
In this case the summation over modes $\nu$
in eq.\ (\ref{eq_Mrpa}) corresponds to a summation over physical decay
modes, $\Delta \rightarrow N + \nu$,
when the $\Delta$ width is calculated from (\ref{eq_DSE2}).

When the $\Delta$ width is included self-consistently in the calculation of the
spin-isospin modes the energies $\omega_\nu$ will be complex,
Im $\omega_\nu < 0$.
This implies that the energy of the spin-isospin mode $\nu$ no
longer is distinct, but instead has a Breit-Wigner-like distribution
with a width 2 Im $\hbar \omega_\nu$ centered around Re $\hbar \omega_\nu$.
To obtain the partial decay width to the mode $\nu$, $\Gamma_\Delta^\nu$,
we therefore  need to ``sum'' over all possible energies of the mode $\nu$,
using that the probability to find the mode $\nu$
in the energy range between $e$ and $e + de$ is given by the factor
\begin{equation}
    \frac{1}{\pi} \: \frac{ \mbox{Im } \hbar \omega_\nu }
                          { (e - \mbox{ Re } \hbar \omega_\nu)^2 +
                            (\mbox{Im } \hbar \omega_\nu)^2 }\ de \ .
\end{equation}
The expression for the partial $\Delta$ width is thus modified to
\begin{eqnarray}
  \Gamma_\Delta^\nu(E_\Delta,\mathbf{p}_\Delta)
     & = &   {\cal M}_\Gamma \int \frac{d^3q}{(2\pi)^3} \: de \;
            | h^l(\Delta N,e) |^2 \, [1-n_N(\mathbf{p}_\Delta - \mathbf{q})]
          \nonumber \\
    & \times &
    \frac{1}{\pi} \: \frac{ \mbox{Im } \hbar \omega_\nu }
                          { (e - \mbox{ Re } \hbar \omega_\nu)^2 +
                            (\mbox{Im } \hbar \omega_\nu)^2 } \,
            2\pi \delta( E_\Delta - e_N - e ) \nonumber \\
     & = & 2  {\cal M}_\Gamma \int \frac{d^3q}{(2\pi)^3} \:
            | h^l(\Delta N,e) |^2 \, [1-n_N(\mathbf{p}_\Delta - \mathbf{q})]
       \nonumber \\
    & \times & \left.
        \frac{ \mbox{Im } \hbar \omega_\nu }
                          { (e - \mbox{ Re } \hbar \omega_\nu)^2 +
                            (\mbox{Im } \hbar \omega_\nu)^2 }
             \right|_{e = E_\Delta - e_N} \ .
\label{eq_GnuCompl}
\end{eqnarray}
Unfortunately the expression
(\ref{eq_GnuCompl}) cannot be used directly since we do not calculate the
amplitudes of different modes with real energy $e$,
but rather the amplitudes of a single mode having a complex energy
$\hbar \omega_\nu$.
Therefore the amplitudes $X^\nu_{mn}$, $Z^\nu$ and $W^\nu$ will be complex
quantities and thus also $h(\Delta N,\omega_\nu)$. However, the squared
amplitudes
\[ \eta^\nu_{mn} X^\nu_{mn} (\tilde{\eta}^\nu_{mn} \tilde{X}^\nu_{mn})^*\ ,
   \quad Z^\nu (\tilde{Z}^\nu)^* \ ,\ {\rm and} \quad
   \quad W^\nu (\tilde{W}^\nu)^*
\]
have very small imaginary parts. We can therefore obtain a good approximation
for $h(\Delta N,e)$ by taking
\begin{equation}
  h(\Delta N,e) \vartheta^l(jk) \approx
  h^l_A(jk,\nu) \vartheta^l(jk) =
                \frac{V_{j \pi k}}{\sqrt{2 \hbar \omega_\pi}}
                \, [\zeta^l(\nu)+\varpi^l(\nu)]                     +
                \sum_{mn}  V_{jk,mn} \, \xi^l_{mn}(\nu) \ ,
\label{eq_hlA}
\end{equation}
with
\begin{eqnarray}
  \xi_{mn}
& = &
  \left[ \mbox{ Re } \eta^\nu_{mn} X^\nu_{mn}
                     (\tilde{\eta}^\nu_{mn} \tilde{X}^\nu_{mn})^*
  \right]^{1/2}\ ,  \\
  \zeta
& = &
  \left[ \mbox{ Re } Z^\nu  (\tilde{W}^\nu)^* \right]^{1/2}\ ,  \\
  \varpi
& = &
  \left[ \mbox{ Re } W^\nu  (\tilde{W}^\nu)^* \right]^{1/2}\ .
\end{eqnarray}
We have numerically compared the total width obtained by summing the partial
widths based on this approximation, with the correct total width based on
eqs.\ (\ref{eq_MbarlDia}), (\ref{eq_MbartDia}), (\ref{eq_DSE2}),
and (\ref{eq_DSEtr}).
We find, in the range of invariant $\Delta$ masses $m$ from
$1000\ \MeV/c^2$ to  $1400\ \MeV/c^2$, that the approximation of the partial
widths somewhat over-predicts the total width at low invariant masses and
somewhat under-predicts it at large invariant mass. The relative error is
between 0\% and 20\% depending on $m$. To improve our
approximation we therefore multiply the approximate partial
widths\footnote{The partial width $\Gamma_\Delta^{N N^{-1}}$ is not multiplied
                by $c_\Delta$ since the amplitudes and energies of the
                nucleon-hole modes are real and do not depend on
                $\Gamma_{\Delta N^{-1}}$.}
by a factor $c_\Delta(E_\Delta,\mathbf{p}_\Delta)$, depending only on the
$\Delta$ energy and momentum, to obtain the correct total width when all
partial widths are summed over. The uncertainty of the partial widths obtained
by this procedure should thus be quite small.

In a transport description where collective modes are propagated as
quasiparticles one will need a cross section for the inverse process
$\nu + N \rightarrow \Delta$. This cross section is obtained analogously to the
partial $\Delta$ width in (\ref{eq_GnuReal}). When a $\Delta$ with mass $m$ is
created in the $\nu + N$ collision the cross section is written
\begin{equation}
  \sigma(\nu + N \rightarrow \Delta) = (\hbar c)^2 \: {\cal M}_{\nu N} \:
       |h^{l,t}(\Delta N,\omega_\nu)|^2 \; \frac{m c^2 \: m_N c^2}
       { e_N \: v_{\rm rel}/c } \; 4 \pi \rho_3(m^2) \ ,
\label{eq_sNnu}
\end{equation}
where the factor $\rho_3(m^2)$, defined in eq.\ (\ref{eq_rhoDmass}), takes
into account the finite width of the $\Delta$, and the spin-isospin factor
${\cal M}_{\nu N}$ is 2/3 in the spin-longitudinal channel and 4/3 in the
spin-transverse channel for the process
$\tilde{\pi}_j^+ + p \rightarrow \Delta^{++}$.

\subsubsection{Inelastic nucleon-nucleon and $\Delta$-nucleon cross sections}

Next we will derive the formulas for calculating cross sections for
the processes
\begin{equation}
  N + N \rightarrow \Delta + N
  \quad \mbox{and} \quad     N + \Delta \rightarrow N + N\ .
\end{equation}
We will start by writing down the $S$-matrix for the processes of interest. The
procedure is very similar to the procedure for deriving $\Sigma_\Delta$. We
will first present the formalism in the spin-longitudinal channel and then
generalize to the spin-transverse channel.

In what we will refer to as the direct term, we consider baryon 1 colliding
with baryon 2.
Baryon 1 will after the collision appear as baryon 3,
while the incoming baryon 2 becomes 4 after the collision,
see fig.\ \ref{fig_EffInt}.
Baryon 2 may be either a $N$ or a $\Delta$ depending on the process,
and in the same way may baryon 3 be a $\Delta$ or a $N$.
We denote the transferred energy and momentum by $\omega_D$ and $\mathbf{q}_D$,
respectively, with
\begin{equation}
  \hbar \omega_D = e_3 - e_1 = e_2 - e_4,\ \qquad
  \mathbf{q}_D = \mathbf{p}_3 - \mathbf{p}_1 = \mathbf{p}_2 - \mathbf{p}_4\ .
\end{equation}
We also take into account the exchange process where baryons 3 and 4 are
interchanged.

The cross section is obtained from
\begin{equation}
  d \sigma  =  w \frac{L^3}{v_{rel}} \left( \frac{L}{2 \pi \hbar} \right)^3
               d^3 p_3 \left( \frac{L}{2 \pi \hbar} \right)^3 d^3 p_4
               \rho_3(m_3^2) d m_3^2\ ,
\end{equation}
where the transition probability per unit time is given by
\begin{equation}
  w = \frac{|S_{fi}|}{\Delta t}\ .
\end{equation}
$S_{fi}$ is the total scattering matrix of the process and $\Delta t$ is here a
finite time interval which will tend to infinity at the end of the calculation.
The function $\rho_3(m_3^2)$ takes into account the finite width of the
$\Delta$ when baryon 3 is a $\Delta$,
\begin{equation}
   \rho_3(m_3^2) = \left\{ \begin{array}{lc}
      \delta(m_3^2 - m_N^2) &,\ 3 = N \\
      \frac{1}{\pi} \,
      \frac{ m_\Delta \Gamma_\Delta(m_3)}
           { (m_3^2 - m_\Delta^2)^2 + m_\Delta^2 \Gamma_\Delta(m_3)^2}
      &,\ 3 = \Delta
      \end{array} \right. \ .
\label{eq_rhoDmass}
\end{equation}

In contrast to the calculation of the $\Delta$ self-energy
the spin-isospin summation is not performed in the $S$-matrix, so
the spin-isospin matrix elements need to be kept. In addition, an overall
energy
and momentum conserving $\delta$-function is included,
and a factor $\sqrt{m_i c^2/e_i}$ for each external baryon.
We also get a contribution from the
short-range interaction, eq.\ (\ref{eq_Vg}), acting directly between the
vertices $3 \leftarrow 1$ and  $4\leftarrow2$. Taking this into account we
can write down the contribution from the direct term in the spin longitudinal
channel to the $S$-matrix
\begin{equation}
 S^l_D =
     \left( \frac{\hbar c}{L} \right)^3 \:
       M^l(34,12) \,
       \prod_{j=1,4} \left[\frac{ m_j c^2 }{e_j}\right]^{1\over2} \
       2 \pi \delta(e_1 + e_2 - e_3 - e_4)\
       \delta_{\mathbf{p}_1 + \mathbf{p}_2, \mathbf{p}_3 + \mathbf{p}_4}\ .
\label{eq_SlD}
\end{equation}
The contribution from the exchange term is obtained by
interchanging baryon 3 and 4, which gives
\begin{equation}
  \hbar \omega_E = e_4 - e_1 = e_2 - e_3, \qquad
  \mathbf{q}_E = \mathbf{p}_4 - \mathbf{p}_1 = \mathbf{p}_2 - \mathbf{p}_3\ .
\end{equation}
The expression for $S^l_E$ is identical to $S^l_D$ with the replacements:
$\omega_D \rightarrow \omega_E$, $\mathbf{q}_D \rightarrow \mathbf{q}_E$,
$ 31 \rightarrow 41$ and $ 24 \rightarrow 23$.

The total $S$-matrix can now be written
\begin{equation}
 S_{fi} = \left( \frac{\hbar c}{L} \right)^3 \:
         M \prod_{j=1,4} \left[\frac{ m_j c^2 }{e_j}\right]^{1\over2}\
         2 \pi \delta(e_1 + e_2 - e_3 - e_4)\
         \delta_{\mathbf{p}_1 + \mathbf{p}_2, \mathbf{p}_3 + \mathbf{p}_4}\ ,
\label{eq_Stot}
\end{equation}
with
\begin{equation}
  M = M^l(34,12) + M^l(43,12) + M^t(34,12) + M^t(43,12)\ ,
\end{equation}
where the expressions for ${M}^t$ in the  spin-transverse channel are
obtained analogously.

Summing over final states and averaging over initial states, we obtain the
differential cross section in the center-of-mass system as
\begin{equation}
\frac{d \sigma}{d \Omega} = (\hbar c)^2 \frac{1}{2s_1 + 1} \frac{1}{2s_2 + 1}
       \frac{c|\mathbf{p}_1| c |\mathbf{p}_3|}{4 \pi^2 I^2}
       \int d m_3^2 \sum_{\mbox{spin}} |M|^2 ( \prod_{j=1,4} m_j c^2)
               \rho_3(m_3^2) \ .
\label{eq_dsdO}
\end{equation}
{}From this differential cross section we can also
write down an invariant cross section,
\begin{equation}
\frac{d \sigma}{d (-t)} = (\hbar c)^2 \frac{1}{2s_1 + 1} \frac{1}{2s_2 + 1}
       \frac{1}{4 \pi I^2}
       \int d m_3^2 \sum_{\mbox{spin}} |M|^2 ( \prod_{j=1,4} m_j c^2)
               \rho_3(m_3^2) \ .
\label{eq_dsdt}
\end{equation}
In these expressions we have used that the relative velocity $v_{rel}$ can be
be expressed by means of the relativistic invariant
\begin{equation}
   I = e_1 e_2 v_{rel}/c =
       \sqrt{ (e_1 e_2 - c \mathbf{p}_1 \cdot c \mathbf{p}_2)^2
                - (m_1 c^2)^2 (m_2 c^2)^2 }\ .
\end{equation}

\section{Parameter values}
\label{sec_Param}
The $\Delta$-hole model with $\pi$ and $\rho$ meson exchange and an effective
$g'$ short-range interaction contains a number of parameters.
Although these can presently not be determined uniquely,
it is possible to use existing experimental information
to put a number of constraints on the parameters,
thus limiting the range of their values.
In this section we will
present our choice of parameter values together with a discussion and a
motivation of this choice.
For convenience,
all the parameter values are summarized in table \ref{tab_param}.

For the $\pi$ and $\rho$ coupling constants,
$f^\pi_{NN}$,  $f^\pi_{N \Delta}$, $f^\rho_{NN}$, $f^\rho_{N \Delta}$,
we choose their values such  that they
are consistent with $\pi$-absorption data on the deuteron, \ie\ according to
the range of possible values in ref. \cite{pidpp}. For the $\pi$ coupling
constants this together with the condition that our model reproduces the value
of the free $\Delta$ width at resonance,
\begin{equation}
   \Gamma_\Delta^{\rm free}(m_\Delta =  1232\ \MeV/c^2)
               \approx 115-120 \mbox{ MeV}\ ,
\end{equation}
fixes the value of $f^\pi_{NN} = 1.0$ and $f^\pi_{N \Delta} = 2.2$.
For the $\rho$ coupling constants we follow ref.\ \cite{pidpp}
and relate $f^\rho_{NN}$ and $f^\rho_{N \Delta}$
according to the quark model relation,
$f^\rho_{N\Delta} = \sqrt{72/25} f^\rho_{NN}$.
However, the value of $f^\rho_{NN}$ is not fixed
by the work in ref.\ \cite{pidpp},
but is only constrained to be in an interval of possible values,
approximately $5.6 \leq f^\rho_{NN} \leq 7.8$.
In our work we choose the value $f^\rho_{NN} = 6.2$ from this interval.
This value has also been used in other studies related to the present work,
\eg\ in ref.\ \cite{Bertsch}.
With this choice we are able to reproduce free cross section for the process
$ {\rm p} + {\rm p} \rightarrow \Delta^{++} + {\rm n}$,
if we adjust the remaining parameters (essentially $g'$) appropriately.
This is not the only choice of $f^\rho_{NN}$ that can reproduce
these cross sections,
but for other choices of $f^\rho_{NN}$ the
remaining parameters would have to be readjusted.

Closely related to the values of the coupling constants are the values of the
cut-off factors, $\Lambda_\pi$ and $\Lambda_\rho$,
in the monopole form factors that are included in the interactions
used at the vertices with a $\pi$ or $\rho$ meson.
In ref.\ \cite{pidpp} the values $\Lambda_\pi = 1.2\ \GeV$ and
$\Lambda_\rho \geq 1.5\ \GeV$ were used.
In our work we take a somewhat lower value of the pion cut off factor,
$\Lambda_\pi = 1.0\ \GeV$,
and we take $\Lambda_\rho = 1.5\ \GeV$,
to achieve a relatively fast cut off in the numerical integrations.

When we calculate the dispersion relations, \ie\ find the energy-momentum
relation $\hbar \omega(\q)$ for the different spin-isospin modes,
we will also find modes with $\hbar \omega(\q)$ close to
$[(c \q)^2 + \Lambda_{\pi,\rho}^2]^{1/2}$. These modes arise from the
singularities in the form factors,
\begin{equation}
   F_{\pi,\rho}(\omega,\q) =
         \frac{ \Lambda_{\pi,\rho}^2 - (m_{\pi,\rho} c^2 )^2 }
              { \Lambda_{\pi,\rho}^2 - (\hbar \omega)^2 + (c \q)^2 }\ .
\end{equation}
In the case of the pion,
these singularities can be seen as the coupling of the pion
to a heavier meson when the pion is spatially close to the nucleon,
like in the electromagnetic case where the pion couples to the photon via a
$\rho$ meson, see \eg\ ref.\ \cite{EricsonWeise}.
However, our model,
with the monopole factors determined in the space-like sector
($c |\q| >  \hbar \omega$),
is not appropriate to describe the physics near the singularities of
the form factors where their true behavior may deviate substantially from the
monopole form.
To avoid this difficulty we exclude the ``form factor" like modes
($\hbar\omega(\q)\sim [(c \q)^2+\Lambda_{\pi,\rho}^2]^{1/2}$)
from the dispersion relations when calculating physical quantities,
like cross sections and the $\Delta$ width,
so that the form factors only contribute by their numerical values
along the other spin-isospin modes.

The short-range interaction contains the correlation parameters
$g'_{NN}$, $g'_{N \Delta}$, and $g'_{\Delta \Delta}$,
and the cut-off factor $\Lambda_g$,
and is an effective interaction that simulates more complicated interactions at
short range,
like exchange of heavier mesons and exchange of two or more mesons.
Taking this interaction according to eq.\ (\ref{eq_Vg}) and
excluding the form factor ($\Lambda_g \rightarrow \infty$),
this interaction becomes a $\delta$ interaction in coordinate space,
and is thus the simplest form of a short-range interaction.
By including the form factor ($\Lambda_g<\infty$),
we take into account the finite size of the interacting particles.
In reality, the effective short-range interaction may have some
additional $q$-dependence via the $g'$-parameters,
and may also depend on the nuclear density.
However, in the present study we neglect such complications
and take constant values for the $g'$-parameters.
The values of $g'_{NN}$, $g'_{N \Delta}$, $g'_{\Delta \Delta}$, and
$\Lambda_g$,
are not well known.
Our preference is therefore to take $\Lambda_g$ equal to either
$\Lambda_\pi$ or $\Lambda_\rho$,
in order to reduce the number of free parameters,
and then vary the $g'$ parameters to fit certain experimental data.
For numerical reasons,
it is more convenient to take $\Lambda_g =\Lambda_\rho$
than  $\Lambda_g = \Lambda_\pi$.
For $\Lambda_g = \Lambda_\pi$ the singularity of $F_g$
would lie close to the $\rho$-like branch in the spin-transverse channel,
and because of the interaction between these two modes
the $\rho$-meson mode would be substantially changed,
and it would be difficult to exclude the ``form factor"-like mode
from the dispersion relations.
To avoid these problems it is therefore convenient to take
$\Lambda_g = \Lambda_\rho$.

The value $\Lambda_g = 1.5\ \GeV$ is a larger value than was used in ref.\
\cite{PaJh} were the $\Delta$ width in nuclear matter and its contribution from
different decay modes was discussed.
In that work $\Lambda_g = 1.0\ \GeV$ was used.
A $g'$ set with a small cut-off factor should approximately correspond to
a set with somewhat smaller $g'$ values with a larger cut-off factor.
This implies that the numerical values of the $g'$ parameters
in this work and ref.\ \cite{PaJh} are only approximately similar.
However, this does not affect how $\Gamma_\Delta^{\rm tot}$ depends on
the $g'$ parameters, as was discussed in ref.\ \cite{PaJh}.

To determine the values of the $g'$ parameters we calculate the
p+p  $\rightarrow \Delta^{++}$+n cross section in vacuum with our model.
Keeping the previously discussed parameters fixed,
we adjust $g'_{N \Delta}$ to reproduce existing data.
In fig.\ \ref{fig_dsdcos0}$a$ we present calculations of
$d\sigma/d\cos(\theta)$ at center-of-mass energy $\sqrt{s} = 2.314\ \GeV$
together with experimental data from ref.\ \cite{ppDn1}.
We find that we reproduce the
angular dependence of the cross section well. In the calculations we have used
the value $g'_{N \Delta} = 0.38$. The cross section depends rather strongly on
$g'_{N \Delta}$, so therefore this value is quite well determined in our model.
In fig.\ \ref{fig_dsdcos0}$b$ we present $d\sigma/dt$
for the same parameter set,
but with $\sqrt{s} = 2.513\ \GeV$, together with experimental data from ref.\
\cite{ppDn2}. We note that in ref.\ \cite{ppDn2} events with $t$ or $u$ equal
to a given value were in the same bin. Thus, the experimental values have in
fig.\ \ref{fig_dsdcos0}$b$ been divided by a factor of two. Also at this energy
the experimental data is reasonably well reproduced, except for the
experimental peak value at $t \approx -0.06$ GeV$^2$. In fig.\ \ref{fig_stot0}
we present the calculated total cross section as a function of $\sqrt{s}$,
together with two parameterizations and experimental data. The solid curve is
our calculation, the dashed is the parameterization from VerWest-Arndt
\cite{VerWA}, and the dash-dotted line is a simple parameterization often used
in BUU calculations \cite{DasGupta}. The data points were estimated from fig.\
2 of ref.\ \cite{VerWA} and can be found in references therein. As can be seen
in fig.\ \ref{fig_stot0}, we reproduce the energy dependence of the total cross
section quite well, and substantially better than the simple parameterization
from ref.\ \cite{DasGupta}. At large $\sqrt{s}$ we under predict the
experimental
points somewhat. Increasing $g'_{N \Delta}$ while keeping all other parameters
fixed will increase the total cross section, but also flatten the angular
distribution in $d\sigma/d\cos(\theta)$. Decreasing  $g'_{N\Delta}$ will reduce
the total cross section and also flatten the angular distribution.

The fit of the $\Delta^{++}$ cross section can be maintained if both $f^\rho$
and $g'_{N \Delta}$ simultaneously are increased. For example, for
$f^\rho_{NN}= 7.2$, $f^\rho_{N \Delta} = 12.2$ and $g'_{N \Delta} = 0.46$ an
equally good fit is obtained. However, the contributions from the
spin-longitudinal and the spin-transverse channels are changed. For
$f^\rho_{NN} = 6.2$ ($g'_{N\Delta} = 0.38$) these two channels contribute
approximately equally, while at $f^\rho_{NN} = 7.2$ ($g'_{N \Delta} = 0.46$)
the contribution to the cross section comes mainly from the transverse channel.

The cross section $\sigma({\rm p}+{\rm p}\rightarrow\Delta^{++}+{\rm n})$ does
not depend on the correlation parameters $g_{NN}'$ and $g_{\Delta\Delta}'$, so
these remain undetermined by fitting $\sigma({\rm p} + {\rm p} \rightarrow
\Delta^{++} + {\rm n})$.
We can obtain some constraints on $g_{NN}'$ and
$g_{\Delta \Delta}'$ by calculating the $\Delta$ width in nuclear matter
at zero temperature,
and identifying $-\Gamma^{\rm tot}_{\Delta}/2$
with the imaginary part of the $\Delta$-nucleus optical potential.
In the $\Delta$-hole model first developed for $\pi$-nucleus
scattering \cite{Hirata} the authors used a $\Delta$-nucleus optical potential
with distinct contributions to the imaginary part,
\begin{equation}
   -2\ {\rm Im}( V^{\rm opt}_\Delta) =
             \Gamma_\Delta^{\rm Free} - \delta \Gamma_\Delta^{\rm Pauli}
             -2\ {\rm Im} (V_{\rm spread})\ .
\end{equation}
The spreading potential was adjusted to fit the experimental results.  For
$^{12}$C $V_{\rm spread}$ was found to be approximately
\begin{equation}
   V_{\rm spread} \approx [ 23 \pm 5 - i(43 \pm 5)]\: \frac{\rho_N}{\rho_N^0}
   \quad \MeV \ ,
\end{equation}
and rather independent of energy in the interval $100\ \MeV \leq T_\pi \leq
250\ \MeV$. Also in microscopic calculations of the $\Delta$-nucleus optical
potential \cite{Lee} the authors have found that the imaginary part of the
spreading  potential is rather independent of energy,
$-40\leq{\rm Im}(V_{\rm spread})\leq -20\ \MeV$,
in the interval $50\ \MeV\leq p_\Delta c \leq400\ \MeV$
for a $\Delta$ on mass shell.
Figure \ref{fig_Vspread} shows the quantity
$[\Gamma_\Delta^{\rm tot} - \Gamma_\Delta^{\rm free} +
\delta \Gamma_\Delta^{\rm Pauli}]/2$,
which corresponds to the spreading potential.
The results for two $g'_{\Delta \Delta}$ values, 0.25 and 0.35 are presented,
using $g'_{NN} = 0.9$.
We have performed the
calculations presented in fig.\ \ref{fig_Vspread} at $\rho_N = 0.75 \rho_N^0$
to compare \cite{Oset,Seki} with the empirical values of ref.\ \cite{Hirata}.
The $\Delta$ energy and momentum are estimated from the relations
\begin{equation}
   E_\Delta = \frac{3}{5} \, \frac{\mathbf{p}_F^2}{2 m_N^*} +
              \hbar \omega_\pi(\mathbf{q})\ , \qquad
   \mathbf{p}_\Delta^2 = \frac{3}{5} \, \mathbf{p}_F^2 +
              \mathbf{q}^2 \ ,
\label{eq_Tpi}
\end{equation}
where $\mathbf{q}$ is the pion momentum determined from the pion kinetic energy
$T_\pi = \hbar \omega_\pi(\mathbf{q}) - m_\pi c^2$.
The calculations shown in fig.\ \ref{fig_Vspread}
also include a binding correction of 20 MeV.
The $\Delta$ width varies only slightly with $g'_{NN}$, but depends quite
strongly on $g'_{\Delta \Delta}$, see ref.\ \cite{PaJh}.
By comparison to the empirical points,
it is seen that $g'_{\Delta \Delta}$ is quite well determined to
be in the approximate interval 0.25 to 0.35.
We have chosen to present our
remaining results for the value $g'_{\Delta \Delta}=0.35$

A value of $g'_{NN} \approx$ 0.5--0.9 is often used in the literature.
Some constraints on $g'_{NN}$ can be obtained from the low-energy ($q\approx0$)
Gamow-Teller response, as seen for example in (p,n) reactions.
Several years ago there were reports that only 60\% of the expected strength
was
found among the low-lying states.
It was suggested that the low-energy strength is due to a strong coupling
with $\Delta N^{-1}$ states,
but the effect may equally well be explained by couplings to
two-particle-two-hole states
(see for example the review by Bertsch and Esbensen \cite{BertschEsbensen}).
A  value of $g'_{NN}\approx0.9$ is consistent with a rather weak coupling
between low-lying Gamow-Teller modes and the $\Delta N^{-1}$ states,
while a smaller value, $g'_{NN}\approx 0.6$,
leads to a renormalization by almost 40\% of the low-lying {\sl GT} strength
due to the coupling to  $\Delta N^{-1}$ states.
The significance of the $g'$ parameters is discussed further
in ref.\ \cite{PaJh} in connection with the calculation of the $\Delta$ width.

Since we perform our calculations at a constant density we will describe the
nucleons by letting them propagate in a constant potential $V_N$.
In addition we take
into account corrections in the nucleon energies, because of their interaction
with the surrounding medium, by taking an effective nucleon mass as
\begin{equation}
  m^*_N = m_N [1 + C {\rho\over\rho_0}]^{-1}\ .
\end{equation}
This density dependent form is obtained in the extended Seyler-Blanchard
model discussed in ref.\ \cite{WDM,RandMed}
and leads to $m_N^* = 0.7718 m_N$ at normal nuclear density, $\rho = \rho_0$.
That model employs an effective nucleon-nucleon interaction,
with a Yukawa force modulated by a quadratic momentum dependence
and an explicit density dependence,
and is solved self-consistently within the Thomas-Fermi approximation.
The model gives a good description
of average properties of standard nuclear matter \cite{WDM,RandMed}.

The $\Delta$ isobars propagate in the same manner in a constant
$\Delta$ potential.
For the modification of the real part of their energies,
due to their interaction with the medium, we follow previous work on
$\pi$-nucleus scattering \cite{Lee,HoThLe}, nuclear photo absorption
\cite{KoMoOh}, and nuclear response in the $\Delta$ region \cite{HoOsUd,Dmit},
and take $V_\Delta - V_N \approx 25$ MeV at normal nuclear density.
Note that in our formalism the
potentials only enter as the difference $V_\Delta - V_N$.
This difference is found to be rather independent of the $\Delta$ energy
and momentum \cite{Lee}.
The quantity $V_\Delta - V_N$ may be expected to vary somewhat with
the nuclear density but in the present study we will use
$V_\Delta - V_N = 25\ \MeV$ also at twice normal nuclear density.

\section{Results and discussion}
\label{sec_Res}

In this section we present and discuss our results.
For various specified densities and temperatures,
we calculate the spin-isospin modes formed in symmetric nuclear matter.
In sec.\ \ref{sec_ResSIM} we exhibit
their dispersion relations, \ie\ the energy-momentum relation $\omega(q)$,
as well as their composition in terms of the unperturbed states.
{}From these quantities,
we calculate the width of the $\Delta$ isobar (sec.\ \ref{sec_GammaD})
and the cross sections for collisions involving a $\Delta$
(in sec.\ \ref{sec_cross-sec}).
Furthermore,
we discuss in sec.\ \ref{sec_transport}
how the results could be incorporated in transport simulations
of heavy-ion collisions, such as those carried out with the BUU model.
In sec.\ \ref{sec_transport} we also discuss previous studies
that have included some in-medium effects in transport calculations.

The hadrons are confined within a periodic cubic box with side length $L$,
and the coordinate system is aligned such that $\q$,
the momentum of the mode considered, is parallel to the $z$ axis,
$\q = q \hat{\mathbf{z}}$.
Since the box is finite,
the momenta take on only discrete values,
\begin{equation}
  \p = \frac{2 \pi \hbar}{L} \bold{j} =
               \frac{2 \pi \hbar}{L} (j_x,j_y,j_z)\ ,
\end{equation}
where $j_x$, $j_y$ and $j_z$ are integers.
The energy of an unperturbed nucleon-hole state is given by
\begin{equation}
  \hbar \omega_{N N^{-1}} =
         E_N(\p + \q) - E_N(\p) =\frac{q^2}{2 m_N^*} +  \frac{q p_z}{m_N^*}\ ,
\end{equation}
and for an unperturbed $\Delta$-hole state
(taking $\Gamma_\Delta$ = 0 for simplicity),
\begin{equation}
  \hbar \omega_{\Delta N^{-1}}=
       E_\Delta(\p + \q) -  E_N(\p) \approx
       \frac{q^2}{2 m_\Delta} + \frac{q p_z}{m_\Delta} + m_\Delta - m_N\ .
\label{eq_wDh-unp}
\end{equation}
Since the energies $\hbar\omega_{N N^{-1}}$ and $\hbar\omega_{\Delta N^{-1}}$
depend only on $p_z$ (and not $p_x$ and $p_y$),
it suffices to specify the quantum number $j_z$
in order to characterize a $N N^{-1}$ or $\Delta N^{-1}$ state.
In reality the $\Delta N^{-1}$ states have also a $p^2$ dependence,
via the $\Delta$ width and the term $p^2/2 m_\Delta - p^2/2 m_N^*$
which is neglected in eq.\ (\ref{eq_wDh-unp}).
We have taken this dependence into account approximately by
making the replacement
\begin{equation}
  p^2\ \rightarrow\ p_z^2 + <p^2_\perp>(p_z;T,\rho_N)\ ,
\end{equation}
with the variance of the transverse momentum given by
\begin{equation}
  <p^2_\perp>(p_z;T,\rho_N)=\int \frac{d^2p_\perp}{(2 \pi)^2}
                                  n_N(\p;T,\rho_N)\ p^2_\perp/
                            \int \frac{d^2p_\perp}{(2 \pi)^2}
                                  n_N(\p;T,\rho_N)\ ,
\end{equation}
where $n_N$ is the nucleon occupation probability given in eq.\ (\ref{eq_n}).

When calculating quantities like the energies and amplitudes of the
spin-isospin modes, the total and partial $\Delta$ widths, and $\Delta$ cross
sections, one can take the width $\Gamma_{\Delta N^{-1}}$
in the Lindhard function $\Phi_\Delta$ (see eq.\ (\ref{eq_PhiD}))
to be either identical to zero or equal to the total $\Delta$ width
calculated self-consistently by an iterative procedure.
In this section we will present results for both cases.
We will refer to the former case as the reference case
and to the latter as the self-consistent case.

\subsection{Spin-isospin modes}
\label{sec_ResSIM}

{}From eq.\ (\ref{eq_det1}) we calculate the energies of the spin-isospin
modes that are formed in the interacting system, \ie\ their dispersion
relations.
Figs.\ \ref{fig_DispT0r10}$a$ and \ref{fig_DispT0r10}$b$ display the
real part of the dispersion relations for the self-consistent case at
normal nuclear density,
$\rho_N = \rho_N^0 = 0.153\ \fm^{-3}$, and zero temperature, $T=0\ \MeV$.
In fig.\ \ref{fig_DispT0r10}$a$ a number of different modes in the
spin-longitudinal ($\pi$-like) channel are apparent.
Some of those are non-collective $NN^{-1}$ modes (solid lines),
which at zero temperature have their energies within the region
\begin{eqnarray}
    0   &   \leq   &   \mbox{Re } \hbar \omega\
            \leq\       \frac{q^2}{2 m_N^*} +  \frac{q p_F}{m_N^*}\ ,
                       \qquad q < 2 p_F                     \ , \nonumber \\
    \frac{q^2}{2 m_N^*} - \frac{q p_F}{m_N^*}
        &   \leq   &   \mbox{Re } \hbar \omega\
            \leq\       \frac{q^2}{2 m_N^*} +  \frac{q p_F}{m_N^*}\ ,
                       \qquad q > 2 p_F\ .
\label{eq_NhCont}
\end{eqnarray}
Since we are presenting our results for a box normalization with a finite side
length $L$, we obtain a discrete number of non-collective $NN^{-1}$ modes. The
total number of spin-isospin modes within the region (\ref{eq_NhCont}) depends
on $L$ and tends towards a continuum in the limit $L \rightarrow \infty$. In
fig.\ \ref{fig_AmpT0r10}$a$ we show the $NN^{-1}$ component of the squared
amplitude for all the modes at a fixed momentum,  $q = 300\ \MeV/c$. It is
clearly seen that all the squared amplitudes of the modes in the $NN^{-1}$
region, $0 \leq \hbar \omega \leq 185\ \MeV$, are dominated by a single
$NN^{-1}$ state (specified by $p_z$) and thus have a non-collective character.
Another feature of the non-collective modes is that their energies are very
close to the energies of the corresponding unperturbed states.

Similarly, a number of non-collective $\Delta N^{-1}$ states emerge in fig.\
\ref{fig_DispT0r10}$a$ which, for a fixed $\q$ and at zero
temperature, have their energies constrained to a band,
\begin{equation}
    m_\Delta - m_N + \frac{q^2}{2 m_\Delta} - \frac{q p_F}{m_\Delta}\
          \leq\     {\rm Re }\ \hbar\omega\
          \leq\     m_\Delta - m_N + \frac{q^2}{2 m_\Delta} +
                       \frac{q p_F}{m_\Delta}.
\label{eq_DhCont}
\end{equation}
The absence of collectivity is seen in fig.\ \ref{fig_AmpT0r10}$b$
where for all modes in the $\Delta N^{-1}$ region,
except a few in the upper half of the region,
a single component (specified by $p_z$) dominates each mode.
The remaining modes show some collective behavior,
but are mostly dominated by only two components.
In fig.\ \ref{fig_DispT0r10} the solid curves show those $\Delta N^{-1}$ modes
that have a contribution larger than 0.5 from one or 0.9 from two components,
while the remaining modes are represented by
dot-dot-dot-dashed lines to indicate that they carry some collective strength.
Also here the perturbed energies are close to the corresponding unperturbed
ones.

In addition, two collective modes are visible in fig.\ \ref{fig_DispT0r10}$a$.
The lower one, represented by a dot-dashed line,
starts at $\hbar \omega = m_\pi c^2$ at $q = 0$ and continues into the $\Delta
N^{-1}$ region around $q \approx 330\ \MeV/c$.
This mode is in the sequel be referred to as $\tilde{\pi}_1$.
The upper collective mode, displayed as the
dot-dot-dashed curve, starts slightly above $\hbar\omega \approx m_\Delta c^2 -
m_N c^2$ at $q = 0$ and approaches $\hbar\omega_\pi = [(m_\pi c^2)^2 +
(cq)^2]^{1/2}$ at large $q$.
This mode is denoted $\tilde{\pi}_2$.
Fig.\ \ref{fig_AmpT0r10}$c$ gives an impression of the structure of the
$\tilde{\pi}_1$ and $\tilde{\pi}_2$ modes: the squared amplitudes of the pion
and the sum of all $NN^{-1}$ and $\Delta N^{-1}$ components are shown for a
fixed momentum, $q = 300\ \MeV/c$.
It is seen that the collective modes have
contributions from all three types of interacting states, while the
non-collective modes have only contribution from one type of state.
Furthermore,
we see from fig.\ \ref{fig_AmpT0r10}$a$ and \ref{fig_AmpT0r10}$b$ that the
total
$NN^{-1}$ and $\Delta N^{-1}$ contributions to the collective modes are made up
from small contributions from all of the individual $NN^{-1}$ and $\Delta
N^{-1}$ states, respectively.

Fig.\ \ref{fig_AmpT0r10}$d$ shows the squared amplitudes of the different
components on the lower collective mode as a function of $q$. For small $q$ the
pion component dominates, while the $\Delta N^{-1}$ component becomes dominant
at around $q \approx 330\ \MeV/c$. At $q\approx500\ \MeV/c$ the lower
collective
mode  has lost almost all of its collective character.

In fig.\ \ref{fig_DispT0r10}$b$ we present the modes in the spin-transverse
channel. The dispersion relations for the non-collective modes are very similar
to the relations in fig.\ \ref{fig_DispT0r10}$a$, while the collective
$\rho$-meson like modes are different. There are two visible collective modes
in fig.\ \ref{fig_DispT0r10}$b$, one dominated by the $\rho$-meson component is
starting at $\hbar \omega = m_\rho c^2$ at $q=0$. This curve is represented by
a dot-dashed curve and is denoted $\tilde{\rho}_1$. The other one of
$\Delta$-hole character is denoted $\tilde{\rho}_2$ and is displayed as a
dot-dot-dashed line. This one starts slightly above $\hbar\omega \approx
m_\Delta c^2 - m_N c^2$ at $q = 0$ and continues into the $\Delta N^{-1}$
region around $q \approx 200\ \MeV/c$ where it gradually looses its collective
character.

The dispersion relations presented in fig.\ \ref{fig_DispT0r10} can also be
calculated with $\Gamma_{\Delta N^{-1}} = 0$,
which gives purely real energies, $\omega_\nu$.
The spin-isospin modes are then built up by stable $\Delta$ isobars
having a fixed mass $m_\Delta$
and so each mode will have a distinct energy.
The self-consistent inclusion of the $\Delta$ width has several consequences.
The existence of a Breit-Wigner like distribution of $\Delta$ masses
implies that the unperturbed $\Delta N^{-1}$ energies no longer
will be distinct, but also have a Breit-Wigner distribution with width
$\Gamma_{\Delta N^{-1}}$.
This in turn leads to the same effect for the spin-isospin modes,
where Re $\hbar \omega_\nu$ represents the centroid of the distribution
and 2 Im $\hbar \omega_\nu$ is its width.
The $\Delta$ width also governs the decay rate of the isobar.
Accordingly, the $\Delta$ isobars will have finite life times
and so will the spin-isospin modes.
The decay out of a mode will then occur by the process
$\nu \rightarrow \Delta N^{-1}$,
where the $\Delta$ will decay further.
Thus Im $\hbar \omega_\nu$ contains information
about the life time of the mode $\nu$,
before it disintegrates due to the decay of one of its constituent
$\Delta$ components.

The dispersion relations for the reference case, $\Gamma_{\Delta N^{-1}}=0$,
are qualitatively similar to the real part of the eigenenergies
obtained in the self-consistent case, Re $\hbar \omega_\nu$,
with differences mainly in the collective modes.
When going from $\Gamma_{\Delta N^{-1}}=0$ to the self-consistent treatment,
the strength of the interaction between the states effectively weakens.
This occurs because the inclusion of
$\Gamma_{\Delta N^{-1}} = \Gamma_{\Delta}^{\rm tot}$
causes the strength of the $\Delta N^{-1}$ states
to be smeared out in the Breit-Wigner like distribution.
As a consequence, for a fixed $q$,
the energy of the pionic mode $\tilde{\pi}_1$ is somewhat raised and,
by the same token, the energy of the upper mode $\tilde{\pi}_2$ is lowered.
The changes in energy is 0--50 MeV,
depending on $q$ and with the largest difference for large $q$.

The imaginary parts of the eigenenergies $\hbar \omega_\nu$
are presented in figs.\ \ref{fig_DispT0r10}$c$ and  \ref{fig_DispT0r10}$d$.
The energies of the non-collective nucleon-hole modes are purely real,
while the non-collective $\Delta$-hole modes have imaginary parts
that are close to half the corresponding widths $\Gamma_{\Delta N^{-1}}$.
For both collective modes,
we find a large imaginary part,  $|\mbox{ Im }\hbar \omega_\nu|$,
when the $\Delta N^{-1}$ components dominate the mode,
and a smaller imaginary part when the mode is dominated by the meson component
(compare fig.\ \ref{fig_DispT0r10}$c$ and \ref{fig_AmpT0r10}$d$).
This implies that the collective modes will have longer life times
when they are more meson-like than when they are $\Delta N^{-1}$-like.

In fig.\ \ref{fig_Dispr20andT25}$a$ we present Re $\hbar\omega_\nu$
for the self-consistent case at twice normal nuclear density at
zero temperature, and in fig.\ \ref{fig_Dispr20andT25}$b$ at normal density and
temperature $T=25\ \MeV$.\footnote{We limit our considerations
        of excited matter to rather moderate temperatures,
        because of the applications we have in mind.}
Comparing the dispersion relations at normal (fig.\ \ref{fig_DispT0r10})
and double density (fig.\ \ref{fig_Dispr20andT25}),
we see that the main differences occur for the two pionic modes.
The enhanced interaction at $2\rho_N^0$
makes the two collective modes repel each other more strongly,
which causes a lowering of the mode $\tilde{\pi}_1$
and a concomitant rise in $\tilde{\pi}_2$.
Another feature in fig.\ \ref{fig_Dispr20andT25}$a$ is that
the non-collective $N N^{-1}$ and $\Delta N^{-1}$ modes
cover a larger region in the $(\omega,q)$ plane since the nucleon
chemical potential is larger at  $\rho_N=2\rho_N^0$ than at $\rho_N=\rho_N^0$.
At $T= 25\ \MeV$ (fig.\ \ref{fig_Dispr20andT25}$b$)
the collective modes $\tilde{\pi}_1$ and $\tilde{\pi}_2$
are very similar to the modes at $T=0\ \MeV$ (fig.\ \ref{fig_DispT0r10}),
but there are many more non-collective modes.
This is because the nucleon occupation probability, eq.\ (\ref{eq_n}),
at finite temperatures allows the occupied nucleon-state to be
found above the Fermi surface.

\subsection{$\Delta$ decay width}
\label{sec_GammaD}

The total $\Delta$ decay width in the nuclear medium, $\Gamma_\Delta^{\rm
tot}$,
can be calculated according to eqs.\ (\ref{eq_DSE2}) and (\ref{eq_DSEtr}).
In fig.\ \ref{fig_GamTot} we show the total $\Delta$ width for various
nuclear densities and temperatures.
The width is presented as a function of the
invariant $\Delta$ mass, $m = [E_\Delta^2 - (c\mathbf{p}_\Delta)^2]^{1/2}/c^2$,
for a fixed $\Delta$ momentum, $p_\Delta= 300\ \MeV/c$.
Fig.\ \ref{fig_GamTot}$a$ displays $\Gamma_\Delta$ for the reference case,
$\Gamma_{\Delta N^{-1}}=0$,
while fig.\ \ref{fig_GamTot}$b$ shows $\Gamma_\Delta^{\rm tot}$
when  $\Gamma_{\Delta N^{-1}}$ is calculated self-consistently,
$\Gamma_{\Delta N^{-1}} =  \Gamma_\Delta^{\rm tot}$.

The most noticeable difference between $\Gamma_\Delta^{\rm tot}$ in the nuclear
medium at normal nuclear density (solid curve) and in vacuum (dotted curve) is
that in vacuum  $\Gamma_\Delta^{\rm tot} = \Gamma_\Delta^{\rm free}$ starts to
increase from zero at the threshold $m=m_N+m_\pi$,
while in the medium $\Gamma_\Delta^{\rm tot}$ can be finite
also for $m<m_N+m_\pi$,
since the $\Delta$ isobar can decay into a nucleon
and a non-collective nucleon-hole mode, $\Delta~\rightarrow~N+N N^{-1}$.
These modes have lower energies
and the decay can thus occur below the free threshold.
These effects are seen in fig.\ \ref{fig_GamPar},
where also the partial contributions to the total $\Delta$ width are presented.
We also see that at $\rho_N=\rho_N^0$, $T=0\ \MeV$,
the total width in the medium, $\Gamma_\Delta^{\rm tot}$,
is larger than $\Gamma_\Delta^{\rm free}$ up to  $m \approx 1400\ \MeV/c^2$.
This effect is also mainly due to the new decay channels present in the medium,
both $\Delta\rightarrow N+NN^{-1}$
and also $\Delta\rightarrow N+\Delta N^{-1}$
which begins to contribute at $m \approx 1200\ \MeV/c^2$.

Another feature seen, in comparing $\Gamma_\Delta^{\rm free}$ with
$\Gamma_\Delta^{\rm tot}$ at $\rho_N = \rho_N^0$,
is that $\Gamma_\Delta^{\rm free}$ increases more steeply
than $\Gamma_\Delta^{\rm tot}$.
This is an effect of the effective nucleon mass $m_N^* < m_N$ at $\rho_N>0$.
A $\Delta$ isobar with energy
$E_\Delta = [(m c^2)^2 + (c \mathbf{p}_\Delta)^2]^{1/2}$
decays into a spin-isospin mode $\nu$ and a nucleon $N$,
and energy conservation yields $E_\Delta = \hbar \omega_\nu + e_N$.
By lowering $m_N^*$ the nucleon energy $e_N$ will be enhanced
and thus the energy $\hbar \omega_\nu$ of the mode $\nu$ is reduced.
This leads to a lower momentum of the mode $\nu$ in the $\Delta$ decay,
which in turn leads to a lower $\Delta$ width.

The large enhancement of $\Gamma_\Delta^{\rm tot}$ in fig.\ \ref{fig_GamTot}
at invariant masses up to $m \approx 1250\ \MeV/c^2$
for $\rho_N = 2 \rho_N^0$ occurs because the contributions
from the nucleon-hole modes are proportional to the nuclear density,
but it also reflects the fact that our parameter set
leads to pion condensation at densities just above $2 \rho_N^0$,
as is manifested by the occurrence of a spin-isospin mode
lying in the region of nucleon-hole modes
with a small real energy and a non-vanishing imaginary energy.
As is also seen in fig.\ \ref{fig_GamPar}$b$,
almost all the contribution to $\Gamma_\Delta^{\rm tot}$
comes from the nucleon-hole channels.
The onset of pion condensation can be pushed up in density
by increasing the values of the $g'$ parameters
$g_{N \Delta}'$ and $g_{\Delta \Delta}'$,
for example by making them density dependent.

Increasing the temperature to $T=25\ \MeV$
has very little effect on the total $\Delta$ width.
However, as seen in in fig.\ \ref{fig_GamPar},
the partial widths are considerably changed.
In fig.\ \ref{fig_GamPar}$a$ we see that at zero temperature there are
mainly four contributions to the total width for $m\leq 1400\ \MeV/c^2$:
non-collective $NN^{-1}$  and  $\Delta N^{-1}$ modes
(long and short dashed curves, respectively)
and the two collective modes $\tilde{\pi}_1$ and $\tilde{\pi}_2$
(dot-dashed and dot-dot-dashed curves, respectively).
At low invariant mass the only energetically possible decay modes
are the non-collective $N N^{-1}$ modes.
For $m \geq 1100\ \MeV/c^2$ the $\Delta$ has enough energy to to decay to
the mode $\tilde{\pi}_1$ and a nucleon above the Fermi surface.
This becomes the dominating contribution to
$\Gamma_\Delta^{\rm tot}$ from  $m \approx 1140\ \MeV/c^2$ up to $m \approx
1320\ \MeV/c^2$.
The non-collective $\Delta N^{-1}$ modes
start to contribute at $m \approx 1200\ \MeV/c^2$
and they become dominant at $m \geq 1320\ \MeV/c^2$.
Accordingly,
at zero temperature and normal nuclear density a $\Delta$ near resonance
($m\approx m_\Delta$)
mainly decays into a nucleon and the pionic mode  $\tilde{\pi}_1$,
and the magnitude of the partial width $\Gamma_\Delta^{\tilde{\pi}_1}$ is
comparable to the free width.
In addition, the partial width for decay into any of the non-collective
nucleon-hole modes is approximately half of the free width,
while the partial widths of the remaining decay channels are small
at $m\approx m_\Delta$.
At the temperature $T=25\ \MeV$ the situation is quite different.
Concentrating still on a $\Delta$ near resonance,
we find that the non-collective $N N^{-1}$ and $\Delta N^{-1}$ modes
give approximately equal and dominant contributions,
while the contribution from the pionic mode is smaller.
So,
even though the probability for the $\Delta$ to decay is the same
at $T=0\ \MeV$ and $T=25\ \MeV$,
there will be very few decays to the pionic mode in the latter case.

There are mainly two reasons for the reduction of
$\Gamma_\Delta^{\tilde{\pi}_1}$ at finite temperature.
The first is that the mode $\tilde{\pi}_1$ loses its collective strength
at a lower $q$ value at finite temperatures, as compared to zero temperature.
The second is that it is energetically possible for the $\Delta$ isobar
to decay into non-collective $\Delta N^{-1}$ modes
at lower $\Delta$ energies at $T=25\ \MeV$ than at $T=0\ \MeV$.
This is because there exits $\Delta N^{-1}$ modes
with lower energy at $T=25\ \MeV$ than at  $T=0\ \MeV$,
and because the nucleon formed in the $\Delta$ decay
can have energy less than $e_F$ at finite temperatures.
With several more decay channels contributing to
$\Gamma_\Delta^{\rm tot}$ at $T=25\ \MeV$ the contribution from
$\Gamma_\Delta^{\tilde{\pi}_1}$ is reduced,
since the total width is almost unaffected by the change in temperature.

The decomposition into partial widths corresponding to specific spin-isospin
modes $\nu$ is unambiguous when $\Gamma_{\Delta N^{-1}}$ vanishes.
However, in some cases it is not possible to determine uniquely
the nature of a specific mode $\nu$.
One such case is when the collective mode $\nu$
enters the non-collective $\Delta N^{-1}$ region
and loses its collective strength.
This transition is of course gradual
and there is some arbitrariness involved in determining
when the mode is no longer collective.
This gives rise to a corresponding (small) uncertainty
in the value of the partial widths $\Gamma_\Delta^{\tilde{\pi}_1}$ and
$\Gamma_\Delta^{\Delta N^{-1}}$.
Furthermore, the interaction between the pionic mode
and the non-collective $\Delta N^{-1}$ modes gives rise to some collective
strength among a few of the $\Delta N^{-1}$ modes
for certain small intervals of the momentum $q$,
see fig.\ \ref{fig_DispT0r10}.
It may be debated whether the contribution from these collective modes
to $\Gamma_\Delta^{\rm tot}$ should be associated with
$\Gamma_\Delta^{\Delta N^{-1}}$, $\Gamma_\Delta^{\tilde{\pi}_1}$,
or something else.
In figs.\ \ref{fig_GamPar} and \ref{fig_GamParSC} we have indicated the
contribution from all such cases by error bars on the partial widths.

{}From the spin transverse channel we get a contribution to $\Gamma_\Delta^{\rm
tot}$  from the non-collective $N N^{-1}$ and $ \Delta N^{-1}$ modes, and two
collective  modes $\tilde{\rho}_1$ and $\tilde{\rho}_2$. However in the range
of invariant masses presented in figs.\ \ref{fig_GamPar} and
\ref{fig_GamParSC} the contributions $\Gamma_\Delta^{\tilde{\rho}_1}$
and $\Gamma_\Delta^{\tilde{\rho}_2}$ are negligible or small and therefore
not displayed in the figures.
The partial quantities $\Gamma_\Delta^{N N^{-1}}$  and
$\Gamma_\Delta^{\Delta N^{-1}}$ contain contributions from both spin
longitudinal and transverse channels.

In fig.\ \ref{fig_GamParSC} we present the partial widths based on
eq.\ (\ref{eq_GnuCompl}) with the approximation (\ref{eq_hlA}),
where the $\Delta$ width is included self-consistently.
In fig.\ \ref{fig_GamParSC}$a$ the  partial widths
are shown at zero temperature and normal nuclear density. Comparing the widths
for a decaying $\Delta$ near resonance ($m\approx m_\Delta$) with
the reference case $\Gamma_{\Delta N^{-1}} =0$ in fig.\ \ref{fig_GamPar}$a$,
we see that $\Gamma_\Delta^{\tilde{\pi}_1}$ is substantially reduced,
$\Gamma_\Delta^{N N^{-1}}$ is rather unaffected,
while we get a substantial contribution from the
$\Gamma_\Delta^{\tilde{\pi}_2}$ partial widths.
The relatively large value of $\Gamma_\Delta^{\tilde{\pi}_2}$
already at $m \approx m_\Delta$ occurs because it is now energetically possible
to form the mode $\tilde{\pi}_2$,
since it may exist at lower energies,
as compared to the case with $\Gamma_{\Delta N^{-1}} \equiv 0$.
The reduction of  $\Gamma_\Delta^{\tilde{\pi}_1}$ is understood
from the facts that the total width is rather unaffected by the inclusion of
$\Gamma_\Delta^{\Delta N^{-1}} = \Gamma_\Delta^{\rm tot}$
and that new competing decay channels have opened up.

\subsection{$\Delta$ cross sections}
\label{sec_cross-sec}
In this section we present results for cross sections for the processes
\begin{equation}
 N + N \rightarrow \Delta + N\ , \qquad N + \Delta \rightarrow N + N
\end{equation}
and
\begin{equation}
 \tilde{\pi}_j + N \rightarrow \Delta\ , \qquad j=1,2\ ,
\end{equation}
where $\tilde{\pi}_j$ denotes the collective spin-isospin mode in the
spin-longitudinal (pion like) channel.

In vacuum the cross sections for the two first processes will depend on the
transferred energy and momentum, ($\omega,\mathbf{q}$), in the relativistically
invariant form $\omega^2 - \mathbf{q}^2$.
It is therefore convenient to perform
the calculations in the center-of-mass system of the colliding particles, where
the total cross section only will depend on the total energy, $\sqrt{s}$,
of the colliding particles.
In the medium, however, the spin-isospin modes,
and thus the effective interaction,
no longer depend on ($\omega,\mathbf{q}$)
in the invariant form $\omega^2 - \mathbf{q}^2$,
see fig.\ \ref{fig_DispT0r10}.
This implies that the total cross sections,
apart from depending on the total center-of-mass energy,
will have additional dependences on the momenta of the colliding particles.
However, since our formalism for calculating the isospin modes
and the effective interaction is non-relativistic,
the calculations should be performed in the rest frame of the medium
(where the spin-isospin modes are calculated).
In order to study the medium effects,
we wish to compare to experimental and calculated cross sections in vacuum.
In this section we therefore calculate and present all cross sections for the
special case that the center-of-mass system of the colliding particles is
identical to the rest frame of the medium.

In fig. \ref{fig_dsdcosNN} we present the differential cross section
$d\sigma/d\cos(\theta_{\rm cm})$
for the process $N+N\to\Delta+N$
at a center-of-mass energy of $\sqrt{s}=2314\ \MeV$.
The figure shows different combinations of density and temperature.
The calculations of fig.\ \ref{fig_dsdcosNN}$a$
correspond to $\Gamma_{\Delta N^{-1}}=0$,
while the results of fig.\ \ref{fig_dsdcosNN}$b$
represent the self-consistent case.
Comparing the results in the nuclear medium with the results in vacuum
(dotted curve),
we find that in normal nuclear matter at zero temperature,
fig.\ \ref{fig_dsdcosNN},
the cross section is somewhat enhanced at forward and backward angles
and slightly suppressed at $\cos(\theta_{\rm cm}) \approx0$.
The enhancement at forward and backward angles occurs because the pionic mode
$\tilde{\pi}_1$ is lowered in the medium, see fig.\ \ref{fig_DispT0r10}.
The scattering matrix contains terms proportional to $[\hbar \omega - \hbar
\omega_\nu(\mathbf{q})]^{-1}$ for each spin-isospin mode,
see eq.\ (\ref{eq_Mrpa}).
The dominant term in the sum over the modes $\nu$ is
the pionic mode $\nu = \tilde{\pi}_1$,
and its energy $\hbar \tilde{\omega}_1(\mathbf{q})$,
keeping $\mathbf{q}$ fixed, decreases when the nuclear density increases.
Hence the difference $\hbar \omega - \hbar \tilde{\omega}_1(\mathbf{q})$
becomes smaller at forward or backward angles,
and the cross section is enhanced in the medium.
This effect is also responsible for the large enhancement of the cross section
at twice normal nuclear density seen in fig.\ \ref{fig_dsdcosNN}.
At the temperature $T=25\ \MeV$ the differential cross section
is only slightly reduced as compared to zero temperature.

Including the $\Delta$ width self-consistently, via $\Gamma_{\Delta N^{-1}}$ in
eq.\ (\ref{eq_PhiD}), reduces the cross section only slightly at normal
density,
fig.\ \ref{fig_dsdcosNN}$b$, but to a significant degree
at twice normal density, fig.\ \ref{fig_dsdcosNN}$b$.
This is because the pionic mode is regularized
by the imaginary part of the pion optical potential.
As a consequence,
the pionic mode has no longer a single well-defined energy,
but instead there is a Breit-Wigner-like distribution
of possible energies centered around Re $\hbar \tilde{\omega}_1$
with a width 2Im $\hbar\tilde{\omega}_1$.
In the former case,
when the pionic mode had a well-defined real energy, $\tilde{\omega}_1$,
the transferred energy and momentum comes close to this energy,
which causes a large enhancement of the cross section,
but in the  latter case the transferred energy and momentum
will only be close to some of the possible energies
around Re $\hbar \tilde{\omega}_1$ and thus only pick up
a fraction of the total strength in the pionic mode.
Technically,
this is taken care of by the imaginary part of $\hbar \tilde{\omega}_1$
which increases the difference $\hbar \omega-\hbar\tilde{\omega}_1(\mathbf{q})$
in the self-consistent case.
The reason why the effect of including
$\Gamma_{\Delta N^{-1}} = \Gamma_\Delta^{\rm tot}$
is so much larger at $\rho_N = 2 \rho_N^0$ than at $\rho_N = \rho_N^0$
is mainly that the transferred energy $\omega$
is closer to Re $\tilde{\omega}_1$ at $2 \rho_N^0$,
but also that Im $\tilde{\omega}_1$ is larger at $2 \rho_N^0$.
The magnitude of Im $\tilde{\omega}_1$ depends on the magnitude of
$\Gamma_\Delta^{\rm tot}$ and at $2\rho_N^0$
the width $\Gamma_\Delta^{\rm tot}$ is larger than at $\rho_N^0$.

In fig.\ \ref{fig_sTotNN} we present the total cross section for the process
$N + N \rightarrow \Delta + N$ as a function of the center-of-mass energy
$\sqrt{s}$ of the two colliding nucleons.
The different curves correspond to different densities and temperatures.
Fig.\  \ref{fig_sTotNN}$a$ displays $\sigma_{\rm tot}$ for
$\Gamma_{\Delta N^{-1}}=0$,
while fig.\ \ref{fig_sTotNN}$b$ shows $\sigma_{\rm tot}$ for the case when
$\Gamma_{\Delta N^{-1}}$ is included self-consistently.
In fig.\ \ref{fig_sTotNN}$a$ we notice that
the total cross section is enhanced by a factor 2--3 at twice
normal nuclear density, as compared to the value in vacuum, while at normal
nuclear density the cross section is similar to the vacuum value with only
minor changes.
As for the differential cross section,
the enhancement of $\sigma_{\rm tot}$ at $\rho_N = 2 \rho_N^0$
originates in the softening of the pionic mode.
The self-consistent inclusion of $\Gamma_{\Delta N^{-1}}$
has two major effects.
At normal nuclear density the cross section is enhanced in the
threshold region ($\sqrt{s}\approx2.1\ \GeV$) and somewhat reduced at energies
$\sqrt{s}\approx2.2$--$2.4\ \GeV$.
This effect arises from the possibility of creating $\Delta$ isobars
with masses lower than $m_N+m_\pi$ in the nuclear medium
and it is technically included by taking the $\Delta$ width in
eq.\ (\ref{eq_rhoDmass}) from the nuclear matter calculations
presented in sec.\ \ref{sec_GammaD}.
At twice normal density the cross section is reduced
at energies  $\sqrt{s}\approx2.2$--$2.4\ \GeV$ by about 10 mb,
as compared to the when case $\Gamma_{\Delta N^{-1}}=0$,
while it is enhanced significantly in the threshold region.
The origin of the reduction has already been discussed above
in connection with the discussion of the differential cross section.
The enhancement at low $\sqrt{s}$ occurs
because the $\Delta$ width is very large at low $\Delta$ energies
(see fig.\ \ref{fig_GamTot}).
The increase of the $\Delta$ width at low $\Delta$ masses
and $\rho_N=2\rho_n^0$ is partly an effect of the system being close to
the onset of pion condensation.
The associated enhancement of $\sigma_{\rm tot}$ is consistent with the picture
that when the system is close to pion condensation
the $\Delta$ isobars having low energy quickly decay
into low-energy pionic modes in the $NN^{-1}$ region.
The effects at $2\rho_N^0$ may be quantitatively changed somewhat
by taking into account Re $\Sigma_\Delta$ and its density dependence.

The total cross section for the reverse process $N + \Delta \rightarrow N+N$
is presented in fig.\ \ref{fig_sTotND}
for a $\Delta$ with mass $m=1230\ \MeV/c^2$.
As in fig.\ \ref{fig_sTotNN},
part ($a$) presents the results for $\Gamma_{\Delta N^{-1}}=0$
and part ($b$) corresponds to $\Gamma_{\Delta N^{-1}}=\Gamma_\Delta^{\rm tot}$.
Similar features as in fig.\ \ref{fig_sTotNN}
are noted in fig.\ \ref{fig_sTotND}.
The vacuum cross sections are only slightly reduced at $\rho_N = \rho_N^0$,
but for $\Gamma_{\Delta N^{-1}}=0$ greatly enhanced at $\rho_N = 2 \rho_N^0$
while  $\Gamma_{\Delta N^{-1}}=\Gamma_\Delta^{\rm tot}$ leads to a cross
section
of similar magnitude at all presented densities for $m=1230\ \MeV$.
Furthermore,
we note that $\sigma(N + \Delta \rightarrow N + N)$ is singular at threshold.
This singularity originates from the factor $|\mathbf{p}_1|/I^2$ in the cross
section, eq.\ (\ref{eq_dsdO}), which in the $N + \Delta$ center-of-mass
system becomes proportional to
$1/|\mathbf{p}_1|$ when $\sqrt{s} \rightarrow m_1 + m_2$.
As this limit is approached,
$|\mathbf{p}_1|$ will tend to zero and hence the cross section grows infinite.
Finally, we note that the cross sections at $\Delta$ masses $m<1230\ \MeV/c^2$
are smaller than at $m=1230\ \MeV/c^2$.

In fig.\ \ref{fig_sNnu} we present, for various nuclear densities and
temperatures, the cross section for the process p+$\tilde{\pi}_j^+
\rightarrow \Delta^{++}$, when $\Gamma_{\Delta N^{-1}}$ is included self
consistently.  In figs.\ \ref{fig_sNnu}$a$ and \ref{fig_sNnu}$b$ we present the
results for the lower pionic mode, $\tilde{\pi}_1$, and in figs.\
\ref{fig_sNnu}$c$ and  \ref{fig_sNnu}$d$ for the upper pionic mode,
$\tilde{\pi}_2$. The factor $\rho_3(m^2)$, defined in eq.\ (\ref{eq_rhoDmass}),
is in figs.\ \ref{fig_sNnu}$a$  and \ref{fig_sNnu}$c$
calculated from the free $\Delta$ width, while the total width in nuclear
matter has been used in figs.\ \ref{fig_sNnu}$b$ and \ref{fig_sNnu}$d$.

The cross sections are presented as a function of the invariant $\Delta$ mass,
$m$, which is equal to the total energy in the p+$\tilde{\pi}_j$
center-of-mass system.
The cross sections show a strong resonance peak at $m \approx m_\Delta$
for all densities and temperatures.
This shape is in eq.\ (\ref{eq_sNnu}) determined by the factor $\rho_3(m^2)$.
The magnitude of the
cross section is determined by this factor and the factors $|h^{l,t}(\Delta
N,\omega_\nu)|^2$ and $v_{\rm rel}/c$.
In vacuum, where only the real pion
dispersion relation exists in the spin longitudinal channel,
the factor
$|h^l|^2$ is proportional to $(c \mathbf{q})^2/\hbar \omega_\pi(\mathbf{q})$.
At finite densities the pion contribution to $|h^l|^2$  will be mixed with
contributions  from $\Delta N^{-1}$ and $ N N^{-1}$ states which depend very
weakly on the momentum and are approximately proportional to
$g_{\Delta \Delta}' \rho_N$ and $g_{N \Delta}' \rho_N$, respectively.
On the lower pionic mode the pion component dominates for low $q$ values,
while the $\Delta N^{-1}$ components dominate at larger $q$,
see fig.\ \ref{fig_AmpT0r10}$d$.
Therefore, the cross section for the lower pionic mode at finite densities will
be similar to the vacuum cross section for small values of $q$ or $m$,
while it will be smaller than the vacuum value for larger $q$.
This is seen in fig.\ \ref{fig_sNnu}$a$,
where the cross sections at finite densities deviates
from the vacuum values for $m \geq 1200\ \MeV/c^2$.
At normal nuclear density
and zero temperature the momentum $q$ is approximately  $270\ \MeV/c$ at the
resonance peak, $m\approx m_\Delta$.
At this momentum the $\Delta N^{-1}$ component in the pionic mode
is quite substantial, see fig.\  \ref{fig_AmpT0r10}$d$.
At twice normal density the contribution from the $\Delta N^{-1}$ component
is approximately doubled,
which is the main reason for the decrease of the cross section at the
resonance peak.
Moreover,
the value of the relative velocity will be changed at finite densities.
There are two competing effects on the lower pionic mode.
The velocity of the pionic mode is lower in the medium,
as compared to a pion in vacuum,
since $\hbar \tilde{\omega}_1$ increases less steeply with $q$ than
$\hbar \omega_\pi$.
On the other hand, the velocity of the nucleon will be larger,
since $m=m_\Delta = e_N(\mathbf{q}) + \hbar \tilde{\omega}_1$ is
obtained at a larger value of $q$,
because $\hbar \tilde{\omega}_1$ is lowered in the medium.
The net results are seen in figure \ref{fig_sNnu}.
At finite  temperatures the cross section is
similar to the case of zero temperature.

In fig.\ \ref{fig_sNnu}$b$ we have used the $\Delta$ width in the nuclear
medium to calculate the factor $\rho_3(m^2)$, rather than  the free width,
used in fig.\ \ref{fig_sNnu}$a$.
In the nuclear medium the $\Delta$ width is
larger than the free width up to invariant masses around $1400\ \MeV/c^2$.
This explains the reduction of all cross sections at the resonance peak in
fig.\ \ref{fig_sNnu}$b$ relative to fig.\ \ref{fig_sNnu}$a$,
since at the resonance peak the factor $\rho_3(m_\Delta^2)$
is proportional to $1/\Gamma_\Delta$.
Note also that the cross sections for the lower pionic mode
in fig.\ \ref{fig_sNnu}$b$ are substantially enhanced for low invariant masses.
Also here the effects at $2\rho_N^0$ may be quantitatively changed somewhat
by taking into account Re $\Sigma_\Delta$ and its density dependence.

On the upper pionic mode the relation of the strength of the pionic and $\Delta
N^{-1}$ components are opposite, and thus the cross section will approach the
vacuum value at large $q$, while it will be smaller at low $q$,
see figs.\ \ref{fig_sNnu}$c$ and \ref{fig_sNnu}$d$.
Note also that the invariant $\Delta$ mass will always be larger
than $m_\Delta$ on this mode,
since $\hbar\tilde{\omega}_2$ starts just above $m_\Delta/c^2 - m_N/c^2$.
In the limit $q\rightarrow0$ the cross section diverges,
since in this limit also the relative velocity approaches zero.

\subsection{Implications for transport descriptions}
\label{sec_transport}
In this section we discuss how our results could be incorporated in a dynamical
transport simulation of a heavy-ion collision, and some of the consequences
this could lead to.
We will carry out this discussion within the framework of a standard
quasiparticle description that propagates nucleons ($N$),
delta isobars ($\Delta$), and pions ($\pi$) as ingredients
and we will refer to this as the ``standard" transport description.
In such a model,
many of the $\Delta$ and $\pi$ properties,
such as decay widths, cross sections, and dispersion relations
are usually taken as the properties in vacuum.
In this section we will discuss how the vacuum properties
can be replaced by in-medium properties in a consistent way.

In a transport simulation of heavy-ion collisions,
a nucleon-hole ($N N^{-1}$) excitation is produced
by promoting a nucleon from below to above the Fermi surface.
This can occur as a result of the nucleon colliding with another particle.
The new  $N N^{-1}$ state is then by construction non-collective
and unperturbed,
\ie\ its energy is given by $E_{\rm particle} - E_{\rm hole}$.
Thus the non-collective $N N^{-1}$ spin-isospin modes
that we have found in nuclear matter
are already incorporated in the standard transport description.
But the energy of the $N N^{-1}$ state is given by the respective
quasiparticle energies and so it is not quite correct
and should in principle be slightly changed,
in accordance with the energies of the non-collective $N N^{-1}$
modes presented in sec.\ \ref{sec_ResSIM}.
However,
the energies of the non-collective $N N^{-1}$ states are only slightly shifted
from the respective unperturbed energies, and it should thus be a good
approximation to neglect this change of the energy.
Similarly,
the non-collective $\Delta N^{-1}$ modes in sec.\ \ref{sec_ResSIM}
correspond in a transport description to
the conversion of an individual nucleon to a $\Delta$ isobar.

The incorporation of the two collective spin-isospin modes is more involved.
These modes can be regarded as separate particles of pionic character,
$\tilde{\pi}_1$and $\tilde{\pi}_2$,
and treated in a manner analogous to the standard treatment of the pion.
Since the pion is then fully included in the description,
it should no longer be treated explicitly.
The propagation of the two collective pionic modes is governed by
the effective Hamiltonians
\begin{eqnarray}
  \tilde{H}_1(\r,\q) & = &
      \mbox{ Re } \hbar \omega_{1}(\q;\rho_N(\r),T(\r))\ \equiv\
                  \hbar \tilde{\omega}_1\ ,
  \nonumber \\
  \tilde{H}_2(\r,\q) & = &
      \mbox{ Re } \hbar \omega_{2}(\q;\rho_N(\r),T(\r))\ \equiv\
      \hbar \tilde{\omega}_2\ ,
\label{eq_Hqpion}
\end{eqnarray}
where $\hbar \omega_{1}$ and $\hbar \omega_{2}$ are the energy-momentum
relations for the lower and upper collective modes discussed
in sec.\ \ref{sec_ResSIM} and
displayed in fig.\ \ref{fig_DispT0r10}$a$ for $\rho_N=\rho_N^0$ and $T=0$. Note
that the spatial dependence of $\tilde{H}_1(\r,\q)$ is incorporated by
representing $\rho_N(\r)$ and $T(\r)$ as local quantities. Moreover, in the
collision term the process for the production and absorption of pions in the
standard description, $\Delta \leftrightarrow N + \pi$, should be replaced by
the two distinct processes
\begin{equation} \Delta
  \leftrightarrow N + \tilde{\pi}_1 \quad \mbox{ and } \quad \Delta
  \leftrightarrow N + \tilde{\pi}_2\ .
\label{eq_Ddecay}
\end{equation}
The $\Delta$ decay is governed by the $\Delta$ decay width in the medium
to these two specific channels, $\tilde{\Gamma}_\Delta^{\tilde{\pi}_j}$.
These partial widths, presented in sec.\ \ref{sec_GammaD},
should be employed in the same manner as the free width,
\ie\ they describe the probability for the $\Delta$ isobar
to decay into a nucleon and a pion.
The only difference is that several collective pionic modes are available
in the final state.
The reverse processes in (\ref{eq_Ddecay}) are characterized by
the cross sections that were discussed in sec.\ \ref{sec_cross-sec}.

The self-consistent inclusion of the $\Delta$ width  in the calculation of the
spin-isospin modes encompasses decay processes like
\[
   \tilde{\pi}_j    \rightarrow    \Delta N^{-1}
                    \rightarrow    (N+\tilde{\pi}_k) N^{-1}\ .
\]
However,
since such processes are already explicitly contained
in the transport simulation by processes like
\[
   \tilde{\pi}_j + N   \rightarrow    \Delta
                       \rightarrow    N+\tilde{\pi}_k \ ,
\]
it would not be correct to include the entire self-consistent $\Delta$ width
when calculating the collective modes to be used in the transport description.
Instead it is more correct to use the results obtained with
$\Gamma_{\Delta N^{-1}}=0$, both for the energies of the modes,
$\hbar \tilde{\omega}_j$, and for the partial $\Delta$ widths
to be used in the decays $\Delta \rightarrow N +\tilde{\pi}_j$.
Still, it is important to use the self-consistent width
when calculating the cross sections for processes like
$ N + N \rightarrow \Delta + N$,
where the spin-isospin modes provide an intermediate effective interaction.

Although the collective pionic modes can thus be effectively treated  as
ordinary particles, the fact that their wave functions contain components from
$\pi$, $NN^{-1}$ and $\Delta N^{-1}$ states makes it difficult to picture them
in a physically simple manner.
Fortunately, their specific structure is irrelevant,
as as long as these quasiparticles remain well inside the nuclear
medium. First when such a quasiparticle penetrates a nuclear surface and
emerges as a free particle is it physically meaningful to determine what
kind of real particle it is.
The gradual transformation of the collective quasiparticle
is automatically taken care of within the formalism,
because as the density is lowered, $\rho_N \rightarrow 0$,
the pionic modes will acquire 100\%
of either the pion component or the $\Delta N^{-1}$ component,
depending on $\omega$ and $q$.
That is to say,
they will turn into either a free pion or an unperturbed $\Delta N^{-1}$ state.
There remains the practical problem of how to represent
an unperturbed $\Delta N^{-1}$ state when $\rho_N \rightarrow0$.
However, we anticipate that only a very small fraction of the pionic modes
will emerge as  unperturbed $\Delta N^{-1}$ states.
For the lower pionic mode,
it is only for large $q$ that the $\Delta N^{-1}$ component will dominate at
low densities, but at large $q$ the lower pionic mode starts gradually
to lose its collective character and there is therefore a low probability
for creating the lower pionic mode at large values of $q$,
see fig.\ \ref{fig_GamParNPB}.
On the other hand, for the upper mode it is at low $q$ that the
$\Delta N^{-1}$-component will dominate.
In cold nuclear matter the $\Delta$ decay at low energies
(which corresponds to low $q$) is strongly reduced,
because the nucleon produced in the decay is Pauli blocked.
In an actual heavy-ion simulation this reduction is smaller,
but we still expect that the number of upper pionic modes at low $q$ values
will be quite small.
These effects will be further investigated
and reported in a subsequent paper.

As was discussed in sec.\ \ref{sec_GammaD} the partial decay width
$\tilde{\Gamma}_\Delta^{\tilde{\pi}_j}$ is reduced at $T=25\ \MeV$ compared to
$T=0$. Thus in a heavy-ion collision there will be very few pionic modes
$\tilde{\pi}_1$ created in the hot region,
while in colder regions the $\Delta$ isobars near resonance
decay mainly into the pionic modes.
In sec.\ \ref{sec_GammaD} we also found, at twice normal nuclear density,
a large enhancement in the $\Delta$ width at $m\approx1200\ \MeV$,
signaling the onset of pion condensation.
Although our parameter set leads to pion condensation at $\rho_N > 2 \rho_N^0$
for an equilibrated infinite system,
it is not clear that the effect will be seen in a nuclear collision,
since the region where such high densities may be created is rather small,
and exists for a only short time.
These effects will also be further investigated.

The calculations of $\Gamma_\Delta$ presented in
figs.\ \ref{fig_GamTot}--\ref{fig_GamParSC}
take account of the Pauli blocking of the nucleon
in the $\Delta$ decay $\Delta \rightarrow N + \nu$.
In a transport description
the Pauli blocking of the nucleon is treated explicitly
and should thus not be included in the width of the $\Delta$.
In fig.\ \ref{fig_GamParNPB} we present the total and partial widths,
without Pauli blocking of the nucleon,
at different temperatures for the reference case ($\Gamma_{\Delta N^{-1}}=0$).

The total $\Delta$ decay width, has apart from the partial contributions
$\Gamma_\Delta^{\tilde{\pi}_j}$, also the partial contributions
$\Gamma_\Delta^{N N^{-1}}$ and $\Gamma_\Delta^{\Delta N^{-1}}$.
The partial width $\Gamma_\Delta^{N N^{-1}}$ gives the probability for the
$\Delta$ to decay into a nucleon and a $N N^{-1}$ state. In a transport
description, this implies that we initially have a $\Delta$ and after the decay
process we have two nucleons above the Fermi surface and a hole left in the
Fermi sea. But this is the same process as if the $\Delta$ would collide with a
nucleon below the Fermi surface to give two nucleons above the Fermi surface.
This process is normally already included in the collision term in a standard
transport description, and the probability for such a collision is given by the
cross section for the process $\Delta + N \rightarrow N + N$. In a transport
description it is therefore not correct to both include a $\Delta$ decay
according to $\Gamma_\Delta^{N N^{-1}}$ and a collision term with $\Delta + N
\rightarrow N + N$. Instead, the correct procedure should be to exclude
$\Gamma_\Delta^{N N^{-1}}$ and modify the cross section
$\sigma(\Delta + N \rightarrow N + N)$ to be the in-medium cross section.
Calculations of such in-medium cross sections is discussed
in sec.\ \ref{sec_cross-sec}.
In the same way,
$\Gamma_\Delta^{\Delta N^{-1}}$ should be excluded in a transport description,
and $\sigma(\Delta + N \rightarrow \Delta + N)$ be the in-medium cross section.

As mentioned in the introduction, the in-medium effects on the cross section
for $N+N\rightarrow\Delta+N$ have previously been investigated in
ref.\ \cite{Bertsch}.
That work corresponds to zero temperature, $T=0$,
and taking $\Gamma_{\Delta N^{-1}}=0$ in $\Phi_\Delta$, eq.\ (\ref{eq_PhiD}).
Our results in figs.\ \ref{fig_dsdcosNN} and \ref{fig_sTotNN}$a$
are in qualitative agreement with those of ref.\ \cite{Bertsch},
\ie\ a large enhancement of the cross section at high nuclear densities
caused by a lowering of the pionic mode.
However, as was also pointed out but not pursued in ref.\ \cite{Bertsch},
the imaginary part of the pion optical potential regularizes this effect.
This is seen by comparing
parts $a$ and $b$ of figs.\ \ref{fig_dsdcosNN} and \ref{fig_sTotNN},
where it is shown that the cross section at twice normal
nuclear density is considerably different if the $\Delta$ width is included
self-consistently in the formalism.
This points towards the importance of incorporating
the $\Delta$ width consistently within the model for obtaining the
correct magnitude of $\Delta$ cross sections at large nuclear densities.

Since the $N + \Delta \rightarrow N + N$ cross section is not experimentally
known one has traditionally in transport descriptions used the  $ N + N
\rightarrow \Delta + N$ vacuum cross section and detailed balance. As pointed
out by previous authors, for example \cite{GiessenDB,DanBertsch} it is then
important to take into account the finite width of the $\Delta$, \ie\ use an
expression according to
\begin{equation}
  \frac{ d\sigma(N + \Delta \rightarrow N + N)}{d\Omega} =
  \frac{1}{N_{\rm f}} \frac{ p_{\rm f}^2 }{ p_{\rm i}^2 }
  \frac{1}{\rho_3(m^2)}
  \frac{ d\sigma(N + N \rightarrow \Delta + N)}{d\Omega d m^2}\ ,
\label{eq_DetBal}
\end{equation}
where $p_{\rm i}$ and $p_{\rm f}$ are magnitudes of the initial and final
momenta in the center-of-mass system of the $N\Delta$ system,
$N_{\rm f}$ is a spin-isospin factor and $\rho_3(m^2)$
is defined in eq.\ (\ref{eq_rhoDmass}).
The condition in eq.\ (\ref{eq_DetBal}) is automatically
satisfied within our formalism.

In refs.\ \cite{Giessen,Texas} the authors included some medium effects of the
pionic modes. In these works a simple form of the pion polarization function
$\Pi_\pi$ were used.
This simple form originates from approximating the
continuum of non-interacting $\Delta N^{-1}$ states to a single state with
energy $\hbar \omega_\Delta = m_\Delta - m_N + (qc)^2/2m_\Delta$,
and neglecting the nucleon-hole states.
When the interaction between the $\Delta N^{-1}$ state and the pion
is turned on two collective states emerge
(but contrary to our model, no non-collective $\Delta N^{-1}$ states are left).
This approximation leads to some differences
as compared to the more complete treatment in this paper.
As seen in fig.\ \ref{fig_DispT0r10}, the lower collective mode $\tilde{\pi}_1$
disappears (\ie\ loses its collective strength)
when it enters the non-collective $\Delta N^{-1}$ region.
This is not the case for the approximation used in
\cite{Giessen,Texas}, where the lower  pionic mode exists for all momenta $q$.
Furthermore, the change in the dispersion relations
$\hbar\omega_\nu(\mathbf{q})$, relative to the relations in vacuum,
are somewhat overestimated
when the approximate form of the polarization function $\Pi_\pi$ is used,
as compared the the more complete treatment in this work.
Finally,
no justification was given for the omission of the $\Delta$ width
in the pion polarization function.

To avoid the problem of how to treat a non-interacting $\Delta N^{-1}$ state
penetrating the nuclear surface (\ie\ when $\rho_N \rightarrow 0$) in the
transport description, the authors in \cite{Giessen} derive effective
dispersion relations corresponding to asymptotically free pions or
$\Delta N^{-1}$ states.
However, the energies in the effective dispersion relations are
quite different from the original ones, especially for large momenta $q$.
At $q\approx700\ \MeV/c$ the difference is as large as 150--200 MeV.

In ref.\ \cite{Giessen} the authors propagate only the collective mode
that corresponds to the asymptotically free pions ($\tilde{\pi}_2$).
The other collective mode is identified with
the propagation of individual $\Delta$ isobars.
This appears to be incorrect, since (as we have argued)
the propagation of individual isobars should be
identified with the (remaining) non-collective $\Delta N^{-1}$ states,
and both collective modes should be treated on an equal footing
in the transport description.
The authors in ref.\ \cite{Giessen} also used the energy of the collective mode
that corresponds to the asymptotically free $\Delta N^{-1}$ states
($\tilde{\pi}_2$)
to estimate the density dependence of the $\Delta$ potential.
This had a large effect on their results.
However,
our calculations show that the non-collective $\Delta N^{-1}$ modes
have their energies very close to the unperturbed energies,
and thus change very little with the nuclear density.
We therefore feel that it is inappropriate to use the properties
of the collective modes to deduce any density dependence
of the $\Delta$ potential to be used for
the explicit propagation of uncoupled $\Delta$ isobars.

The authors of ref.\ \cite{Texas} do treat the two collective modes
on an equal footing, propagating both of them as quasipions
in accordance with the dispersion relation implied by their model,
and their model also contains
explicit propagation of uncoupled $\Delta$ isobars.
They found that only a small fraction of the quasipions
approach free $\Delta$-hole states as they penetrate the nuclear surface.
These quasipions were approximated by on-shell pions at the surface,
by changing the momenta of the quasipions.

Furthermore,
in ref.\ \cite{Texas} the authors calculated partial widths
for the $\Delta$ decay to a nucleon and a quasipion.
The authors state that there are also
contributions to the total $\Delta$ width from decay processes as
$\Delta \rightarrow N + N N^{-1}$ and $\Delta \rightarrow N + \Delta N^{-1}$,
which are not taken into account since these processes are already included in
the transport simulations.
However, by using the simple form of the pion polarization function,
the $\Delta N^{-1}$ continuum is compressed to a single state.
But this state contains all the strength of the $\Delta N^{-1}$ continuum,
and therefore the partial $\Delta$ width for this channel will also contain
some contribution from the $\Delta \rightarrow N + \Delta N^{-1}$ decay.

Neither of the works \cite{Giessen} nor \cite{Texas}
have taken into account any modifications of cross sections
for processes involving a $\Delta$.

\section{Summary}
\label{sec_Summary}
We have investigated the properties of spin-isospin modes in an infinite system
of interacting nucleons, $\Delta$ isobars, and $\pi$ and $\rho$ mesons
at various densities and temperatures.
The aim has been to derive and discuss quantities
that can be incorporated in a transport description of a heavy-ion collision,
by use of a local density and temperature approximation.

Within the random-phase approximation we have derived dispersion relations for
the spin-isospin modes and the amplitudes of the their components.
While the dispersion relations yield the energy-momentum relation of each mode,
the character of the modes is determined by the amplitudes of the different
components.
In both the spin-longitudinal and spin-transverse channels,
we find two collective modes,
while the remaining modes are non-collective in  their nature
(except in limited regions in $q$ for a few specific modes).
The non-collective modes correspond in a transport description
to propagation of uncoupled nucleons and $\Delta$ isobars,
while the collective modes correspond to propagation of quasimesons.
These quasimesons can be incorporated in a transport description
in a manner analogous to how real pions have been incorporated
in standard treatments based on vacuum properties.
One notable feature of the lower pionic mode in the spin-longitudinal channel
is that it gradually loses its collective character when it enters the
region of of non-collective modes.
This implies that a $\Delta$ isobar with sufficiently
high energy cannot decay to this mode.
Previous works \cite{Giessen,Texas} that included some in-medium effects
employed a simpler model for the collective modes.
In that simpler model the lower collective mode exists for all momenta $q$
and can thus be excited by a decaying $\Delta$ at any energy,
in contrast to our more refined results.

The decay of a $\Delta$ isobar into a nucleon and a spin-isospin mode
is governed by the partial $\Delta$ decay widths.
Therefore, we have calculated total and partial $\Delta$ decay widths
within the model.
At twice normal density the total $\Delta$ width is significantly enhanced
at low $\Delta$ energies.
However,
this enhancement is mainly associated with the decay
to non-collective nucleon-hole modes,
while instead the partial width for the decay to the lower pionic mode
$\tilde{\pi}_1$ is reduced.
This effect is also different from the previous works \cite{Giessen,Texas}
where the nucleon-hole channel was neglected
in the calculation of $\Gamma_\Delta$.
At finite temperatures up to $T=25\ \MeV$ the
total $\Delta$ width is almost unaffected,
while the partial widths change somewhat with $T$.
The dependence is stronger when
$\Gamma_{\Delta N^{-1}}=0$ than for the self-consistent case,
$\Gamma_{\Delta N^{-1}} = \Gamma_{\Delta}^{\rm tot}$.

The partial $\Delta$ widths representing decay to non-collective modes
correspond in transport models to processes like
$\Delta + N  \rightarrow N+N$.
Since these processes are already explicitly included in the
transport description, these partial widths should be ignored, while the
corresponding cross sections should contain the in-medium modifications.
Examples of such in-medium cross sections have been presented.
At center-of-mass energies $\sqrt{s} \approx 2.3\ \GeV$,
we have for the reference case $\Gamma_{\Delta N^{-1}}=0$
found a large increase in
$\sigma(N+N \rightarrow \Delta + N)$ at $\rho_N = 2 \rho_N^0$
as compared with $\rho_N=\rho_N^0$,
in agreement with ref.\ \cite{Bertsch}.
However, this effect is substantially reduced in the self-consistent treatment,
 $\Gamma_{\Delta N^{-1}} = \Gamma_{\Delta}^{\rm tot}$.
Instead,
the cross section is significantly enhanced at low center-of-mass energies,
($\sqrt{s} \sim 2.0-2.2\ \GeV$).
This was not investigated in  ref.\ \cite{Bertsch}.
Furthermore, the cross section of the process $\Delta + N  \rightarrow N+N$
is reduced when  $\Gamma_{\Delta N^{-1}}$ is included self-consistently,
compared to when $\Gamma_{\Delta N^{-1}}$ is zero.
We have found that all the calculated cross sections are almost independent
of temperature up to $T=25\ \MeV$.

In a forthcoming paper we will incorporate the in-medium effects presented in
this paper into a microscopic transport model. We will study the importance of
the these in-medium properties in heavy-ion collisions and compare to the
previous treatments \cite{Bertsch,Giessen,Texas}.
We expect that the $\Delta$ production  will be enhanced,
especially at low center-of-mass energies of the colliding nucleons,
but we also expect that the $\Delta$ decay to a pionic mode
will be somewhat reduced.
This could lead to a faster thermalization of the system
and possibly slightly less real pions produced.
The net effect, however,
is difficult to predict without an explicit transport simulation,
since an average over different densities and temperatures will be taken,
and the pionic modes that will escape the system as real pions will be produced
mainly at the nuclear surface where the in-medium effects are small.

The model presented in this paper constitutes a more consistent way of
obtaining and incorporating in-medium effects in transport descriptions of
heavy-ion collisions than previous works \cite{Bertsch,Giessen,Texas}.
However,
also in the model presented in this paper some approximations and assumptions
have been made, and there is room for further improvements.
We expect that our model will be applicable in transport models
up to moderately large bombarding energies of about 1 GeV per nucleon,
since it is assumed within that the density of $\Delta$ isobars and pions
is relatively small.
Furthermore,
we have so far not included any density dependence of the
coupling constants, $g'$ correlation parameters, form factors, and meson
masses,
such as may result from a possible partial chiral restoration,
since we feel that such effects are not very well known,
at the present stage.\\

This work was supported by the Swedish Natural Science Research Council,
by the National Institute for Nuclear Theory at the University of Washington
in Seattle,
and by the Director, Office of Energy Research, Office of High Energy
and Nuclear Physics, Nuclear Physics Division of the U.S. Department of
Energy under Contract No.\ DE-AC03-76SF00098.

\appendix
\section{Solution to the RPA equations for spin-isospin interaction}
\label{sec_RPAsolu}

In this section we show how to obtain the eigenenergies of the RPA equations
(\ref{eq_RPAMx}) in the spin longitudinal channel using the interactions
defined by eqs.\ (\ref{eq_Vnpn}) to (\ref{eq_Vg}). The
eigenenergies in the spin transverse channel is obtained analogously.

We wish to find an spin-isospin excitation propagating with momentum
$\mathbf{q}$ and isospin $\lambda$. To such a spin-isospin mode that has a
momentum $\mathbf{q}$, only the baryon pairs that has the relative
momentum $\mathbf{q}$ will contribute. For this purpose we need to restrict the
summation over all baryon states in eqs.\ (\ref{eq_Qrpal}) and
(\ref{eq_Qrpat}) to those that will have the relative momentum $\mathbf{q}$.
In the same way we restrict the sum over meson states to those with momentum
$\mathbf{q}$ and isospin $\lambda$. Therefore we take
\beqar\label{eq_Xdeltaq}
  X_{jk} &\to& X_{jk}(\omega,\mathbf{q},\lambda)\
                     \delta_{\mathbf{p}_j,\mathbf{p}_k + \mathbf{q}}\ , \\
  Z_r &\to&
        Z(\omega,\mathbf{q},\lambda)\ \delta_{\mathbf{q}_r,\mathbf{q}}\
                                     \delta_{\lambda_r,\lambda}\ ,\\
  W_r &\to&
        W(\omega,\mathbf{q},\lambda)\  \delta_{\mathbf{q}_r,-\mathbf{q}}\
                                      \delta_{\lambda_r,-\lambda}\ .
\eeqar
In $\mathbf{q}_{cm}$ (eq.\ (\ref{eq_qcm})) we will neglect the term
$\mathbf{p}_N \, (\hbar \omega)/(m_N c^2 + \hbar \omega) $ which is small in
$N N^{-1}$ or $\Delta N^{-1}$ states,
since the hole momentum $\mathbf{p}_N$ is small, and take
\begin{equation}
  \mathbf{q}_{cm} \approx \mathbf{q}_i = \frac{m_N c^2}
                          {m_N c^2 + \hbar \omega} \mathbf{q}\ .
\label{eq_qcmI}
\end{equation}
Furthermore we take
\begin{equation}
  \sqrt{s} = (E_N(\mathbf{p}_N) + \hbar \omega)^2 -
             (c \mathbf{q} + c \mathbf{p}_N)^2
          \approx  (m_N c^2 + \mbox{ Re } \hbar \omega)^2 - c(\mathbf{q})^2
         \equiv \sqrt{s_i}\ .
\end{equation}
With these approximations we can make the ansatz,
\begin{equation}
  X_{jk}(\omega,\mathbf{q},\lambda) =
     \left( \frac{\hbar c}{L} \right)^{3/2}
     \frac{x(t_j,t_k; \, \mathbf{q}; \, \lambda, \, \omega)}
          {e_{t_j}(\mathbf{p}_k + \mathbf{q}) - e_{t_k}(\mathbf{p}_k)
           + \Sigma_{jk} -
           \hbar \omega} \; \vartheta_{j k}(\hat{q},-\lambda)\ ,
\label{eq_Xansatz}
\end{equation}
We consider the case when  $x(3/2,3/2; \, \mathbf{q}; \, \lambda, \, \omega)
\equiv 0$. We then write  $x(1/2,1/2) \equiv x^N$, $x(1/2,3/2) = x(3/2,1/2)
\equiv x^{\Delta}$, and analogously for other quantities.
The RPA equations can then be written in the matrix form
\begin{equation}
  \left(
         \begin{array}{cc}
              1 + {\cal W}^{NN} {\cal M}^N \Phi^N
         &    {\cal W}^{N \Delta} {\cal M}^\Delta \Phi^\Delta       \\
              {\cal W}^{\Delta N} {\cal M}^N \Phi^N
         &    1 + {\cal W}^{\Delta \Delta} {\cal M}^\Delta \Phi^\Delta
         \end{array}
  \right)
  \left(
         \begin{array}{c}
           x^N \\ x^\Delta
         \end{array}
  \right) = 0\ ,
\label{eq_Xx}
\end{equation}
and
\begin{eqnarray}
  Z & = &
      \frac{1}{\hbar \omega_\pi - \hbar \omega}
      \left[ {\cal M}^N v_\pi^N \Phi^N x^N +
             {\cal M}^\Delta v_\pi^\Delta \Phi^\Delta x^\Delta \right]
      \label{eq_Zx}  \\
  W & = &
      \frac{1}{\hbar \omega_\pi + \hbar \omega}
      \left[ {\cal M}^N v_\pi^N \Phi^N x^N +
             {\cal M}^\Delta v_\pi^\Delta \Phi^\Delta x^\Delta \right]\ ,
      \label{eq_Wx}
\end{eqnarray}
with
\begin{eqnarray}
  {\cal W}^{\alpha \beta} & = &
       \frac{f^\pi_{N \alpha} f^\pi_{N \beta}}{(m_\pi c^2)^2}
       [ \: |F_g|^2 g_{\alpha \beta}' +
         R^\alpha_i R^\beta_i
         |F_\pi|^2
         (c \mathbf{q}_i)^2 D^0_\pi\: ], \\
  v_{\pi}^\alpha(\mathbf{q},\omega)
 & = &
       i \, R^\alpha_i
       F_{\pi}
       \frac{ f^\pi_{N \alpha} }{m_{\pi} c^2}
       \frac{|c \mathbf{q}_{cm}|}
            {\sqrt{2 \hbar \omega_{\pi}(\mathbf{q})}}      \\
  R_\alpha^i(q)^2    & = & \frac{ 2m_\alpha c^2}{m_\alpha c^2 + \sqrt{s_i} }\ ,
\label{eq_vpi}
\end{eqnarray}
where $\alpha,\beta = N,\Delta$, and where we have defined the Lindhard
functions
\begin{eqnarray}
  \Phi^N(\omega, \mathbf{q}) & = &
     \left( \frac{\hbar c}{L} \right)^3 \sum_{\mathbf{p}}
     \frac{n(\mathbf{p}) - n(\mathbf{p} + \mathbf{q})}
          { (\mathbf{p} + \mathbf{q})^2/ 2 m_N^*  -
             \mathbf{p}^2/2 m_N^* - \hbar \omega}
\label{eq_PhiN} \\
  \Phi^\Delta(\omega, \mathbf{q}) & = &
     \Phi(\frac{1}{2},\frac{3}{2}; \: \omega, \mathbf{q}) +
     \Phi(\frac{3}{2},\frac{1}{2}; \: \omega, \mathbf{q}) =
     \left( \frac{\hbar c}{L} \right)^3 \sum_{\mathbf{p}}
     \left\{ \frac{n(\mathbf{p})}{\delta e^+_{\Delta N} } +
             \frac{n(\mathbf{p})}{\delta e^-_{\Delta N} } \right\}
\label{eq_PhiD}
\end{eqnarray}
with
\begin{eqnarray}
  \delta e^+_{\Delta N} & = &
          \frac{ (\mathbf{p}_\Delta^+)^2 }{ 2 m_\Delta } -
            \frac{ \mathbf{p}^2}{ 2 m_N^* } + \Delta m
            + V_\Delta^{\rm eff}(\varepsilon_\Sigma^+, \mathbf{p}_\Delta^+) -
            i \Gamma_{\Delta N^{-1}}(\varepsilon_\Sigma^+,
                                     \mathbf{p}_\Delta^+)/2 -
            \hbar \omega  \\
  \delta e^-_{\Delta N} & = &
           \frac{ (\mathbf{p}_\Delta^-)^2 }{ 2 m_\Delta } -
            \frac{ \mathbf{p}^2}{ 2 m_N^* } + \Delta m
            + V_\Delta^{\rm eff}( \varepsilon_\Sigma^-, \mathbf{p}_\Delta^-) -
            i \Gamma_{\Delta N^{-1}}(\varepsilon_\Sigma^-,
                                     \mathbf{p}_\Delta^-)/2 +
            \hbar \omega
\end{eqnarray}
\begin{equation}
  \Delta m = m_\Delta - m_N; \qquad
  \varepsilon_\Sigma^\pm =
        m_N + \frac{\mathbf{p}^2}{2 m_N^*} \pm \hbar \omega; \qquad
  \mathbf{p}_\Delta^\pm = \mathbf{p} \pm\ \mathbf{q}\ .
\end{equation}
The
numerical factors ${\cal M}^N = 4$ and ${\cal M}^\Delta = 16/9$ originates from
the spin-isospin summation.

The eigenenergies are obtained from
\begin{equation}
  \det \left(
         \begin{array}{cc}
              1 + {\cal W}^{NN} {\cal M}^N \Phi^N
         &    {\cal W}^{N \Delta} {\cal M}^\Delta \Phi^\Delta       \\
              {\cal W}^{\Delta N} {\cal M}^N \Phi^N
         &    1 + {\cal W}^{\Delta \Delta} {\cal M}^\Delta \Phi^\Delta
         \end{array}
  \right) = 0\ ,
\label{eq_det1}
\end{equation}
with the approximation $\Gamma_{\Delta N^{-1}}(\varepsilon_h + \hbar \omega)
\approx  \Gamma_{\Delta N^{-1}}(\varepsilon_h + \mbox{Re } \hbar \omega)$. The
eigenenergies of the auxiliary equations are obtained from
\begin{equation}
  \det \left(
         \begin{array}{cc}
            1 + \tilde{{\cal W}}^{NN} {\cal M}^N \tilde{\Phi}^N
         &  \tilde{{\cal W}}^{N \Delta} {\cal M}^\Delta \tilde{\Phi}^\Delta
         \\
            \tilde{{\cal W}}^{\Delta N} {\cal M}^N \tilde{\Phi}^N
         &  1 + \tilde{{\cal W}}^{\Delta \Delta} {\cal M}^\Delta
            \tilde{\Phi}^\Delta
         \end{array}
  \right) = 0\ .
\label{eq_det2}
\end{equation}
Since the normal RPA equations (\ref{eq_RPAMx}) and the auxiliary RPA equations
differ only in the matrix $A^{(1)}$,
we can use the relationships $\Phi^N(\omega^*) = \Phi^N(\omega)^*$,
${\cal W}^{\alpha \beta}(\omega^*) = {\cal W}^{\alpha \beta}(\omega)^*$,
and $\tilde{\Phi}^\Delta(\omega^*) = \Phi^\Delta(\omega)^*$
to show that if $\omega$ is a solution of eq.\ (\ref{eq_det1}),
then $\tilde{\omega} = \omega^*$ is a solution of eq.\ (\ref{eq_det2}).

{}From the normalization condition (\ref{eq_Normz}) we obtain
\begin{eqnarray}
  \pm 1 & = &
           {\cal M}^N \eta_N  \tilde{x}_N^* x_N +
           {\cal M}^\Delta \eta_\Delta  \tilde{x}_\Delta^* x_\Delta +
           \tilde{Z}^* Z - \tilde{W}^* W = \nonumber \\
        & = &
  \tilde{x}_N^* x_N
  \left( {\cal M}^N \eta_N +
          \left[ \frac{ 1 + {\cal W}^{NN} {\cal M}^N \Phi^N }
               { {\cal W}^{N \Delta} {\cal M}^\Delta \Phi^\Delta }
          \right]^2 {\cal M}^\Delta \eta_\Delta
         - \right. \nonumber \\
  & - &
      \qquad
      \left[ \frac{1}{(\hbar \omega_\pi - \hbar \omega)^2} -
             \frac{1}{(\hbar \omega_\pi + \hbar \omega)^2} \right]
      \nonumber \\ &  &  \qquad \left. \times
      \left\{ {\cal M}^N v_\pi^N \Phi^N -
              \frac{1 + {\cal W}^{NN} {\cal M}^N \Phi^N}{{\cal W}^{N \Delta}}
               v_\pi^\Delta
      \right\}^2 \right)\ ,
\label{eq_SoNormzFi}
\end{eqnarray}
with
\begin{eqnarray}
  \eta_N(\omega, \mathbf{q}) & = &
     \left( \frac{\hbar c}{L} \right)^3 \sum_{\mathbf{p}}
     \frac{n(\mathbf{p}) - n(\mathbf{p} + \mathbf{q})}
          { [ \frac{ (\mathbf{p} + \mathbf{q})^2 }{ 2 m_N^* } -
              \frac{ \mathbf{p}^2}{ 2 m_N^* } - \hbar \omega]^2 } =
     \frac{\partial}{\partial \hbar \omega} \Phi_N(\omega, \mathbf{q}) \\
  \eta_\Delta(\omega, \mathbf{q}) & = &
     \left( \frac{\hbar c}{L} \right)^3 \sum_{\mathbf{p}} \left\{
     \frac{n(\mathbf{p})
         [1 + i \partial \Gamma_{\Delta N^{-1}}(\varepsilon_h +
          \mbox{ Re } \hbar \omega, \mathbf{p}_\Delta)/
          2 \partial \hbar \omega]}
          { [ \delta e^+_{\Delta N} ]^2} \right.   \nonumber \\
  & - &  \left.
     \frac{n(\mathbf{p})
         [1 + i \partial \Gamma_{\Delta N^{-1}}(\varepsilon_h -
          \mbox{ Re } \hbar \omega, \mathbf{p}_\Delta)/
          2 \partial \hbar \omega]}
          { [ \delta e^-_{\Delta N} ]^2} \right\}   \nonumber \\
  & = &
     \frac{\partial}{\partial \hbar \omega} \Phi_\Delta(\omega, \mathbf{q})\ ,
\end{eqnarray}
where $\partial V_\Delta^{\rm eff}/\partial \hbar \omega \approx 0$.

The quantities of the auxiliary equations are related to the ordinary
quantities by
\begin{equation}
   \tilde{x}_b(\omega^*)^* = x_b(\omega)\ , \quad b = N,\Delta\ ;  \qquad
   \tilde{Z}(\omega^*)^* = - Z(\omega)\ , \qquad
   \tilde{W}(\omega^*)^* = - W(\omega)\ .
\end{equation}

It is straightforward to show that the solutions defined by eqs.\
(\ref{eq_Xansatz}), (\ref{eq_Zx}), (\ref{eq_Wx}), (\ref{eq_det1}) and
(\ref{eq_SoNormzFi}) satisfies eqs.\ (\ref{eq_RPAMx}), and that the
corresponding auxiliary solutions solves the auxiliary equations.

\section{The effective spin-isospin interaction}
\label{sec_AppEffI}
A spin-isospin mode exchanged between the two-baryon states 31 and 24,
as in fig.\ \ref{fig_EffInt},
acts like an effective interaction.
In this appendix,
we derive an expression for the effective spin-isospin interaction $M(34,12)$,
which consists of exchange of all the spin-isospin modes
present in our RPA approximation.
This effective interaction then appears naturally
in the calculations of the $\Delta$ width and $\Delta$-cross sections,
as presented in sec.\ \ref{sec_DeltaW}.
$M(34,12)$ is of the form
\begin{equation}
     M^{l,t}(34,12) =
        \vartheta^{l,t}(31)\ [\vartheta^{l,t}(24)]^*\ \bar{M}^{l,t}(34,12)\ ,
\end{equation}
where the spin-isospin matrix elements, $\vartheta^{l,t}(31)$,
are defined in eqs.\ (\ref{eq_spinMxl}) and (\ref{eq_spinMxt}).

As before we will present two different expressions for $M(34,12)$. The first
originates from the summation of non-interacting Green's functions according to
the diagrams in fig.\ \ref{fig_GsimGraph}, while the second expression
originates from the RPA expansion in eq.\ (\ref{eq_GrpaEx}). The first
expression becomes
\begin{eqnarray}
  \bar{M}^l(34,12) & = &
    \frac{ f^\pi_{31} f^\pi_{24} }{(m_\pi c^2)^2}
    \left[ D_\pi F_\pi^2 c^2 q^2_{\rm eff}(34,12) + F_g^2 g'_{\rm eff}(34,12)
    \right] \\
  \bar{M}^t(34,12)  & = &
    \frac{ f^\rho_{31} f^\rho_{24} }{(m_\rho c^2)^2}
    D_\rho F_\rho^2 c^2 q^2_{\rm eff}(34,12) +
    \frac{ f^\pi_{31} f^\pi_{24} }{(m_\pi c^2)^2} F_g^2 g'_{\rm eff}(34,12) \ .
\end{eqnarray}
With
\begin{equation}
  \hbar \omega = e_3 - e_1 = e_2 - e_4, \qquad
  \mathbf{q} = \mathbf{p}_3 - \mathbf{p}_1 = \mathbf{p}_2 - \mathbf{p}_4\ ,
\end{equation}
the dressed pion, $D_\pi$, and $\rho$-meson, $D_\rho$, propagators are written
\begin{equation}
  D_{\pi,\rho}(q)  =  [(D_\pi^0)^{-1} - \Pi_{\pi,\rho}]^{-1} =
                      [ (\hbar \omega)^2 - (c \mathbf{q})^2 -
                        (m_{\pi,\rho} c^2)^2
                        - \Pi_{\pi,\rho}]^{-1}
\end{equation}
with the polarization functions
\begin{equation}
  \Pi_\pi(q) = - (c \mathbf{q}_i)^2
                        \frac{F_\pi^2}{f_{\rm det}}
                        \left[ (R^N_i)^2 \, \chi_N +
                        (R^\Delta_i)^2 \, \chi_\Delta +
                        F_g^2 \: \delta g'_2
                        \: \chi_{N} \, \chi_{\Delta} \right]
\end{equation}
and
\begin{eqnarray}
  \Pi_\rho(q) & = &
       - (c \mathbf{q}_i)^2 \: \frac{F_\rho^2}{f_{\rm det}} \:
         \frac{m_\pi^2}{m_\rho^2} \:
       [\: \left(R^N_i \frac{f^\rho_{NN}}{f^\pi_{NN}} \right)^2 \,   \chi_N +
        \left(R^\Delta_i \frac{f^\rho_{N \Delta}}{f^\pi_{N\Delta}} \right)^2 \,
         \chi_\Delta\,   \nonumber \\
              & ~&
       \qquad \qquad \qquad \quad  +\ F_g^2 \: \delta g'_3
                        \: \chi_{N} \, \chi_{\Delta} ]\ .
\end{eqnarray}
The form factors $F_{\pi,\rho}$ and $F_g$, depend on the transferred energy
and momentum $q = (\hbar \omega,c \mathbf{q})$, and are defined in eqs.\
(\ref{eq_Fpi}) and (\ref{eq_Fg}). The relativistic corrections appearing in
(\ref{eq_Vnpn}), (\ref{eq_Vnpd}), (\ref{eq_Vnrn}), and \ref{eq_Vnrd})
are here denoted $R^\alpha$, with
\begin{equation}
  R^\alpha_i(q)^2  =  \frac{ 2m_\alpha c^2}{ m_\alpha c^2 + \sqrt{s_i} }
                      \qquad \alpha = N,\Delta \ .
\end{equation}
The index $i$ denotes that center-of-mass energy $\sqrt{s}$ originates from an
internal vertex, and is thus approximately taken
\begin{equation}
  s_i(q) = (m_N c^2 + \hbar \omega)^2 - (c \mathbf{q})^2\ .
\end{equation}
Furthermore $\mathbf{q}_i$ is the $N \pi$ center-of-mass momentum at an
internal vertex, and is approximated by
\begin{equation}
  \mathbf{q}_i =  \frac{ m_N c^2 } { m_N c^2 + \hbar \omega } \mathbf{q}\ .
\end{equation}
The susceptibilities $\chi_\alpha$ are defined from the Lindhard functions in
eqs.\ (\ref{eq_PhiN}) and (\ref{eq_PhiD}),
\begin{equation}
  \chi_\alpha(q) = {\cal M}_\alpha \left( \frac{f^\pi_{N \alpha}}{m_\pi c^2}
                                   \right)^2
  \Phi^\alpha(q) \qquad \alpha = N,\Delta \qquad
  {\cal M}_\alpha = \left\{ \begin{array}{cc}   4 & \alpha = N \\
                                             16/9 & \alpha = \Delta
  \end{array} \right. \ .
\end{equation}
In the expressions of $D_{\pi,\rho}$ there also appears the renormalization
factor
\begin{equation}
  f_{\rm det}(q) =  1 + F_g^2 \, g_{NN}' \, \chi_{N} +
                           F_g^2 \, g_{\Delta \Delta}' \, \chi_{\Delta} +
                           F_g^4 \: \delta g'_1 \: \chi_{N} \,
                           \chi_{\Delta}\ ,
\end{equation}
and we have used the short hand notation
\begin{eqnarray}
  \delta g'_1   & = & g_{NN}' g_{\Delta \Delta}' - (g_{N \Delta}')^2       \\
  \delta g'_2   & = & (R_\Delta^i)^2 g_{NN}' +
                      (R_N^i)^2 g_{\Delta \Delta}' -
                      2 R_N^i R_\Delta^i g_{N\Delta}'        \\
  \delta g'_3   & = & (R_\Delta^i)^2 g_{NN}'
                      \left( \frac{f^\rho_{N \Delta}}{f^\pi_{N\Delta}}
                      \right)^2             +
                      (R_N^i)^2 g_{\Delta \Delta}'
                      \left( \frac{f^\rho_{NN}}{f^\pi_{NN}} \right)^2  -
                      2 R_N^i R_\Delta^i g_{N \Delta}'
                      \frac{f^\rho_{NN} f^\rho_{N \Delta}}
                           {f^\pi_{NN}  f^\pi_{N \Delta} } \ .
\end{eqnarray}

The quantity $q^2_{\rm eff}$ consists of four terms
\begin{equation}
  q^2_{\rm eff}(34,12) = t_0(34,12) + t_{01}(34,12) +
                 t_{10}(34,12) + t_{11}(34,12)\ ,
\end{equation}
with
\begin{eqnarray}
 t_0(34,12)    & = &  R^{31} \, R^{24} \mathbf{q}_{31}
                                 \cdot \mathbf{q}_{24}\ ,  \\
 t_{10}(34,12) & = &
     - ( \mathbf{q}_{31} \cdot \mathbf{q}_i )
       \frac{R^{31} \, F_g^2}{f_{\rm det}}
       \left[ g'_{N,24} \, R^N_i \, \chi_N +
              g'_{\Delta,24} \, R^\Delta_i \, \chi_\Delta + \right.
              \nonumber \\ & + &  \left.
              \delta g'_1 \: R^{24}_i \, F_g^2
                        \, \chi_{N} \, \chi_{\Delta} \right]\ ,   \\
 t_{01}(34,12) & = & t_{10}(21,43)\ ,             \\
 t_{11}(34,12) & = &
       \frac{F_g^4 \, \mathbf{q}_i^2 }{f_{\rm det}^2}
       \left[ g'_{N,31} \, g'_{N,24} \, (R^N_i)^2 \, \chi_N^2 +
              g'_{\Delta,31} \, g'_{\Delta,24} \, (R^\Delta_i)^2 \,
              \chi_\Delta^2 +
         \right. \nonumber \\ & + & \left.
              ( g'_{N,31} \, g'_{\Delta,24} +
                g'_{\Delta,31} \, g'_{N,24} ) \, R^\Delta_i \,
              R^N_i \, \chi_\Delta  \, \chi_N  +
         \right. \nonumber \\ & + & \left.
              \delta g'_1 \: (g'_{N,31} \, R^{24}_i +
                              g'_{N,24}  \, R^{31}_i)
              R^N_i \, F_g^2  \, \chi_{N}^2 \, \chi_{\Delta}
         \right. \nonumber \\ & + & \left.
              \delta g'_1 \: (g'_{\Delta,31} \, R^{24}_i +
                              g'_{\Delta,24}  \, R^{31}_i)
              R^\Delta_i \, F_g^2  \, \chi_{N} \, \chi_{\Delta}^2
         \right. \nonumber \\ & + & \left.
              (\delta g'_1)^2 \: R^{31}_i \, R^{24}_i \,
              F_g^4  \, \chi_{N}^2 \, \chi_{\Delta}^2 \right]\ ,
\end{eqnarray}
where the relativistic correction at the external $\pi$-$jk$ or $\rho$-$jk$
vertex is taken as
\begin{equation}
  R^{jk}(q)^2  =  \frac{ 2m_{jk} c^2}{ m_{jk} c^2 + \sqrt{s_{jk}} }
\end{equation}
with
\begin{equation}
  s_{jk}(q) = e_k^2 - (c \mathbf{p}_k)^2\ .
\end{equation}
In this paper we only take into account $\pi$ or $\rho$ vertices with either
two nucleons or one nucleon and one $\Delta$, and there is no ambiguity of
which mass to substitute for $m_{jk}$. The notation $R^{jk}_i$ means that the
relativistic correction originates from an internal vertex, but the baryon mass
appearing in $R^{24}_i$ is determined from the baryons at the external $jk$
vertex,
\begin{equation}
  R^{jk}_i(q)^2  =  \frac{ 2m_{jk} c^2}{ m_{jk} c^2 + \sqrt{s_i} } \ .
\end{equation}
The $N \pi$ ($N \rho$) center-of-mass momentum in the external $\pi$-$jk$
($\rho$-$jk$) vertex is given
by
\begin{equation}
  \mathbf{q}_{jk} =
      \frac{m_k c^2}{\hbar \omega + m_k c^2} \mathbf{p}_j - \mathbf{p}_k \ .
\end{equation}
Finally, the quantity $g'_{\rm eff}(34,12)$ is given by
\begin{equation}
  g'_{\rm eff}(34,12) = g'_{31,24} - \frac{F_g^2}{f_{\rm det}}
     [g'_{N,31} g'_{N,24} \chi_N + g'_{\Delta,31} g'_{\Delta,24} \chi_\Delta
      + F_g^2 g'_{31,24} \: \delta g'_1 \: \chi_N \chi_\Delta ] \ .
\end{equation}

Alternatively,
$\bar{M}(34,12)$ can be expressed using an expansion in RPA eigenstates,
\begin{eqnarray}
     \bar{M}^{l,t}(34,12) & = &
     \sum_{\mbox{Re $\omega_\nu^{l,t} > 0$}}
     \left\{ \frac{ h^{l,t}(31;\nu) h^{l,t}(24;\nu) }
                  { \hbar \omega_D - \hbar \omega_\nu^{l,t} + i \eta }
       -     \frac{ h^{l,t}(31;\nu)  h^{l,t}(24;\nu) }
                  { \hbar \omega_D + \hbar \omega_\nu^{l,t} - i \eta }
       \right\} \nonumber \\ & ~ &
      +\ \frac{ f^\pi_{31} f^\pi_{24} }{(m_\pi c^2)^2} F_g^2 g'_{34,12} \ .
\end{eqnarray}
The factors $h^{l,t}(jk;\nu)$ are motivated in eq.\ (\ref{eq_hMotiv}) and can,
using the RPA solution in appendix \ref{sec_RPAsolu},
be explicitly expressed as
\begin{eqnarray}
 h^{l,t}(jk,\nu;\omega,\mathbf{q})  & = &
   \sum_{\alpha=N,\Delta}
   {\cal M}_\alpha \Phi_\alpha(\omega_\nu^{l,t}) x_\alpha(\omega_\nu^{l,t})
    \left[ (\hat{\mathbf{q}}_{jk} \mathbf{\cdot} \hat{\mathbf{q}}_i)
            v_B^{\alpha,jk}(\omega)
    \right. \nonumber \\ & & \left.
            -\ 2 \hbar \omega_{\pi,\rho}
            D_{\pi,\rho}^0(\omega_\nu^{l,t})
            v_{\pi,\rho}^{jk}(\omega) v_{\pi,\rho}^{\alpha}(\omega_\nu^{l,t})
            \right]\ ,
\end{eqnarray}
where the quantities $x_\alpha$ and $v_\pi$ are defined in appendix
\ref{sec_RPAsolu}, and
\begin{equation}
  v_B^{\alpha,jk}(\omega,\mathbf{q}) = g'_{\alpha,jk}
      \frac{f^\pi_{N,\alpha} f^\pi_{N,jk} }{(m_\pi c^2)^2}\
      F_g^2(\omega,\mathbf{q}) \ .
\end{equation}

\newpage
\begin{table}[h]
\begin{tabular}{||l|l|l|l||}
  \hline
        $m_N = 940\ \MeV/c^2$
&       $g'_{NN} = 0.9$
&       $f^{\pi}_{NN} = 1.0$
&       $f^{\rho}_{NN} = 6.2$           \\
        $m_{\Delta} = 1230\ \MeV/c^2$
&       $g'_{N \Delta} = 0.38$
&       $f^{\pi}_{N \Delta } = 2.2$
&       $f^{\rho}_{N \Delta } = 10.5$   \\
        $m_{\pi} = 140\ \MeV/c^2$
&       $g'_{\Delta \Delta} = 0.35$
&       $f^{\pi}_{\Delta \Delta } = 0$
&       $f^{\rho}_{\Delta \Delta } = 0$ \\
        $m_{\rho} = 770\ \MeV/c^2$
&       $\Lambda_g = 1.5\ \GeV$
&       $\Lambda^{\pi} = 1.0\ \GeV$
&       $\Lambda_{\rho} = 1.5\ \GeV$                            \\
  \hline
        $\rho_0 = 0.153\ \fm^{-3}$
&       $V_\Delta - V_N = 25.0\ \MeV$
&       \multicolumn{2}{|l||}{$m_N^*= m_N/[1+0.4049(\rho/\rho_0)]$}  \\
  \hline
\end{tabular}
\caption{ Parameter values used in the numerical calculations.
          \protect\label{tab_param}
}
\end{table}

\newpage

\bfig\caption{}\label{fig_GsimGraph}
Diagrammatic illustration of the microscopic structure
of the Green's function $G^{\rm RPA}$ for a spin-isospin mode.
A given spin-isospin mode (wiggly line)
consists of dressed $\pi$ or $\rho$ mesons (long-dashed)
and dressed baryon-hole excitations.
The center row shows how a dressed meson is built from
the corresponding free meson (short dashed)
and dressed baryon-hole excitations.
The double (solid-dashed) line in the dressed baryon-hole bubble
can represent either a nucleon or a $\Delta$ isobar,
as illustrated in the bottom row,
where a single solid line represents a nucleon
and a double solid line represents a $\Delta$ isobar.
\efig

\bfig\caption{}\label{fig_EffInt}
Diagrammatic representation of the effective spin-isospin
interaction $M(34,12)$.
\efig

\bfig\caption{}\label{fig_DseGraph}
Diagrammatic representations of the $\Delta$ self energy $\Sigma_\Delta$
(left-hand side) and the partial width $\Gamma_\Delta^\nu$ (right-hand side).
\efig

\bfig\caption{}\label{fig_dsdcos0}
Differential cross sections in vacuum for the process p+p $ \rightarrow
\Delta^{++} $+n, in the pp center-of-mass system.
In ($a$) is shown $d\sigma/d\cos(\theta)$ for $\sqrt{s} = 2.314$ GeV,
while ($b$) shows $d\sigma/dt$ for $ \sqrt{s} = 2.513  $ GeV.
The solid curve is the full cross section, $d\sigma/d\cos(\theta_{\rm cm})$,
the dashed curves are the direct and exchange contributions
from the spin-longitudinal channel,
and the dash-dotted curves are the contributions
from the spin-transverse channel.
There are also contributions from mixed direct and exchange terms;
these are included in the full cross section
but not displayed since they are small.
The data points originate from refs.\ \cite{ppDn1,ppDn2},
but have here been estimated from figs.\ 5 and 8  of ref.\ \cite{ppnD}.
\efig

\bfig\caption{}\label{fig_stot0}
The total cross section in vacuum for the process
p+p $\rightarrow\Delta^{++}$+n,
in the pp center-of-mass system as a function of $\sqrt{s}$.
The solid curve is our calculation, the dashed is the parameterization from
VerWest-Arndt \cite{VerWA}, and the dash-dotted line is a simple
parameterization often used in BUU calculations \cite{DasGupta}.
The data points were estimated from fig.\ 2 of ref.\ \cite{VerWA}
and can be found in references therein.
\efig

\bfig\caption{}\label{fig_Vspread}
The quantity $[\Gamma_\Delta^{\rm tot} - \Gamma_\Delta^{\rm free} + \delta
\Gamma_\Delta^{\rm Pauli}]/2$,
calculated at the density $0.75 \rho_N^0$
and for two different values of the parameter $g'_{\Delta\Delta}$;
the $\Delta$ energy and momentum are related to the pion kinetic energy $T_\pi$
by eq.\ (\ref{eq_Tpi}).
The displayed quantity may be compared with the spreading potential
in ref.\ \cite{Hirata}.
The empirical points were originally determined in ref.\ \cite{Hirata},
but have here been estimated from fig.\ 12 of ref.\ \cite{Oset}.
\efig

\bfig\caption{}\label{fig_DispT0r10}
The dispersion relations for the spin-isospin modes in infinite nuclear matter
at normal nuclear density and zero temperature.
Parts ($a$) and ($b$) show the real part of $\hbar\omega_\nu$
in the spin-longitudinal and spin-transverse channel, respectively.
The corresponding imaginary parts are presented in ($c$) and ($d$).
The non-collective modes are shown by solid curves,
while collective modes are represented by either
a dot-dashed curve ($\tilde{\pi}_1$ or $\tilde{\rho}_1$),
a dot-dot-dashed curve ($\tilde{\pi}_2$  or $\tilde{\rho}_2$),
or a dot-dot-dot-dashed curve (see text).
As a reference, the free pion dispersion relation
$\hbar \omega_\pi(q) = [(m_\pi c^2)^2 + (cq)^2]^{1/2}$
is included as a dotted curve.
\efig

\bfig\caption{}\label{fig_AmpT0r10}
Squared amplitudes for the spin-isospin modes displayed in fig.\
\ref{fig_DispT0r10}$a$, for $q = 300$ MeV/c:
the individual $NN^{-1}$ components ($a$),
the individual $\Delta N^{-1}$ components ($b$),
and the pion component together with
the sum of all $NN^{-1}$ and $\Delta N^{-1}$ components ($c$).
Moreover,
for the lower collective mode $\tilde{\pi}_1$ is shown the $q$ dependence of
the pion component and the total $NN^{-1}$ and $\Delta N^{-1}$ components
($d$).
\efig

\bfig\caption{}\label{fig_Dispr20andT25}
Same as fig.\ \ref{fig_DispT0r10}$a$,
but for the density $\rho_N = 2\rho_N^0$
and the temperature $T=0\ \MeV$ in ($a$)
and for $\rho_N = \rho_N^0$ and $T=25\ \MeV$ in ($b$).
\efig

\bfig\caption{}\label{fig_GamTot}
The total $\Delta$ width $\Gamma_\Delta^{\rm tot}$
calculated either with $\Gamma_{\Delta N^{-1}}=0$ in eq.\ (\ref{eq_PhiD}) ($a$)
or with $\Gamma_{\Delta N^{-1}}$ included self-consistently ($b$),
for a variety of scenarios:
$\rho_N=0$, $T=0$ (dotted),
$\rho_N = \rho_N^0$, $T=0$ (solid),
$\rho_N = 2 \rho_N^0$, $T=0$ (short-dashed),
$\rho_N = \rho_N^0$, $T=25$ MeV (long-dashed), and
$\rho_N = 2 \rho_N^0$, $T=25$ MeV (dash-dotted).
\efig

\bfig\caption{}\label{fig_GamPar}
The total $\Delta$ width $\Gamma_\Delta^{\rm tot}$ and its partial
contributions from different spin-isospin modes. The solid curve represents the
total width, the long-dashed line is the contribution from the non-collective
$N N^{-1}$ modes, the short-dashed line is the contribution from the
non-collective $\Delta N^{-1}$ modes, the dot-dashed line is the contribution
from the lower pionic mode $\tilde{\pi}_1$,
and  the dot-dot-dashed line is the contribution from
the upper pionic mode $\tilde{\pi}_2$.
The calculations have been made with
$\Gamma_{\Delta N^{-1}}=0$ in eq.\ (\ref{eq_PhiD}) and for:
$\rho_N = \rho_N^0$, $T=0$ ($a$),
$\rho_N = 2 \rho_N^0$, $T=0$ ($b$),
$\rho_N = \rho_N^0$, $T=25$ MeV ($c$), and
$\rho_N = 2 \rho_N^0$, $T=25$ MeV ($d$).
The error bars indicate the estimated uncertainty associated with
the classification procedure (see text).
\efig

\bfig\caption{}\label{fig_GamParSC}
Similar to fig.\ \ref{fig_GamPar},
but with $\Gamma_{\Delta N^{-1}}$ included self-consistently
in eq.\ (\ref{eq_PhiD}).
\efig

\bfig\caption{}\label{fig_dsdcosNN}
Differential cross section $d\sigma/d\cos(\theta_{\rm cm})$
for the process p+p$\rightarrow \Delta^{++}$ + n in the nuclear medium,
in the pp center-of-mass system at the energy $\sqrt{s}=2314$ MeV,
calculated either with $\Gamma_{\Delta N^{-1}}=0$ ($a$)
or with $\Gamma_{\Delta N^{-1}} =\Gamma_\Delta^{\rm tot}$ ($b$),
for the following scenarios:
$\rho_N= 0$, $T=0$ (dotted),
$\rho_N = \rho_N^0$, $T=0$ (solid),
$\rho_N = 2 \rho_N^0$, $T=0$ (short-dashed),
$\rho_N =\rho_N^0$, $T=25$ MeV (long-dashed), and
$\rho_N = 2 \rho_N^0$, $T=25$ MeV (dash-dotted).
\efig

\bfig\caption{}\label{fig_sTotNN}
Total cross section $\sigma(\sqrt{s})$ for the process
p+p $\rightarrow \Delta^{++}$ + n
for either $\Gamma_{\Delta N^{-1}}=0$ in eq.\ (\ref{eq_PhiD}) ($a$)
or with $\Gamma_{\Delta N^{-1}}$ included  self-consistently ($b$).
The notation is the same as in fig.\ \ref{fig_dsdcosNN}.
\efig

\bfig\caption{}\label{fig_sTotND}
Same as in fig.\ \ref{fig_sTotNN},
but for the reverse process n +$\Delta^{++} \rightarrow$ p + p,
for a $\Delta$ with mass $m=1230$ MeV/c$^2$,
with either $\Gamma_{\Delta N^{-1}}=0$ in eq.\ (\ref{eq_PhiD}) ($a$)
or with $\Gamma_{\Delta N^{-1}}$ included self-consistently ($b$).
\efig

\bfig\caption{}\label{fig_sNnu}
Total cross section $\sigma(\sqrt{s}=m)$
for the process p+$\tilde{\pi}_j \rightarrow \Delta^{++}$
with $\Gamma_{\Delta N^{-1}}$ included self-consistently
and for the following scenarios:
$\rho_N= 0$, $T=0$ (dotted),
$\rho_N = \rho_N^0$, $T=0$ (solid),
$\rho_N = 2 \rho_N^0$, $T=0$ (short-dashed),
$\rho_N =\rho_N^0$, $T=25$ MeV (long-dashed), and
$\rho_N = 2 \rho_N^0$, $T=25$ MeV (dash-dotted).
Parts ($a$) and ($b$) are for the lower pionic mode $\tilde{\pi}_1$,
while parts ($c$) and ($d$) are for the upper pionic mode $\tilde{\pi}_2$.
The open squares represent
the empirical total cross section for $\pi^+$p scattering
and are estimated from fig.\ 2.2 in ref.\ \cite{EricsonWeise}.
The factor $\rho_3(m^2)$ in eq.\ (\ref{eq_rhoDmass})
was in ($a$) and ($c$) calculated from the free $\Delta$ width,
while the total width in nuclear matter was used in ($b$) and ($d$).
\efig

\bfig\caption{}\label{fig_GamParNPB}
Same as in fig.\ \ref{fig_GamPar}$a$ and \ref{fig_GamPar}$c$,
but the intermediate nucleon in the $\Delta$ decay is not Pauli blocked.
\efig

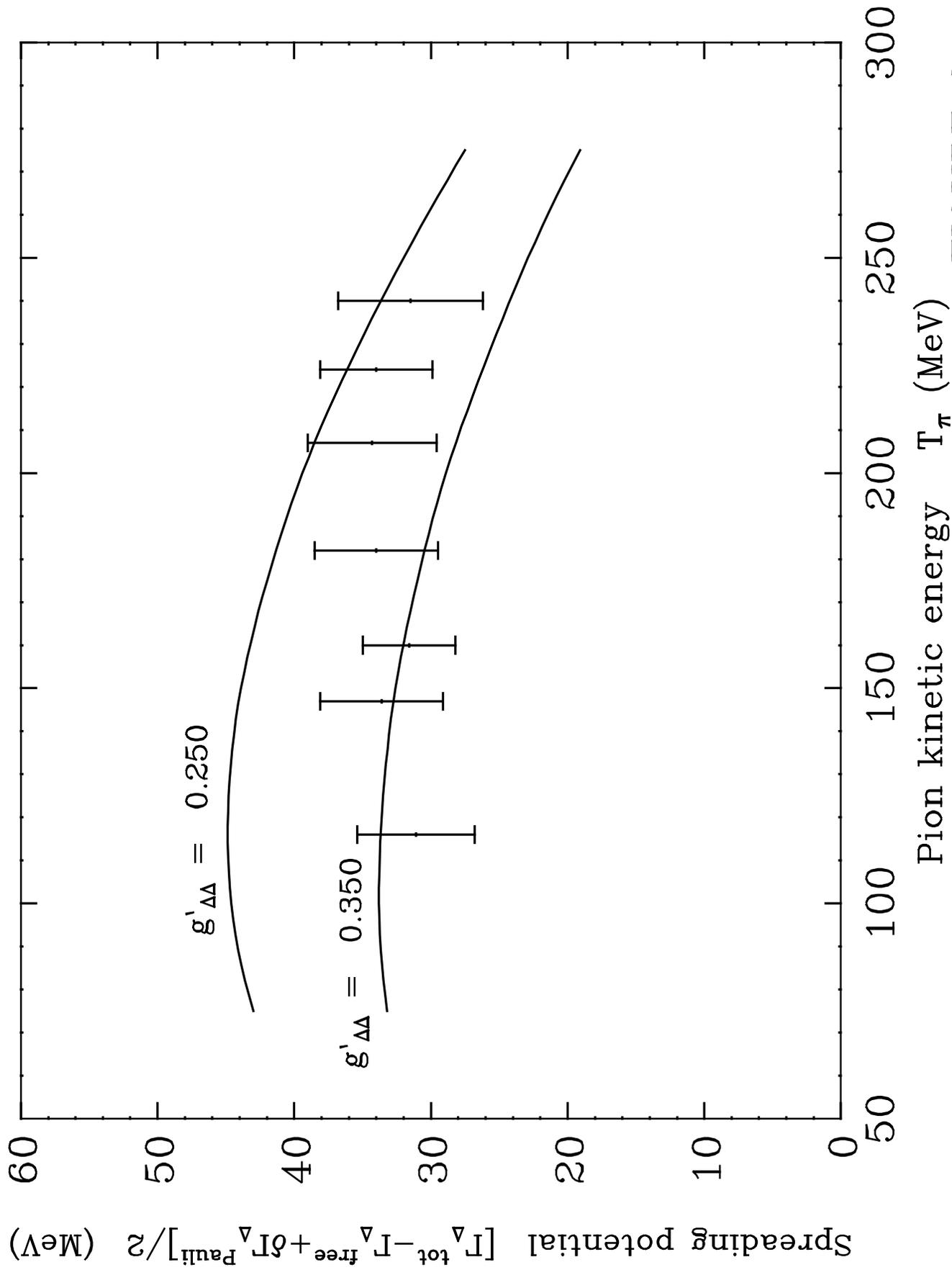
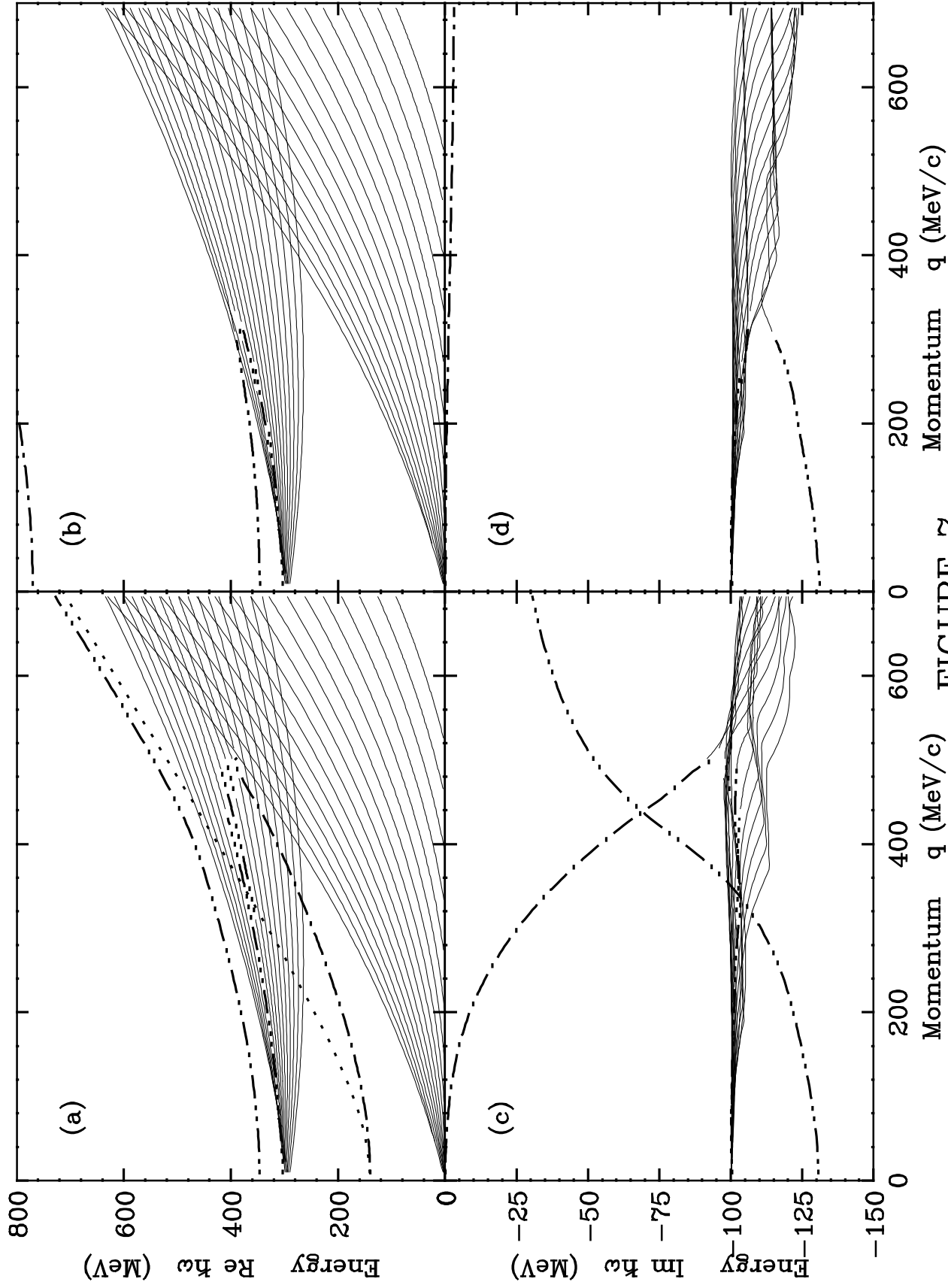
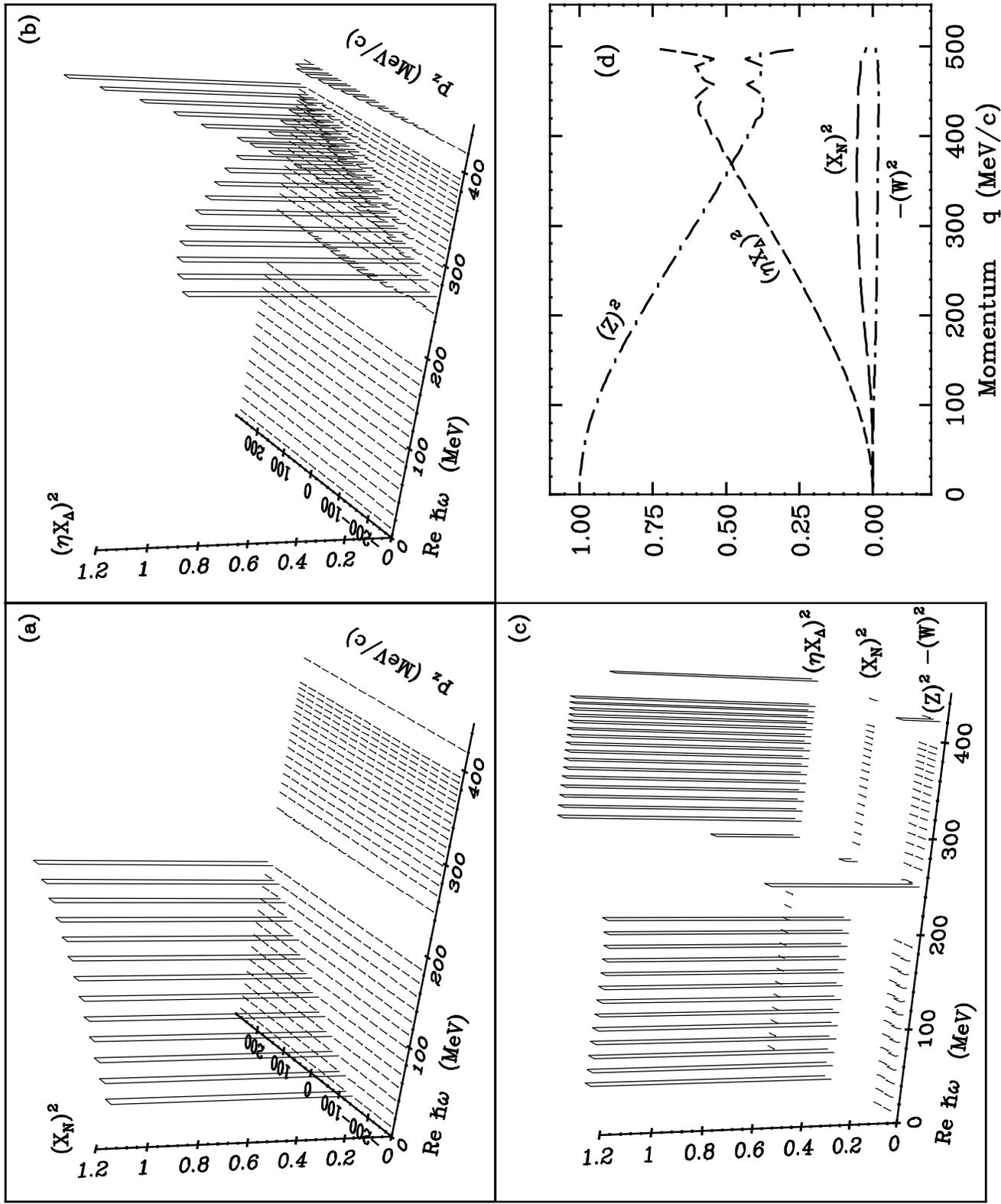
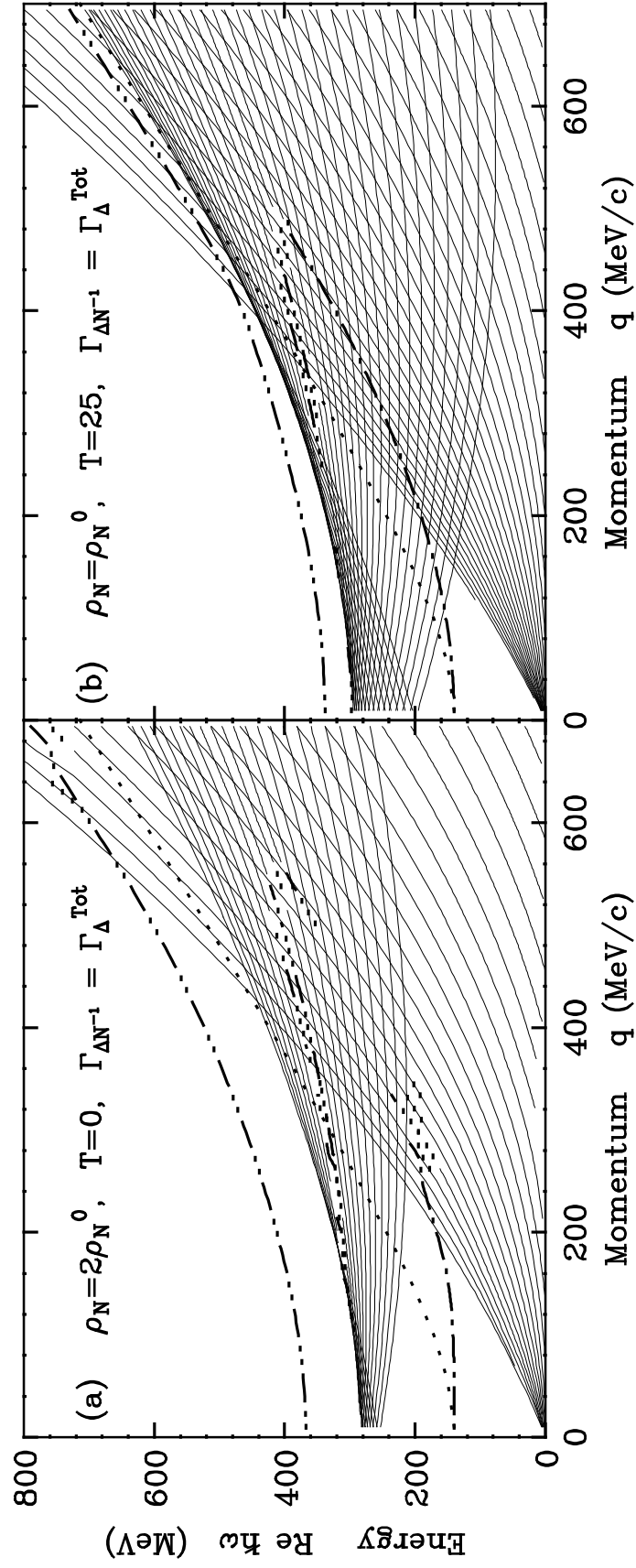
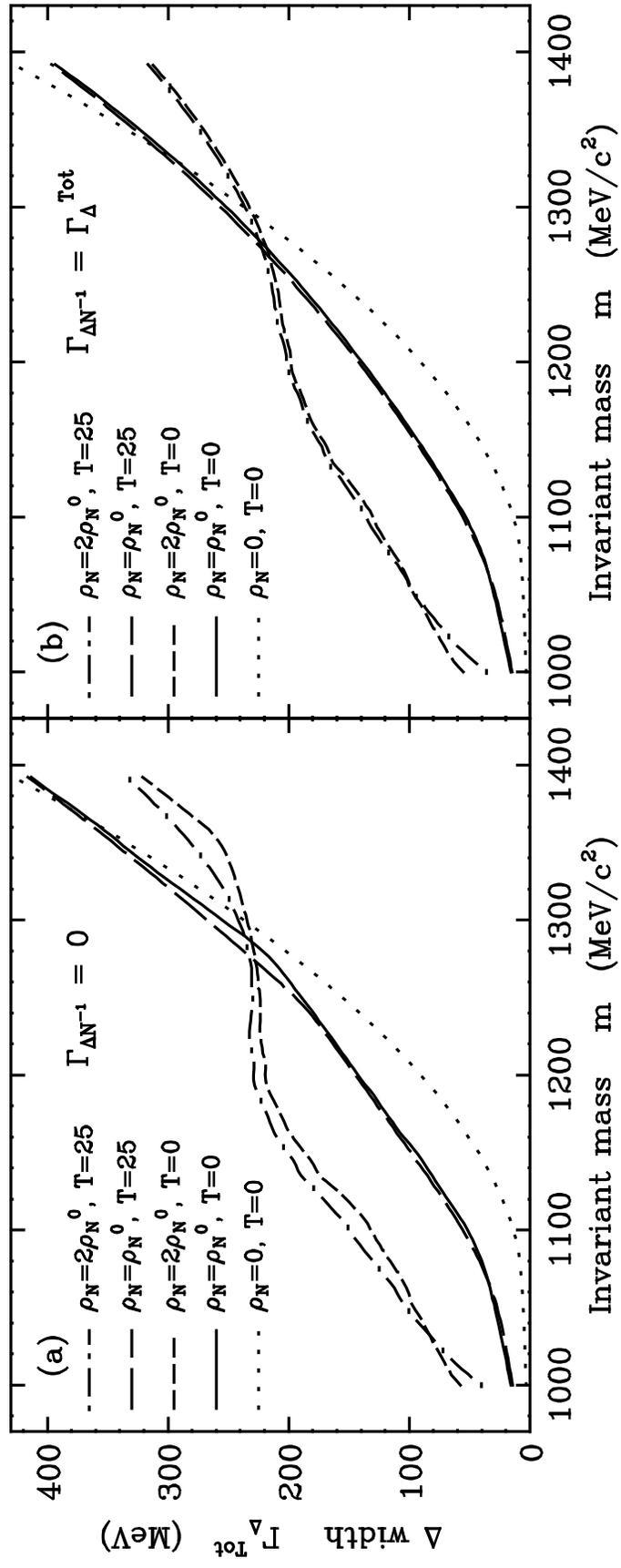
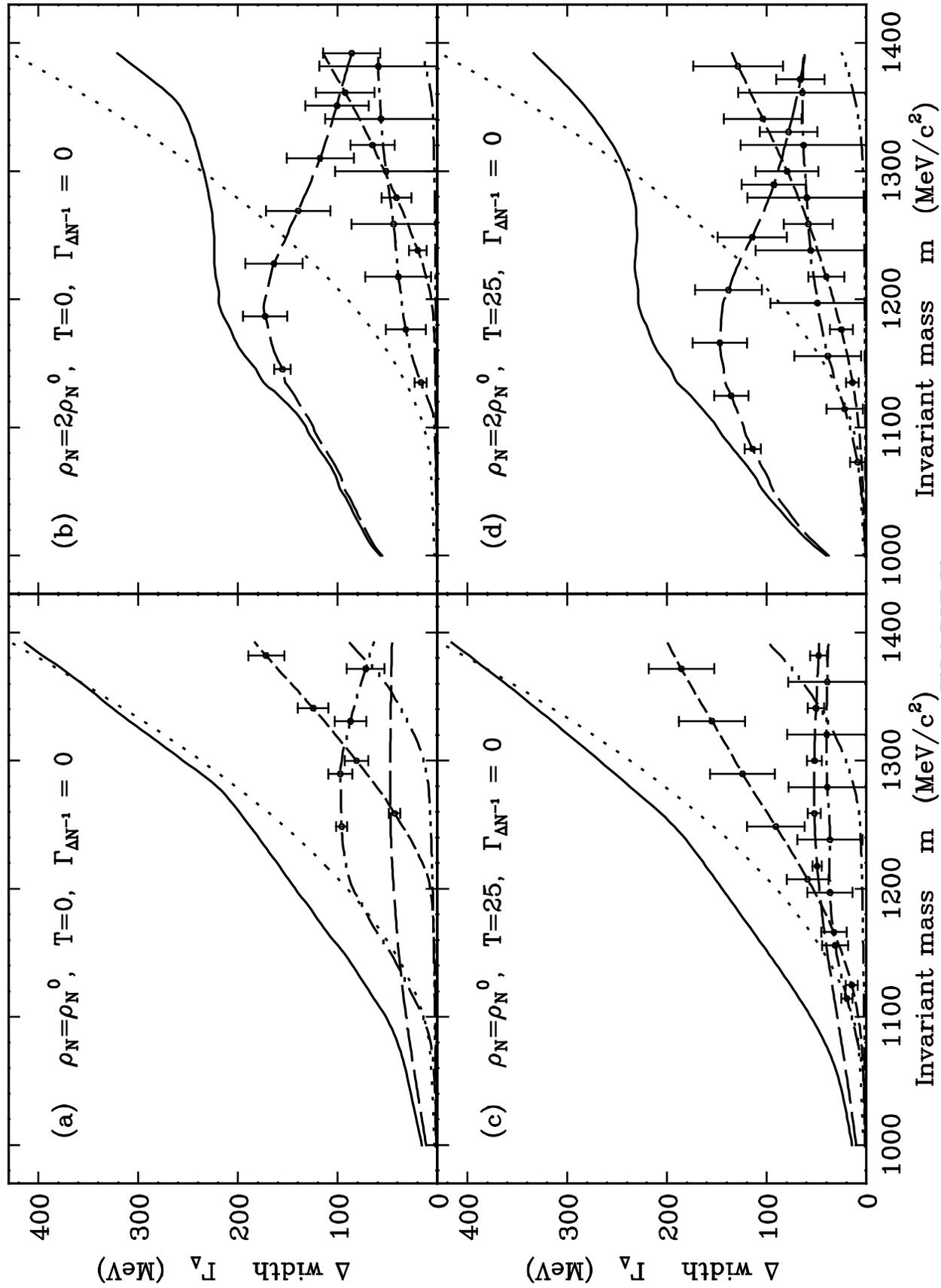
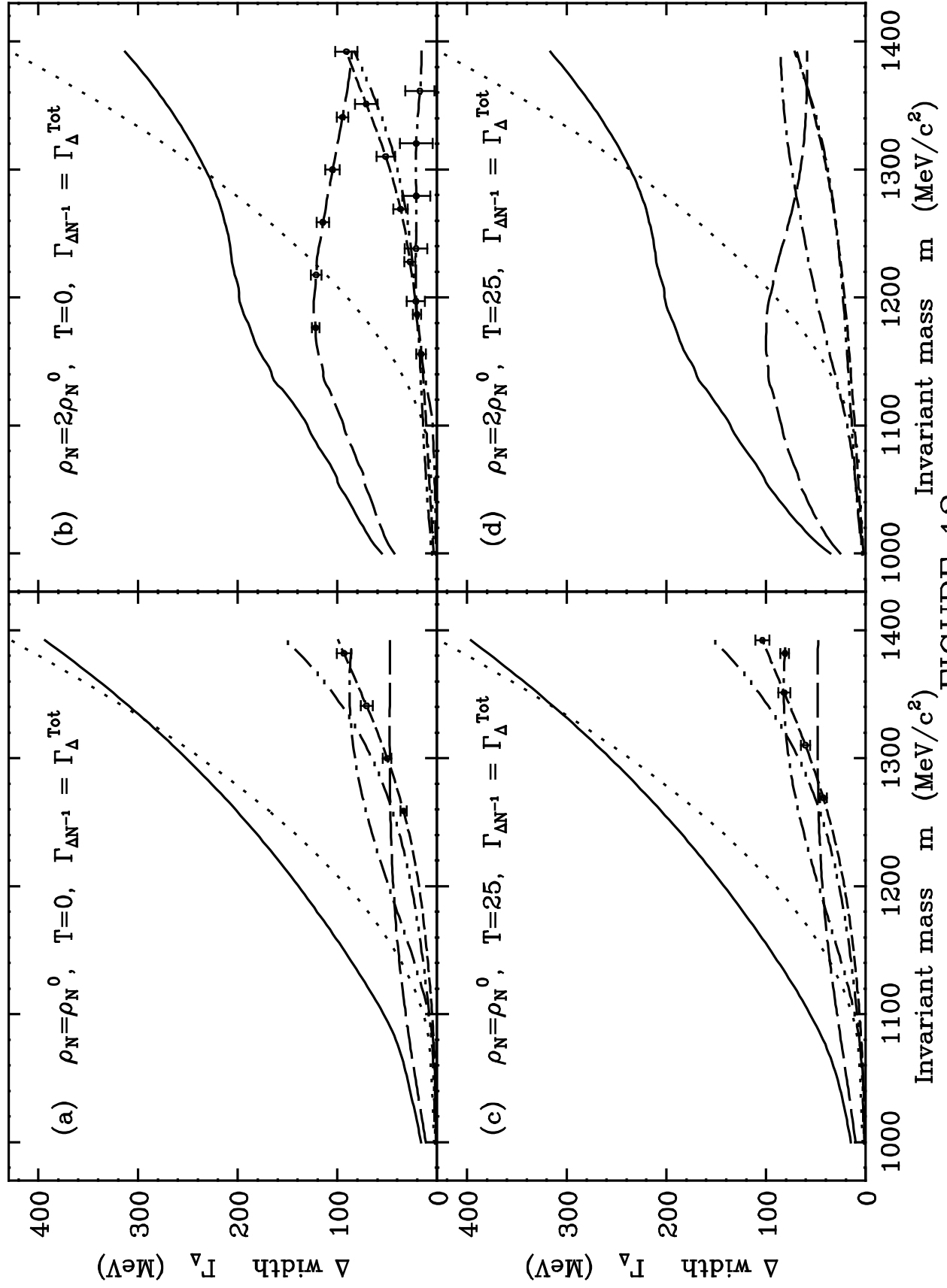
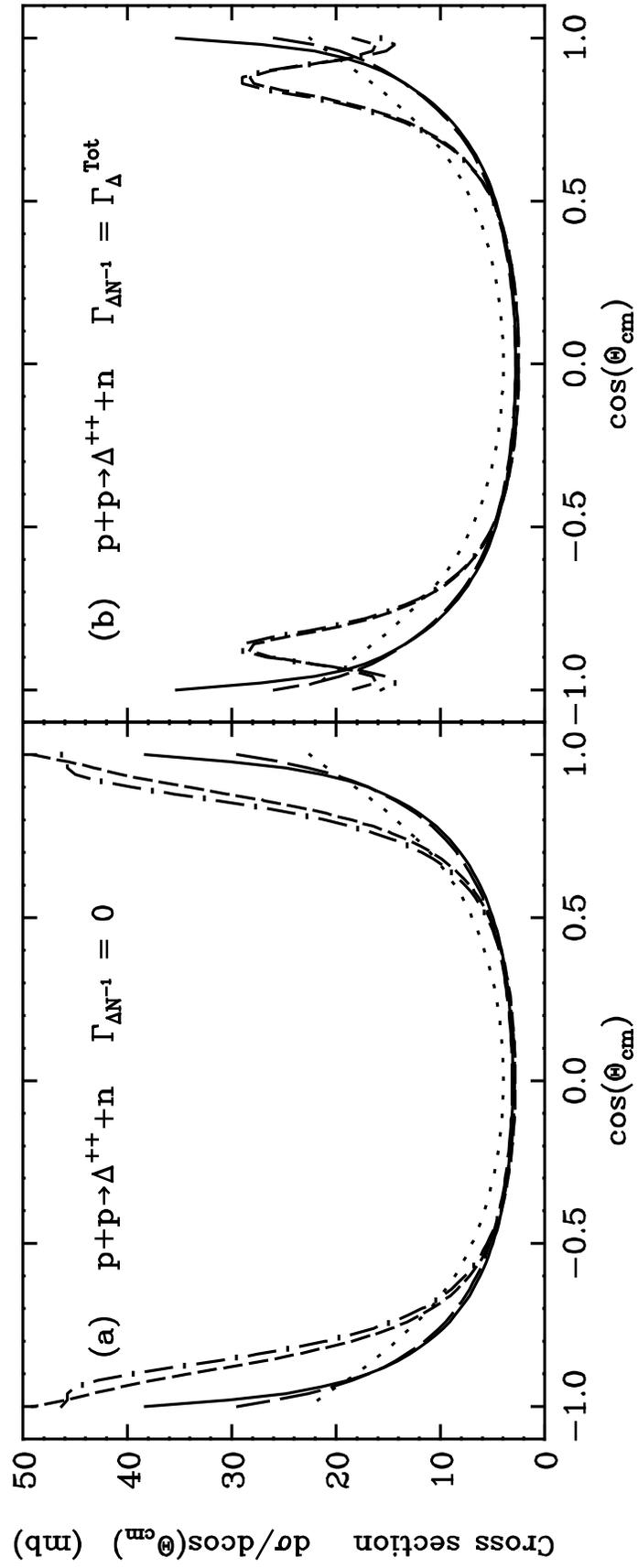
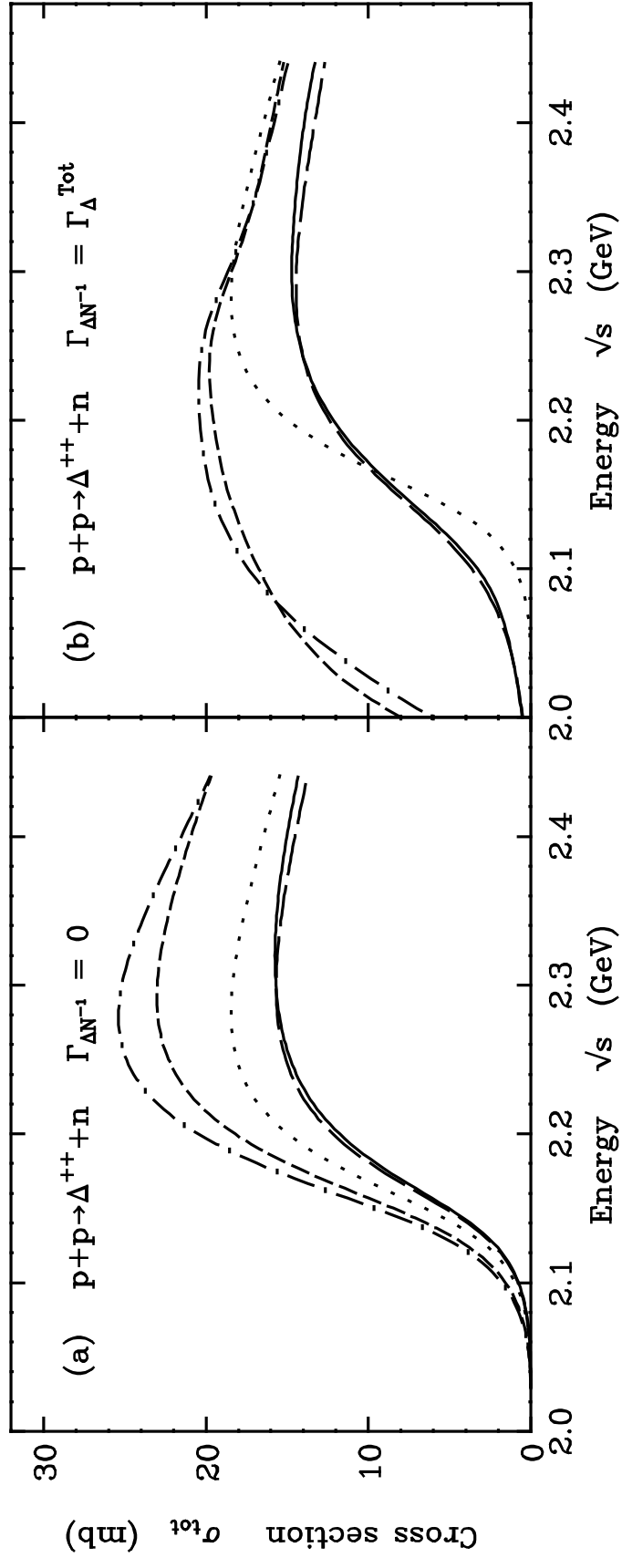
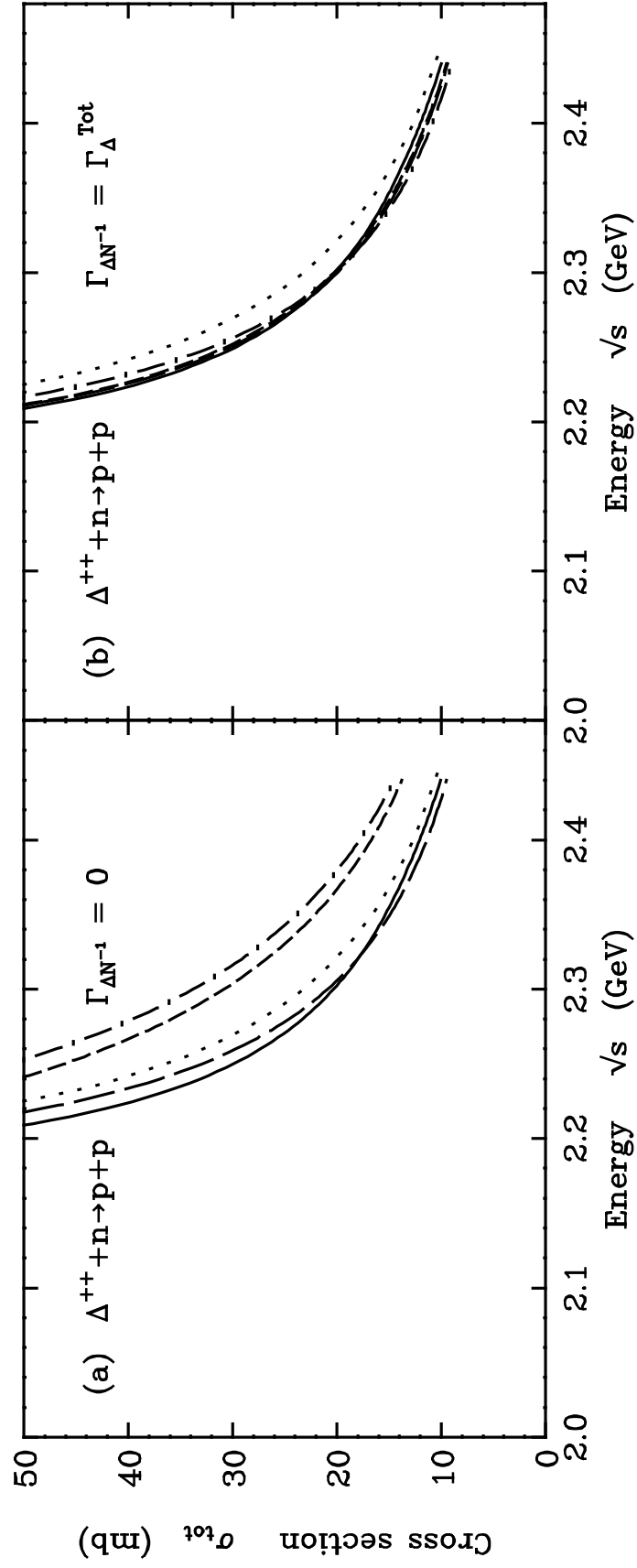
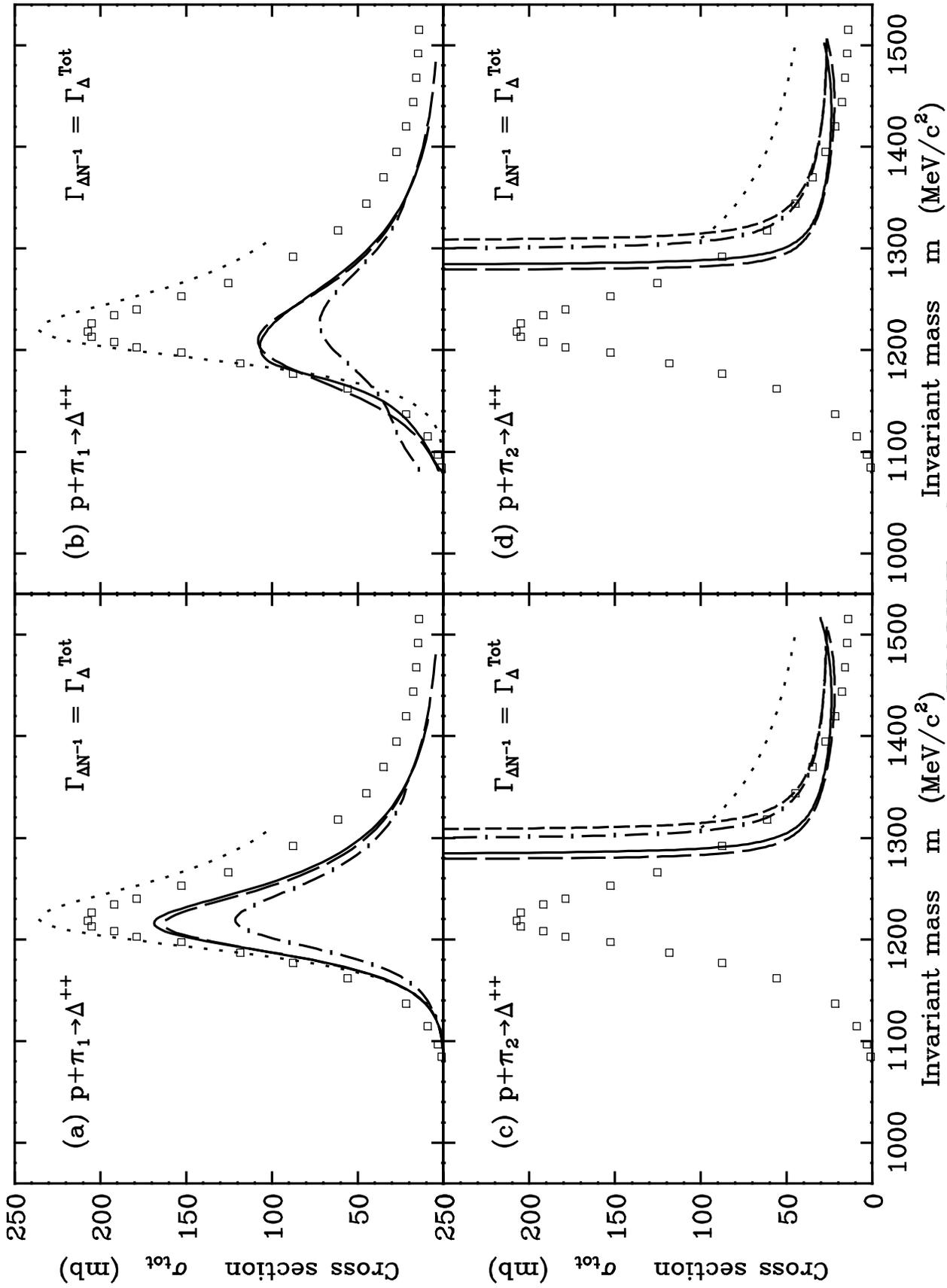
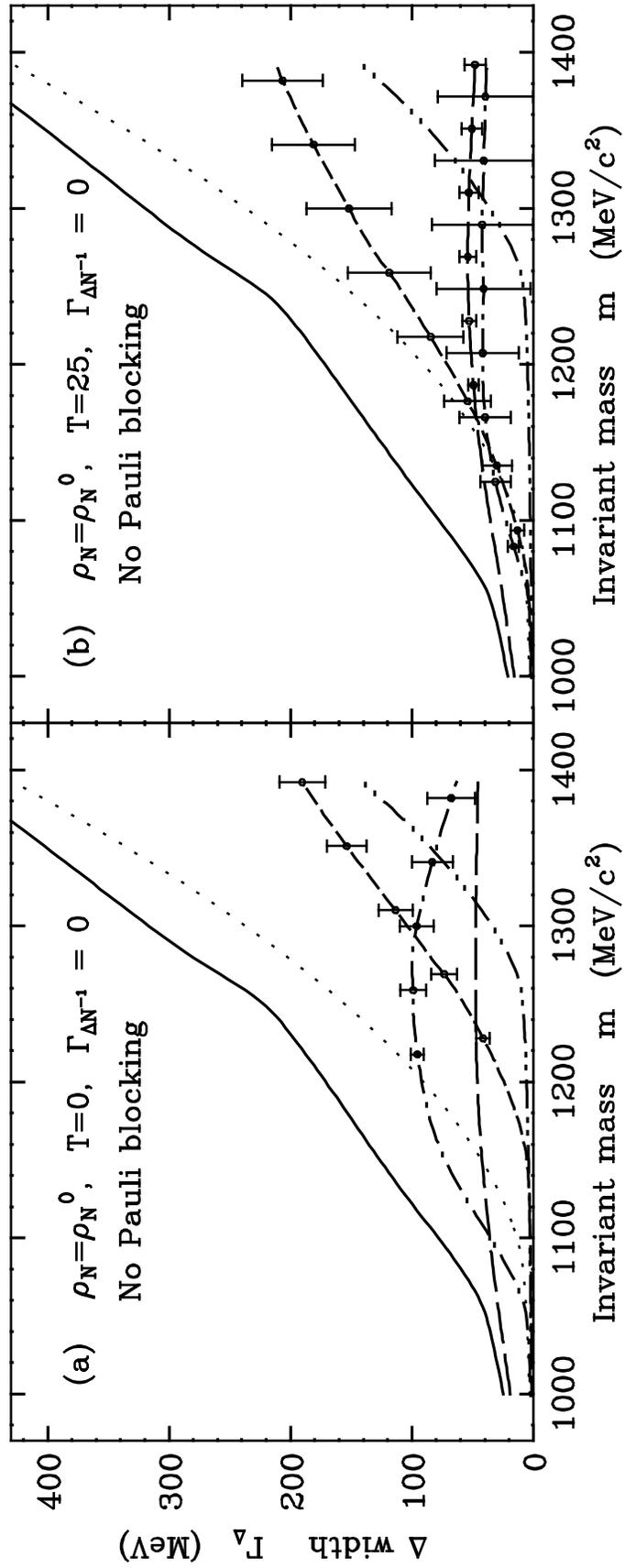
\end{document}